\documentclass[12pt]{article}
\pdfoutput=1
\usepackage{jheppub}
\usepackage{amsmath}
\usepackage{bm}
\usepackage{slashed}
\usepackage{graphicx}

\newlength{\dummysp}
\newcommand{\beq}{\begin{eqnarray}}
\newcommand{\eeq}{\end{eqnarray}}

\newcommand{\e}{{\epsilon}}

\newcommand{\gappeq}{\mathrel{\rlap {\raise.5ex\hbox{$>$}}
{\lower.5ex\hbox{$\sim$}}}}
\newcommand{\lappeq}{\mathrel{\rlap{\raise.5ex\hbox{$<$}}
{\lower.5ex\hbox{$\sim$}}}}

\newcommand{\ben}{\begin{enumerate}}
\newcommand{\een}{\end{enumerate}}

\newcommand{\bit}{\begin{itemize}}
\newcommand{\eit}{\end{itemize}}

\def\[{\left [}
\def\]{\right ]}
\def\({\left (}
\def\){\right )}

\title{Gauging the Standard Model $1$-form symmetry via gravitational instantons}
   
 \author{Mohamed M. Anber}
  
\affiliation{Centre for Particle Theory, Department of Mathematical Sciences, Durham University, South Road, Durham DH1 3LE, UK}

\emailAdd{mohamed.anber@durham.ac.uk}  

\abstract{

{\flushleft{W}}e investigate the fate of the Standard Model (SM) $\mathbb Z_6^{(1)}$ electric  $1$-form global symmetry in the background of gravitational instantons, focusing on Eguchi--Hanson (EH) geometries. We show that EH instantons support quantized $\mathbb Z_6^{(1)}$  fluxes localized on their $S^2$ bolt, inducing fractional topological charge without backreacting on the geometry. The requirement that quark and lepton wavefunctions be globally well-defined under parallel transport imposes boundary conditions, removing ill-defined fermion zero modes; the surviving spectrum is confirmed by an explicit solution of the Dirac equation and by the Atiyah--Patodi--Singer index theorem.  The Euclidean path integral in the EH background can be interpreted as a transition amplitude from an entangled state between two identical halves of space to the vacuum. Summing over all $\mathbb{Z}_6^{(1)}$ flux sectors in the path integral gauges the SM $1$-form symmetry; thus,  it cannot persist as an exact global symmetry in the semiclassical limit of gravity. We further show that these fluxes induce baryon- and lepton-number violating processes, which are exponentially suppressed due to the smallness of the hypercharge coupling constant. 
}

\begin{document}

\maketitle

\flushbottom

%%%%%%%%%%%%%%%%%%%%%%%%%
\section{Introduction}
%%%%%%%%%%%%%%%%%%%%%%%%%

Generalized global symmetries \cite{Gaiotto:2014kfa} have advanced our understanding of various aspects of quantum field theory, particularly by providing a rigorous mathematical foundation for long-standing ideas and offering new, sometimes exact, nonperturbative statements of what to expect, especially when systems become strongly coupled; see the reviews \cite{Brennan:2023mmt,Shao:2023gho}. However, the simple old textbook notion of symmetries acting at a point, now extended to higher-form and noninvertible symmetries, poses more puzzles and questions when gravity is taken into account. It is expected that global symmetries are absent in quantum theories of gravity \cite{Hawking:1975vcx,Zeldovich:1976vq,Harlow:2018tng}. The swampland program puts this statement into scrutiny by seriously examining the possible class of low-energy effective field theories that are compatible with UV-complete theories of gravity; see \cite{Brennan:2017rbf,Palti:2019pca,Grana:2021zvf} for reviews.

The Standard Model (SM) of particle physics is believed to be the low-energy effective field theory of some, yet-to-be-constructed, UV-complete theory of gravity. However, in its present form with the known matter content, the SM admits a global electric $\mathbb{Z}_6^{(1)}$ $1$-form symmetry~\cite{Tong:2017oea,Wan:2019gqr,Anber:2021upc}. 
It is conceivable that this symmetry is only approximate at the energy scales probed so far in terrestrial accelerators, and that as-yet-undiscovered heavy particles could explicitly break it at higher energies \cite{Alonso:2024pmq,Li:2024nuo,Koren:2025utp}. Another possibility is that there exists a ``desert'' between the electroweak and Planck scales, or that any additional heavy degrees of freedom preserve $\mathbb{Z}_6^{(1)}$, or at least a nontrivial subgroup thereof.

Even if the low-energy $1$-form symmetry is not explicitly broken by new particles, quantum gravity considerations strongly suggest that it cannot remain exact. A UV-complete theory must either gauge the symmetry---that is, promote it to part of the dynamical gauge structure---or provide dynamical objects (or gravitational processes) that explicitly violate it. 
Several mechanisms have been proposed for how quantum gravity forbids exact higher-form global symmetries. For instance, Wilson lines charged under $1$-form symmetries may terminate on dynamical branes, thereby invalidating the symmetry; such objects arise naturally in string theory \cite{Polchinski:1995mt}. Also, black hole physics indicates that exact $1$-form global symmetries are inconsistent. Indeed, a line operator that appears admissible in the low-energy theory would correspond to a charged black hole, but the consistency of black hole evaporation rules out exact conservation of such charges, forcing the symmetry to be either gauged or explicitly broken \cite{Banks:2010zn}.  Furthermore, gravitational instantons and wormholes can induce processes that violate charge conservation~\cite{Kallosh:1995hi,Hsin:2020mfa}. See \cite{Yonekura:2020ino,Rudelius:2020orz,Heidenreich:2021xpr,Chen:2020ojn,McNamara:2021cuo} for a sample list of recent works on the violation of global symmetries in quantum gravity.

In this work, we investigate means of gauging the SM
$\mathbb{Z}_{6}^{(1)}$ $1$-form symmetry in the Eguchi--Hanson (EH) background, 
within the framework of the semiclassical Euclidean path integral. 
Our analysis treats the EH space purely as a Euclidean classical background, 
without addressing fluctuations of the spacetime itself. 
Although this approach does not resolve fundamental questions of full quantum gravity, 
it nevertheless provides valuable lessons. 
In the remainder of this introduction, we summarize the main points of our work, 
while deferring the technical details to the body of the text.

EH instantons are (anti-)self-dual solutions to the Euclidean Einstein equations with vanishing cosmological constant \cite{Eguchi:1978xp,Eguchi:1978gw}. These geometries are said to be asymptotically locally Euclidean (ALE), since instead of approaching the 3-sphere $S^3$ at large radial distance $r \rightarrow \infty$ (which is a globally Euclidean space), they asymptotically approach the Lens space $S^3/\Gamma$ (only locally Euclidean), where $\Gamma$ is a finite discrete subgroup of $SU(2)$. In their original formulation, $\Gamma$ was taken $\mathbb{Z}_2$, so that the asymptotic boundary becomes $S^3/\mathbb{Z}_2 \equiv \mathbb{RP}^3$, the real projective 3-space. The (anti)self-dual ALE metrics are the best understood class of gravitational instantons; see the reviews \cite{Eguchi:1980jx,Dunajski:2024pkf}. 

Pure EH instantons (i.e., in the absence of gauge fields) possess self-dual curvature, which implies a vanishing Ricci tensor and Ricci scalar. Consequently, they contribute zero to the Einstein--Hilbert action. Topologically, their second cohomology group is 
$
H^2(\text{EH},\mathbb{Z}) \;\cong\; \mathbb{Z},
$
indicating that EH spaces can support nontrivial fluxes associated with $1$-form symmetries. Geometrically, this arises because the EH manifold contains an $S^2$ bolt, located at radial distance $r=a$, through which such fluxes can thread. The scale $a$ is taken to be constant in a semi-classical treatment. 

A key feature of the EH geometry is that it admits a unique normalizable harmonic self-dual $2$-form ${\cal K}$. The normalizability condition ensures that ${\cal K}$ is localized near the bolt, while self-duality guarantees that the associated energy--momentum tensor vanishes identically. As a result, turning on fluxes supported on ${\cal K}$ does not backreact on the background geometry, making such configurations stable semiclassical backgrounds. 
One may therefore consistently turn on a $U(1)$ gauge field with field strength
\begin{equation}\nonumber
F = -2\pi\, {\cal C}\, {\cal K}, 
\end{equation}
where the flux is localized at the bolt. Flux quantization requires 
\begin{equation}\nonumber
\frac{1}{2\pi}\int_{S^2} F = -{\cal C} \;\in\; \mathbb{Z}\,,
\end{equation}
so the integer ${\cal C}$ labels distinct flux sectors supported by the EH background. 
Moreover, a small deformation $S'^{2}$ of the bolt $S^2$ intersects itself transversely, leading to a nontrivial $U(1)$ topological charge. Explicitly, one finds
\begin{equation}\nonumber
Q =\frac{1}{8\pi^2}\int_{\text{EH}} F \wedge F 
  = \frac{{\cal C}^2}{4}\,,
\end{equation}
thus, the flux threading the bolt induces a fractional topological charge, quadratic in the integer 
${\cal C}$. Unlike the pure EH space (without fluxes), now the self-dual gauge fields come with non-zero actions. In general, $S=\frac{8\pi^2|Q|}{g^2}$, where $g$ is the gauge field coupling constant.

The Standard Model (SM) gauge group is based on the Lie algebra 
$\mathfrak{su}(3)\times \mathfrak{su}(2)\times \mathfrak{u}(1)$, 
and, as mentioned above, it admits a global $\mathbb{Z}_6^{(1)}$ $1$-form symmetry. 
If this symmetry is not explicitly broken all the way up to the Planck scale, 
one natural way to eliminate the global symmetry is to gauge it. 
In this case, the true gauge group of the SM is
\begin{equation}\nonumber
\frac{SU(3)\times SU(2)\times U(1)}{\mathbb{Z}_6}\,.
\end{equation}
Operationally, gauging the symmetry amounts to turning on background fluxes of the 
$\mathbb{Z}_6^{(1)}$ $1$-form symmetry and summing over all such flux sectors in the path integral. 
Concretely, this is achieved by introducing fractional fluxes along the Cartan subalgebras of 
$SU(3)\times SU(2)$; consistency of the theory then requires these fractional fluxes to be 
accompanied by compensating fractional fluxes in the $U(1)$ hypercharge sector. Thus, we have at the bolt
\begin{eqnarray}\nonumber
\int_{S^2}F_{(3)}&=&-2\pi m_{(3)}\,\mathrm{diag}\!\left(\tfrac{2}{3},-\tfrac{1}{3},-\tfrac{1}{3}\right)\;\in\;2\pi\mathbb{Z}_3\,, \\ \nonumber
\int_{S^2}F_{(2)}&=&-2\pi m_{(2)}\,\mathrm{diag}\!\left(\tfrac{1}{2},-\tfrac{1}{2}\right)\;\in\;2\pi\mathbb{Z}_2\,,\\\nonumber
\int_{S^2}F_{(1)}&=&-2\pi\left({\cal C}+\tfrac{3m_{(2)}+2m_{(3)}}{6}\right)\;\in\;2\pi\mathbb{Z}_6\,,
\end{eqnarray}
where the subscripts $(1)$, $(2)$, $(3)$ denote the field strength of the corresponding gauge group, and the diagonal matrices are proportional to the Cartan generators. Here, $m_{(2)}$ and $m_{(3)}$ are integers that determine the amount of the $\mathbb Z_6^{(1)}$ flux. The topological charges corresponding to these fluxes are also fractional
\begin{eqnarray}\nonumber
Q_{(3)}=\frac{m_{(3)}^2}{6}\,, 
\qquad Q_{(2)}=\frac{m_{(2)}^2}{8}\,, 
\qquad Q_{(1)}=\frac{1}{4}\left({\cal C}+\frac{3m_{(2)}+2m_{(3)}}{6}\right)^2\,.
\end{eqnarray}

Embedding quarks and leptons in the background of EH instantons with $\mathbb{Z}_6^{(1)}$ flux supported at the bolt requires additional consistency checks. In particular, one must ensure that the fermions are globally well-defined. A spinor must be continuous along any contractible path connecting two points in the manifold, which imposes nontrivial boundary conditions near the bolt in the presence of gauge fields. These boundary conditions encode topological data that is transmitted to the asymptotic boundary $\mathbb{RP}^3$. For a Dirac spinor $\Psi$ in the presence of a $U(1)$ gauge field, it is sufficient to impose the boundary condition near the bolt:
\begin{equation}\label{boltBC}\nonumber
\Psi(r\approx a,\theta,\varphi,\psi+2\pi) = (-1)^{\cal C}\,\Psi(r\approx a,\theta,\varphi,\psi)\,.
\end{equation}
This condition transmits information about the bulk gauge field to the asymptotic boundary. Consequently, at the boundary $\mathbb{RP}^3$, the spinor satisfies
\begin{equation}\label{boundaryBC}\nonumber
\Psi(r\to \infty, \theta, \varphi, \psi+2\pi) = (-1)^{\mathcal{C}+1}\, \gamma_5\, \Psi(r\to\infty, \theta, \varphi, \psi)\,.
\end{equation}
Here, $(\theta,\varphi,\psi)$ are the Hopf coordinates on $S^3$, and $\psi\rightarrow \psi+2\pi$ amounts to the parallel transport of the fermion around a closed path. Such conditions are generalized accordingly in the case of a background gauge field of the $\mathbb Z_6^{(1)}$ $1$-form symmetry. 

It is well known that EH instantons do not support fermion zero modes in the absence of fluxes. However, once fluxes are turned on, this picture changes: the gauge field background induces normalizable fermion zero modes. At the same time, the boundary conditions project out certain modes, since otherwise the zero modes would fail to be globally well-defined. To examine this structure, we explicitly solve for the fermion zero modes in EH instantons with fluxes, and we further confirm the result by an independent computation of the Atiyah--Patodi--Singer (APS) index.

The identification of the asymptotic region under the discrete $\mathbb{Z}_2$ symmetry, which gives rise to the boundary $\mathbb{ RP}^3$, implies that the space surrounding the EH instanton is effectively a half-space. In this framework, the EH instanton  can be interpreted as mediating the transition amplitude between a specially prepared initial state entangled between the two identical halves of the space (left $L$ and right $R$; see Footnote \ref{discretized} for a discretized version of such state):
\begin{equation}\nonumber
|\mathrm{EH}\rangle\sim \prod_{\bm x, \mu=1,..,4}\int  dA_\mu \;\; |A_\mu(\bm{x})\rangle_R \otimes |A_\mu(-\bm{x})\rangle_L\,,
\end{equation}
and the vacuum $|\Omega\rangle$. Here, $A_\mu$ stands as a short-hand notation for the gauge fields in the $U(1)$ hypercharge sector as well as those along the Cartan generators of the $SU(3)\times SU(2)$ groups. We describe how the corresponding path integral is constructed in the semiclassical limit, where the dynamics reduce to those of a local quantum field theory. In particular, we show that the transition amplitude in a given $\mathbb Z_6^{(1)}$ flux, designated by the pair $(m_{(2)}, m_{(3)})$, is given by the path integral (we omit the fermions for notational convenience)
\begin{equation}\label{semifinalamplintro}\nonumber
\langle \Omega|\mathrm{EH}\rangle^{m_{(3)}, m_{(2)}} 
\;\sim\; e^{-S_{(3)}-S_{(2)}} \sum_{{\cal C}\in \mathbb{Z}} e^{-S_{(1)}} 
\int [\mathcal D A_\mu] \Bigg|_{\substack{A_\mu(\bm x, t_0)=A_\mu(\bm x)\\A_\mu(-\bm x, t_0)=-A_\mu(\bm x)}} 
e^{-S_{\scriptsize\rm eff}}\,,
\end{equation}
with the specified initial conditions on the gauge field $A_\mu$. The actions $S_{(1)}$, $S_{(2)}$, $S_{(3)}$ are the actions of the self-dual gauge fields in the $U(1)$, $SU(2)$, and $SU(3)$ sectors, and $S_{\scriptsize\rm eff}$ is the quadratic action about the background flux. 
 Furthermore, we argue that summing over all $\mathbb{Z}_6^{(1)}$ flux sectors has the effect of gauging the associated $1$-form symmetry in this semiclassical regime. Finally, assuming that the Higgs VEV is much smaller than $a^{-1}$, we find that these fluxes induce baryon- and lepton-number violating processes, which remain extremely suppressed due to the smallness of the $U(1)$ hypercharge gauge coupling as $S_{(1)}\sim\frac{|Q_{(1)}|}{g_Y^2}$.
 
Although the EH space is asymptotic to $\mathbb{R}^4/\mathbb{Z}_2$
rather than strictly $\mathbb{R}^4$, its asymptotic region is {\em locally}
indistinguishable from flat space: on scales small compared to the
curvature radius, and within a simply connected neighbourhood that does not
wind the nontrivial $1$-cycle, the induced metric and SM background fields
coincide with those of $\mathbb{R}^4$. Thus no local observer confined to
such a region can distinguish $S^3$ from $S^3/\mathbb{Z}_2$, and ordinary
correlators of point operators with all insertions in that region agree
with those in flat space. The difference lies only in the global topology,
which, once the SM $1$-form symmetry is gauged, modifies the allowed
topological sectors of the gauge bundle and thereby the structure of
instanton-induced multi-fermion operators. In this sense EH space is a
perfectly admissible Euclidean saddle contributing to the path integral
with asymptotically locally flat (ALE) boundary conditions. In fact, EH
space is the minimal and most analytically tractable ALE space, and thus
can be viewed as a representative of a broader class of configurations
that, outside a compact region, locally coincide with flat $\mathbb{R}^4$
but have nontrivial topology in the interior.

Recent works have exploited powerful kinematical constraints to extract lessons from the global structure of the SM and its implications for low-energy physics~\cite{Choi:2023pdp,Reece:2023iqn,Cordova:2023her}. A key message of our study is that what once looked like a mere consistency condition now finds a concrete nonperturbative realization in gravitational dynamics in the semi-classical limit. We return to this point in the discussion section of this paper.

This paper is organized as follows. Section~\ref{Eguchi-Hanson instantonsbulk} forms the foundation of our analysis. We begin by reviewing the key geometrical and topological features of EH instantons, emphasizing the role of the unique normalizable harmonic $2$-form. This structure allows us to introduce fractional fluxes supported on the bolt. We then explain how fermions can be consistently defined in the presence of $U(1)$ backgrounds, paying particular attention to global aspects. The Dirac equation is solved explicitly, and the resulting fermion zero modes are compared with the predictions of the Atiyah--Patodi--Singer (APS) index theorem.  

In Section~\ref{The background of UN gauge bundle and}, we extend the analysis to $U(N)$ gauge fields. We show how the global consistency conditions for fermions constrain the admissible zero modes and discuss the generalization of the APS index in this setting. Section~\ref{Backgrounds for the Standard Model} applies these constructions to the SM: we turn on fluxes associated with the $\mathbb{Z}_6^{(1)}$ $1$-form symmetry, study the corresponding fermion zero modes, and analyze baryon- and lepton-number violating processes. We demonstrate that such processes exist but are exponentially suppressed. Section~\ref{EH spaces as saddles in the path integral} interprets EH instantons as saddles of the Euclidean path integral in semiclassical gravity, with an initial state entangling the left and right halves of space, and explains how summing over $\mathbb{Z}_6$ fluxes dynamically gauges the SM $1$-form symmetry.  We conclude in Section \ref{Outlook and future directions} by giving an outlook on general lessons learned as well as proposals for future studies. 

Several technical developments are relegated to appendices. Appendix~\ref{The Eguchi-Hanson geometry} collects the necessary mathematical background needed to study EH spaces. Appendix~\ref{Fermions in the EH background fermion zero modes} contains a detailed solution of the zero modes of the Dirac equation, including how to project out globally ill-defined modes. Appendix~\ref{The Atiyah-Patodi-Singer index theorem} reviews the APS index theorem in the context of $U(1)$ gauge fields, extends it to non-Abelian bundles, and presents an explicit computation of the $\eta$-invariant, showing how the boundary twist induced by the bulk gauge bundle determines its value. Appendix~\ref{Fermions in the background of PSU(2) gauge field} explores fundamental and adjoint fermions in a $PSU(2)$ bundle aligned along the third color direction; while somewhat tangential, this example provides useful lessons on fermionic behavior in simple bundles. Finally, Appendix~\ref{Phase space path integral} analyzes the EH path integral in phase-space variables, illustrating how Gauss’s law is enforced.

%%%%%%%%%%%%%%%%%%%
\section{Eguchi-Hanson instantons, fractional fluxes, and spinors}
\label{Eguchi-Hanson instantonsbulk}
%%%%%%%%%%%%%%%%%%%%
Eguchi-Hanson (EH) instantons are (anti-)self-dual solutions to the Euclidean Einstein equations with vanishing cosmological constant \cite{Eguchi:1978xp,Eguchi:1978gw}.
The EH metric admits a particularly convenient expression when written in terms of the left-invariant $1$-forms $\sigma_x, \sigma_y, \sigma_z$ on $SU(2)\cong S^3$. In this parametrization, the line element takes the form  
\begin{eqnarray}\label{EHsigmaBULK}
ds^2_{\text{EH}} \;=\; f^2(r)\, dr^2 \;+\; r^2 \left( \sigma_x^2 + \sigma_y^2 + f^{-2}(r)\, \sigma_z^2 \right),
\end{eqnarray}
where the deformation function is given by
\begin{equation}
f^{-2}(r) \;=\; 1 - \left( \frac{a}{r} \right)^4,
\end{equation}
and the left-invariant $1$-forms $\{\sigma_x,\sigma_y,\sigma_z\}$ are
\begin{eqnarray}\label{sigmasinewh} \nonumber
\sigma_x &=& \tfrac{1}{2}\left( \sin\psi \, d\theta \;-\; \sin\theta \cos\psi \, d\varphi \right), \\\nonumber
\sigma_y &=& \tfrac{1}{2}\left(-\cos\psi \, d\theta \;-\; \sin\theta \sin\psi \, d\varphi \right), \\
\sigma_z &=& \tfrac{1}{2}\left( d\psi \;+\; \cos\theta \, d\varphi \right).
\end{eqnarray}
Here, $r$ is the radial coordinate, $(\theta,\varphi,\psi)$ are the Hopf coordinates (Euler angles) on $S^3$, and $a$ is a real, positive parameter with dimensions of length. It plays the role of a resolution scale for the geometry: when $a=0$ the metric reduces to flat $\mathbb{R}^4$, while for $a>0$ the space develops a smooth two-sphere ($2$-cycle or ``bolt'') at $r=a$. Thus, the apparent singularity at $r=a$ is a coordinate artifact; the EH space is geodesically complete.
 For semiclassical consistency, $a$ is typically taken larger than the inverse Planck length so that Einstein gravity remains valid.  
The coordinate ranges are
\begin{eqnarray}
a \leq r < \infty, \qquad 
0 \leq \theta \leq \pi, \qquad 
0 \leq \varphi \leq 2\pi, \qquad 
0 \leq \psi \leq 2\pi. 
\end{eqnarray}
Note in particular that the coordinate $\psi$ has period $\Delta\psi = 2\pi$, which is half of the usual range $\Delta\psi = 4\pi$ encountered in the standard coordinates on $S^3$. This reduced periodicity, which is needed to ensure smoothness of the metric at $r=a$, reflects the fact that the EH space is asymptotically locally Euclidean (ALE), with its asymptotic boundary not being the three-sphere $S^3$, but rather the real projective  space
\begin{equation}
\partial (\text{EH}) =\; S^3 / \mathbb{Z}_2 \;\cong\; \mathbb{RP}^3.
\end{equation}
Thus, at infinity the EH manifold looks like flat $\mathbb{R}^4$ modulo a $\mathbb{Z}_2$ identification, making it the simplest nontrivial ALE space. 

To reiterate the understanding of the identifications of the opposite points on $S^3$, which will be of great importance when discussing spinors in EH space as well as the physical interpretation of EH instantons in a path integral, we also note that one can use the coordinates $(x,y,z,t)$ given by:
\begin{eqnarray}\label{hoopfxyzt}
x+iy=r\cos\left(\frac{\theta}{2}\right)e^{\frac{i}{2}(\psi+\varphi)}\,, \quad z+it=r\sin\left(\frac{\theta}{2}\right)e^{\frac{i}{2}(\psi-\varphi)}\,,
\end{eqnarray}
with $r^2=x^2+y^2+z^2+t^2\geq a^2$.
Under $\psi\rightarrow\psi+2\pi$ we have $(x,y,z,t)=-(x,y,z,t)$, and thus, restricting the range of $\psi$ to $0\leq \psi\leq 2\pi$ implies that we are identifying $(x,y,z,t)$ with $-(x,y,z,t)$. See Appendix \ref{The Eguchi-Hanson geometry} for all the conventions used in this paper, an elementary exposition of the tools needed to analyze the EH geometry, as well as more computational details. 

 The exact self-duality of the EH space immediately implies that the Ricci scalar vanishes, and hence the Einstein-Hilbert action vanishes identically 
\begin{eqnarray}
S_{\scriptsize\mbox{EH}}=0\,.
\end{eqnarray}
Thus, in the absence of any background gauge fields, there is no small parameter that enables us to perform reliable semi-classical calculations\footnote{Higher-curvature terms have been argued to ameliorate this problem. In particular, the authors in \cite{Dvali:2024dlb} proposed supplementing the action with the Euler characteristic, $\alpha_E\chi$, where a positive coefficient $\alpha_E\gg1$ restores semiclassical control. A similar problem is also encountered in wormholes \cite{Kallosh:1995hi}, where higher-curvature corrections were proposed to solve it.}. The situation here, in some sense, is worse than in pure Yang-Mills theory. Fortunately, this problem is resolved when self-dual background gauge fields are turned on, as we discuss next.

%%%%%%%%%%%%%%%%%%%%%%%%
\subsection{Turning on a $U(1)$ gauge field}
%%%%%%%%%%%%%%%%%%%%%%%%

An important feature of the EH space is that it can support abelian as well as nonabelian gauge fields without altering the underlying gravitational background. This is possible because certain gauge configurations can be chosen to be self-dual, which ensures that their stress-energy tensor vanishes identically. As a result, these fields constitute nontrivial but non-backreacting backgrounds for charged matter.
The abelian $U(1)$ gauge potential is \cite{Eguchi:1978gw} given by
\begin{eqnarray}\label{form of ABUL}
A_{U(1)} \;=\; {\cal C}\, \frac{a^2}{r^2}\, \sigma_z\,,
\end{eqnarray}
where ${\cal C}$ is the charge that determines the $U(1)$ flux, as we shall show momentarily. This gauge field is globally well-defined on the EH manifold and is regular at the bolt located at $r=a$. 
The corresponding field strength $2$-form 
\begin{eqnarray}\label{feildstrngth}
F_{(1)} = dA_{U(1)}=-\frac{{\cal C} a^2}{r^3}dr\wedge (d\psi+\cos\theta d\varphi)+\frac{{\cal C}a^2}{2r^2}\sin\theta d\varphi\wedge d\theta
\end{eqnarray}
 is easily shown to be anti-self-dual, 
$
\star F_{(1)} = -F_{(1)}
$; see Appendix \ref{Turning on a U1 gauge field} for more details. The subscript $(1)$ is used to indicate that this is the field strength associated with a $U(1)$ gauge field. This distinction in notation will be important later, since we will encounter multiple gauge field configurations of different types, and it will be useful to keep track of their origins explicitly.
 The self-duality condition guarantees that the Maxwell stress tensor vanishes identically, so the metric (\ref{EHsigmaBULK}) remains an exact solution of the coupled Einstein--Maxwell system. Other possible abelian self-dual configurations exist, but they are singular, and thus the form (\ref{form of ABUL}) provides a unique smooth realization\footnote{See the discussion below regarding the fact that the EH space admits a unique normalizable self-dual harmonic $2$-form.}.

The $U(1)$ flux through the two-sphere at the bolt, located at $r=a$, is obtained by integrating $F_{(1)}$ over $S^2 \subset \text{EH}$ (covered by the coordinates $\theta$ and $\varphi$):
\begin{eqnarray}
\int_{S^2} F_{(1)} 
= -2\pi {\cal C}\,.
\end{eqnarray}
Thus, the constant ${\cal C}$ controls the quantized flux through the bolt, and in the quantum theory ${\cal C}$ must be chosen so that this flux is an integer multiple of $2\pi$.  
The gauge field also carries a topological charge, or instanton number, defined by
\begin{eqnarray}
Q_{(1)} \;=\; \frac{1}{8\pi^2}\int_{\scriptsize\mbox{EH}} F_{(1)} \wedge F_{(1)}\,,
\end{eqnarray}
and explicit evaluation yields
\begin{eqnarray}
Q_{(1)} = \frac{{\cal C}^2}{4}\,,
\end{eqnarray}
which shows that the configuration is topologically nontrivial. In fact, this is the abelian analogue of a Yang-Mills instanton on the EH space, and it represents a localized flux threading the bolt. As a bonus, the background gauge field now comes with its own action. Using the self-duality of the gauge field, we readily find
\begin{eqnarray}
S_{(1)}=\frac{8\pi^2 |Q_{(1)}|}{g_1^2}\,,
\end{eqnarray}
 where $g_1$ is the $U(1)$ gauge field coupling constant. Because the flux is localized near the bolt, the dominant contribution to the action arises from the region in its vicinity. For small enough $g_1$, the action is large and one can perform reliable semi-classical computations in the background of the EH instanton with a flux.
 
 Before concluding this section, let us clarify the geometric origin of the factor $1/4$ appearing in the topological charge. 
The EH space admits a unique normalizable harmonic\footnote{A form ${\cal K}$ is said to be harmonic if it is annihilated by the Laplacian: $\Delta{\cal K}=0$, where $\Delta=d\delta+\delta d$ and $\delta=-*d*$. Thus, ${\cal K}$ is harmonic if and only if it is both closed and co-closed, i.e.\ $d{\cal K}=\delta{\cal K}=0$. That the $U(1)$ gauge field $F_{(1)}$, and hence ${\cal K}$, is harmonic is evident from the fact that we are discussing free Maxwell theory, so $dF_{(1)}=d*F_{(1)}=0$.} $2$-form, denoted by ${\cal K}$, 
and the $U(1)$ field strength can be written as\footnote{Here, it helps to mention a few important geometric facts about the EH space. The EH space is a hyper-K\"ahler manifold with Hirzebruch signature $\tau=1$ and Euler characteristics $\chi=\tau+1=2$. Given the self-dual nature of the space, the number of the normalizable harmonic $2$-forms is equal to $\tau=1$; see, e.g., the discussion in \cite{Hawking:1978ghb}. }
\begin{equation}
    F_{(1)} = -2\pi\,{\cal C}\,{\cal K}.
\end{equation}
We normalize ${\cal K}$ such that its integral over the bolt satisfies
\begin{equation}\label{normalizK}
    \int_{S^2} {\cal K} = 1 \,.
\end{equation}
Now, consider a small deformation $S^{2\prime}$ of the bolt $S^2$ such that $S^{2\prime}$ intersects $S^2$ transversely. 
The self-intersection number\footnote{The self-intersection number of the bolt is $\pm 2$ \cite{Yuille:1987vw,Kronheimer:1989zs}. Here, it is fractional because of the parametrization of the EH metric (\ref{EHsigmaBULK}) and the way we choose to normalize ${\cal K}$ in (\ref{normalizK}).} of the bolt is then captured by the wedge product of ${\cal K}$ \cite{Bianchi:1996zj}:
\begin{equation}
    \int_{\scriptsize\mbox{EH}} {\cal K}\wedge{\cal K} = \tfrac{1}{2}.
\end{equation}
It follows that the topological charge of the $U(1)$ field is\footnote{It also helps to compare this result with the topological charge of a $U(1)$ gauge field in the background of the complex projective space $\mathbb {CP}^2$, a nonspin manifold. Here, the space admits a normalizable harmonic $2$-form ${\cal K}$ such that the integral of ${\cal K}$ over the $2$-cycle, in this case it is $\mathbb {CP}^1\subset \mathbb {CP}^2 $, is $\int_{\mathbb {CP}^1}{\cal K}=1$. The self-intersection number is given by the wedge product $\int_{\mathbb{CP}^2}{\cal K}\wedge {\cal K}=1$. A self-dual $U(1)$ gauge field is given by $  F_{(1)} = -2\pi\,{\cal C}\,{\cal K}$, and thus, its associated topological charge is \begin{eqnarray}Q_{(1)}^{\mathbb{ CP}^2}=\frac{{\cal C}^2}{2}\,.\end{eqnarray} See, e.g., the appendices of \cite{Anber:2020gig} for the details of the calculations. We see that the different self-intersection numbers of EH and $\mathbb{CP}^2$ spaces explain the extra factor of $1/2$ that multiplies the topological charge in EH space.}
\begin{equation}
    Q_{(1)} \;=\; \frac{(2\pi{\cal C})^2}{8\pi^2}\int_{\scriptsize\mbox{EH}} {\cal K}\wedge{\cal K}
    \;=\; \frac{{\cal C}^2}{2}\times \frac{1}{2}.
\end{equation}

In summary, the EH space admits smooth, self-dual $U(1)$ gauge fields that do not alter the background geometry but carry nontrivial flux and topological charge. These will play a crucial role in determining the boundary conditions and spectrum of charged fermions on the EH background.

%%%%%%%%%%%%%%%%%%%%%%%%%%%%%%%%%%%%%%%%%%%
\subsection{Turning on a background field of $\mathbb Z_N^{(1)}$ $1$-form symmetry}
\label{backgroundZN}
%%%%%%%%%%%%%%%%%%%%%%%%%%%%%%%%%%%%%%%%%%%

We can generalize the simple $U(1)$ gauge field configuration 
introduced in (\ref{form of ABUL}) in order to turn on a background 
for the $\mathbb{Z}_N^{(1)}$ one-form symmetry. The construction is achieved by turning on a gauge field valued in the Cartan subalgebra of $SU(N)$, so that the resulting configuration effectively behaves abelian while still living inside the non-abelian group \cite{Anber:2019nze,Anber:2020gig}. 
The corresponding field strength is a $2$-form 
that represents the background of the $\mathbb{Z}_N^{(1)}$ $1$-form symmetry.  
Concretely, we take the $1$-form gauge potential to be
\begin{eqnarray}\label{gaugesun}
A_{SU(N)} \;=\; m_{(N)} \, \bm{H}\cdot\bm{\nu} \, \frac{\sigma_z\,a^2}{r^2}\,,
\end{eqnarray}
where $m_{(N)} \in \mathbb{Z}$ is an integer parameter, 
$\bm{H} = (H_1,H_2,\dots,H_{N-1})$ are the Cartan generators of $SU(N)$, 
and $\bm{\nu}_a$ (with $a=1,\dots,N$) denotes a weight vector 
of the fundamental representation. By construction, the associated field strength,
\begin{equation}
F_{(N)} \;=\; dA_{SU(N)}\,,
\end{equation}
is self-dual, and hence does not backreact on the EH geometry.  

Without loss of generality, let us choose the weight $\bm{\nu}_1$. 
Using the standard identity for the weights,
$
\bm{\nu}_a \cdot \bm{\nu}_b \;=\; \delta_{ab} - \tfrac{1}{N}
$
(we use a convention where the square of the simple roots is $\bm\alpha^2=2$ and the Dynkin index of the fundamental representation is $T_\Box$=1), we find that the flux of $F_{(N)}$ through the bolt two-sphere at $r=a$ is
\begin{eqnarray}\label{flux QZN}
\int_{S^2} F_{(N)} 
= -2\pi\, m_{(N)} \,
\mathrm{diag}_{N\times N}\!\left(1-\tfrac{1}{N},\, -\tfrac{1}{N},\, \dots,\,-\tfrac{1}{N}\right) \in 2\pi \,\mathbb{Z}_N \,.
\end{eqnarray}
This result demonstrates that the flux of $F_{(N)}$ is fractional, 
quantized in units of $2\pi/N$. In other words, the configuration 
(\ref{form of ABUL}) when embedded into $SU(N)$ provides precisely 
the $2$-form background gauge field, $F_{(N)}$, of the $\mathbb{Z}_N^{(1)}$ one-form symmetry.  

The corresponding topological charge of the $2$-form background field is
\begin{eqnarray}
Q_{(N)} 
= \frac{1}{8\pi^2}\int_{\scriptsize\mbox{EH}} \mathrm{tr}\,F_{(N)}\wedge F_{(N)} 
= \frac{m_{(N)}^2}{4}\,\bm{\nu}_1\cdot \bm{\nu}_1 
= \frac{m_{(N)}^2}{4}\left(1-\frac{1}{N}\right)\,.
\end{eqnarray}
Thus, the topological charge is, in general, fractional. It is important to note, however, that the appearance of the additional  $1$ alongside the $1/N$ term originates from the fact that we are working with an abelianized configuration, obtained by turning on a gauge field along the Cartan subalgebra of $SU(N)$  
\footnote{It remains unclear whether employing a fully non-abelian self-dual background gauge field could remove this extra contribution, thereby yielding a minimal topological charge of exactly $m_{(N)}^2/(4N)$. This possibility is suggested by analogy with the well-studied case of self-dual fractional instantons on the four-torus $\mathbb{T}^4$, where the minimal charge $1/N$ can only be realized when non-abelian fields and/or transition functions are used \cite{tHooft:1981nnx,GarciaPerez:2000aiw,Gonzalez-Arroyo:2019wpu,Anber:2023sjn,Anber:2025yub}. By extension, it is suspected that a similar mechanism may be operative in EH space.  In this regard, we point out that a systematic construction of all Yang-Mills EH instantons based on $SU(N)$ Lie groups, as well as those with abelian gauge bundles, was presented in the pioneering work of Kronheimer and Nakajima \cite{PeterBKronheimer:1990zmj}; see also \cite{Bianchi:1995xd,Bianchi:1996zj}.  Although, to the best of our knowledge, an explicit construction of $\mathbb Z_N^{(1)}$ fluxes starting from a full nonabelian parent theory has not been established.}.

One might object that the flux threading the bolt (\ref{flux QZN}) is not quantized in integral units. Indeed, probing this flux with a unit charge would violate the Dirac quantization condition. We will show later how to resolve this issue by introducing additional $U(1)$ flux that restores compliance with the Dirac condition. 

%%%%%%%%%%%%%%%%%%%%%%%%%%%%%%%%%%
\subsection{Spinors in EH space in the presence of abelian fields}
\label{Spinors in EH space in the presence of abelian fields}
%%%%%%%%%%%%%%%%%%%%%%%%%%%%%%%%%%

While the study of fermion zero modes in instanton backgrounds is a classic subject, the EH geometry introduces subtleties that make it worthwhile to review the analysis carefully. In particular, the boundary of EH space is $\mathbb{RP}^3$, whose nontrivial $\mathbb{Z}_2$ cycle affects the definition of spinor boundary conditions in the presence of background gauge fields, as first emphasized by ’t~Hooft~\cite{tHooft:1988wxy}. As a result, the allowed holonomies and the selection rules for fermion representations differ from the more familiar $\mathbb{R}^4$ or BPST instanton case. The explicit derivation of boundary conditions, holonomies, and the subsequent zero-mode counting will be essential for our later analysis of $\mathbb{Z}_N^{(1)}$ backgrounds.

In flat $\mathbb{R}^4$, the angular coordinate $\psi$ on $S^3$ has the full range $0 \leq \psi \leq 4\pi$. In EH space, however, one identifies
\begin{equation}\label{xtomx}
(x,y,z,t) \sim (-x,-y,-z,-t),
\end{equation}
which reduces the allowed range to $0 \leq \psi \leq 2\pi$. A spinor $\Psi(x)$ (taken to be a Dirac spinor) must be continuous along any contractible path connecting $(x,y,z,t)$ and $(-x,-y,-z,-t)$; this path has to define an element of the covering group of $\mathrm{SO}(4)$, namely $\mathrm{Spin}(4)$, unambiguously \cite{Back:1978zf}. This forces us to choose the correct boundary conditions on  $\Psi(x)$, in the background of the gauge field, under the identification (\ref{xtomx}). Let us explain this statement carefully.  A contractible path is one that can be continuously shrunk to a point. When you parallel transport around such a path (in the absence of a gauge field), you should get the same result as if you had transported around the trivial path (no motion at all). When parallel transporting a boson (in the absence of a gauge field), the holonomy of the motion is valued in $\mathrm{SO}(4)$, and the trivial loop corresponds to the identity transformation. When considering the parallel transport of a fermion (again in the absence of gauge field), one then must consider the lifting of $\mathrm{SO}(4)$ to $\mathrm{Spin}(4)$, and the identity in $\mathrm{SO}(4)$  has two possible lifts: $(+1)$ and $(-1)$. However, for contractible loops, you must pick the identity element ($+1$), because the holonomy must depend continuously on the loop. If you allowed $(-1)$, then shrinking the loop smoothly to zero length would suddenly flip the sign of the spinor, which is discontinuous and inconsistent. The identification (\ref{xtomx}) amounts to relating the spinors at $\psi$ and $\psi+2\pi$, and one must ensure that the spinors obey the correct boundary conditions upon parallel transporting along contractible paths connecting $\psi$ and $\psi+2\pi$. When we switch on a gauge field, we shall see that there are two inequivalent boundary conditions, depending on whether the charge ${\cal C}$ is even or odd.  

Consider the parallel transport of a Dirac fermion $\Psi$, carrying a positive unit electric charge under $U(1)$, along the contractible path $\ell_{r\approx a}$ connecting $(r \approx a,\theta=0,\varphi=0,\psi)$ and $(r \approx a,\theta=0,\varphi=0,\psi+2\pi)$ (or any path with constant $\theta,\varphi$) near the bolt\footnote{We evaluate the holonomy along a small contractible loop 
$\ell_{r=a+\epsilon}$  and then take the limit 
$\varepsilon\to 0$. In our coordinates, the frame becomes singular 
exactly at $r=a$ (i.e. the metric degenerates to $S^2$), so it is technically more convenient to work 
"near the bolt" and understand all conditions as limits 
$r\to a$ from above.} $S^2$ (the metric near $r\approx a$ is given by (\ref{EHRA})). Along this path, the fermion acquires both $U(1)$ and gravitational holonomies:
\begin{align}\label{PARALLTABUL}
\Psi(r,\theta,\varphi,\psi+2\pi) &= \exp\left[i\int_{\ell_{r \approx a}} A\right]\,
\exp\left[-\tfrac{1}{2}\int_{\ell_{r \approx a}}\omega_{ab}\Sigma^{ab}\right]\Psi(r,\theta,\varphi,\psi)\,,
\end{align}
where $\omega_{ab}$ are spin connection $1$-forms (see their definition in Appendix \ref{The Eguchi-Hanson geometry}), $\Sigma^{ab}=[\gamma^a,\gamma^b]/4$, and $\gamma^a$ are the Dirac matrices. The Latin letters $a,b$ run over $0,1,2,3$.
Simple calculations show (the calculations are performed in Appendix \ref{Fermions in the EH background fermion zero modes}, and in particular see the treatment starting at Eq. (\ref{PARALLTA}))
\begin{eqnarray}
\exp\left[-\tfrac{1}{2}\int_{\ell_{r \approx a}}\omega_{ab}\Sigma^{ab}\right] 
= I_4\,,\quad
\exp\left[i\int_{\ell_{r \approx a}} A\right] = e^{i\pi \mathcal{C}}.
\end{eqnarray}
Thus, the gravitational holonomy is trivial along $\ell_{r \approx a}$, reflecting the fact that EH spaces are spin. Only the gauge holonomy is generally nontrivial, so that
\begin{align}\label{PARALLTA2BUL}
\text{at } r \approx a:\qquad \Psi(r,\theta,\varphi,\psi+2\pi) =(-1)^{\cal C}\,\Psi(r,\theta,\varphi,\psi)\,.
\end{align}
Thus, the fermions must obey periodic or anti-periodic boundary conditions under $\psi\rightarrow \psi+2\pi$, depending on whether the charge ${\cal C}$ is even or odd.

Practically, one finds that the solution of the Dirac equation yields normalizable states with total angular momentum $j$ quantized either in integers or half-integers. The boundary condition (\ref{PARALLTA2BUL}) selects the integer values of $j$ when ${\cal C}$ is even and the half-integer values of $j$ when ${\cal C}$ is odd. This is verified by explicitly solving for the zero modes of the Dirac equation, as we discuss in the next section. 

We now study the behavior of the fermions as $r\rightarrow\infty$  where the EH space approaches its boundary $\mathbb{RP}^3$. 
Let $\ell_{\infty}$ denote a path that takes us from $\psi$ to $\psi + 2\pi$ at $r\rightarrow\infty$ and fixed angular coordinates $(\theta,\varphi)$. This loop is noncontractible, since $\mathbb{RP}^3$ is not simply connected. 
Its first homology group is 
$
H_1(\mathbb{RP}^3, \mathbb{Z}) \cong \mathbb{Z}_2,
$
so $\mathbb{RP}^3$ has a single nontrivial $1$-cycle, which is $\mathbb{Z}_2$-torsion.
  Under parallel transport along $\ell_{\infty}$,  we find (see the detailed treatment near Eq. (\ref{holatinfty}) in Appendix \ref{Fermions in the EH background fermion zero modes})
\begin{eqnarray}
\exp\left[-\tfrac{1}{2}\int_{\ell_{\infty}}\omega_{ab}\Sigma^{ab}\right] 
= -\gamma^5\,, \quad \gamma^5\equiv \gamma^0\gamma^1\gamma^2\gamma^3\,.
\end{eqnarray}
Notice here that, unlike the case of parallel transporting near the bolt, the gravitational holonomy is non-trivial.
The computation of the gauge holonomy $e^{i\int_{\ell_{\infty}} A}$ requires more care. Naively, because $A$ is rapidly decaying, one would conclude that this holonomy is trivial, $(+1)$. This, however, is not correct. The idea is that such a quick conclusion overlooks the possibility that topological data may be transmitted from the interior to the boundary, which is indeed our case. To correctly compute the gauge holonomy, we use Stokes' theorem $\int_{\ell_{\infty}} A=\int_{\Sigma} F$, where $\Sigma$ is a $2$-surface that interpolates between the loop $\ell_{\infty}$ on the boundary $\mathbb{RP}^3$ and the bolt in the interior. Topologically, it is a cigar shape stretching from the boundary, capping off at the bolt (the $\psi$-circle shrinks smoothly at the bolt\footnote{See the EH metric (\ref{EHRA}) in the limit $r\rightarrow a$, which helps with the visualization.}). Using the expression of $F_{(1)}$ in (\ref{feildstrngth}), fixing $\theta,\varphi$, and integrating over $0\leq\psi\leq2\pi$, $a\leq r\leq\infty$, we readily find
\begin{eqnarray}\label{mainhol}
\exp\left[i\int_{\ell_\infty} A\right]=\exp\left[i\int_{\Sigma} F\right]=(-1)^{{\cal C}}\,.
\end{eqnarray}
Thus, even though the boundary line bundle is flat (the field strength decays rapidly and reaches $0$ at infinity), its holonomy is nontrivial when ${\cal C}$ is odd, contrary to the naive expectation. 
We conclude that the boundary condition on $\Psi$ at $r\rightarrow\infty$ is
\begin{eqnarray}\label{inftyBC}
\text{at}\, \partial\,(\mbox{EH}):\,
\Psi(r\rightarrow \infty, \theta, \varphi, \psi + 2\pi) 
=(-1)^{\mathcal{C}+1} \, \gamma_5 \, \Psi(r\rightarrow\infty, \theta, \varphi, \psi)\,.
\end{eqnarray}

The fact that the holonomy (\ref{mainhol}) is in  $\mathbb Z_2$ encodes an essential feature of the EH space. While the bolt admits arbitrary integer flux, the non-contractible Wilson loop at the boundary is sensitive only to its value modulo $2$. 

Depending on whether ${\cal C}$ is even or odd, the sign of the r.h.s of (\ref{inftyBC}) plays a crucial role in computing the $\eta$-invariant, which is required for applying the Atiyah--Patodi--Singer (APS) index theorem to count the zero modes of the Dirac operator on EH space in the presence of a $U(1)$ gauge field, as we explain in Appendix \ref{The Atiyah-Patodi-Singer index theorem}. In particular, when ${\cal C}$ is even, then the line bundle at the boundary has a trivial holonomy and the $\eta$-invariant gives $(-1/8)$. When ${\cal C}$ is odd, the boundary line bundle carries a holonomy of $(-1)$ and the $\eta$-invariant is $(+1/8)$.

Before concluding this section, let us also note that a scalar $h$, which carries a unit electric charge under $U(1)$, acquires a gauge holonomy upon parallel transport along a contractible path near the bolt. Similarly, the holonomy of the flat line bundle at infinity picks a $(-1)$ sign when ${\cal C}$ is odd. Thus,  the boundary conditions for the scalar are:
\begin{eqnarray}\nonumber
&&\text{at } r \approx a:\qquad h(r,\theta,\varphi,\psi+2\pi) = (-1)^{\cal C}\,h(r,\theta,\varphi,\psi)\,,\\
&&\text{at}\, \partial\,(\mbox{EH}):\qquad h(r\rightarrow \infty,\theta,\varphi,\psi+2\pi) = (-1)^{\cal C}\,h(r\rightarrow \infty,\theta,\varphi,\psi)\,.
\end{eqnarray}
%

%%%%%%%%%%%%%%%%%%%%%%%%%%%
\subsection{Dirac fermion zero modes in EH space}
%%%%%%%%%%%%%%%%%%%%%%%%%%%

Finding the chiral zero modes in instanton backgrounds is an important step, since their chirality has important consequences. These modes can signal the presence of anomalies or act as mediators for physical processes, such as baryon- and lepton-number violating interactions. 
In this section, we aim to solve the Dirac equation on EH spaces in the presence of a $U(1)$ gauge field. This analysis will be straightforwardly generalized by considering a more general $\mathbb{Z}_N^{(1)}$ background field.

The detailed solution is presented in Appendix \ref{Fermions in the EH background fermion zero modes}. Here, we emphasize only the key points necessary to understand the solution and its physical significance. We are solving for the normalizable solutions of  the Dirac equation
\begin{eqnarray}
\slashed D\Psi=0\,, 
\end{eqnarray}
(again, see Appendix \ref{Fermions in the EH background fermion zero modes} for the explcit form of the covariant derivative $\slashed D$) along with the boundary condition (\ref{PARALLTA2BUL}). We assume that the Dirac fermion has a positive unit charge under $U(1)$, and without loss of generality, we assume that ${\cal C}\geq0$. The normalizability condition demands
\begin{eqnarray}
\int_{\scriptsize\mbox{EH}}|\Psi|^2<\infty\,.
\end{eqnarray}
 Writing the Dirac spinor in terms of the Weyl components, the left-handed $\lambda_\alpha$ and the right-handed $\chi^{\dot\alpha}$:
\begin{eqnarray}
\Psi=\left[\begin{array}{c}\lambda_\alpha\\\bar\chi^{\dot\alpha}\end{array}\right]\,,
\end{eqnarray}
where $\alpha,\dot\alpha=1,2$ is the spin index, we find that there exist no normalizable zero modes of $\bar\chi$ on EH space in the absence or presence of a $U(1)$ gauge field. On the other hand, the solutions of $\lambda_\alpha$ take the form
\begin{eqnarray}
\lambda_1=g_1(r)|j,m'=j,m\rangle\,, \quad \lambda_2=g_2(r)|j,m'=-j,m\rangle\,.
\end{eqnarray}
The ket $|j,m',m\rangle$ is the Wigner-D matrix \cite{Sakurai:2011zz}, which is the generalization of spherical harmonics on $\mathbb {S} ^3$. Here, $j$ is the total angular momentum, which can be a positive half-integer, positive integer, or $0$: $j=0,\frac{1}{2},1,...$. The values of $m', m$ are in the range $|m|,|m'|\leq j$. The functions $g_1,g_2$ satisfy the first-order differential equations
\begin{eqnarray}\nonumber
\left[ \frac{1}{f}\frac{\partial}{\partial r}+\frac{2f}{r}+\frac{1}{fr}+\frac{a^2{\cal C}f}{r^3}+\frac{2f j}{r}\right]g_1&=&0\,,\\
\left[ \frac{1}{f}\frac{\partial}{\partial r}+\frac{2f}{r}+\frac{1}{fr}-\frac{a^2{\cal C}f}{r^3}+\frac{2f j}{r}\right]g_2&=&0\,.
\end{eqnarray}
Because the positive sign in front of the term $\frac{a^2{\cal C}f}{r^3}$ in the first equation (remember that we assume ${\cal C}\geq0$), the solutions $g_1$ are not normalizable on EH space. This leaves $g_2$ as the only normalizable solutions. Before applying the boundary condition (\ref{PARALLTA2BUL}), the allowed values of $j$ and $m$ are
\begin{equation}
j = 0, \frac{1}{2}, 1, \ldots, \frac{{\cal C} - 1}{2}\,, \quad |m|\leq j\,.
\end{equation}
We also note that when ${\cal C}=0$, there exists no normalizable solution for either $g_1,g_2$. The absence of normalizable zero modes in EH space in the absence of a background gauge field is a well-known result \cite{Eguchi:1980jx}.  

After applying the boundary condition (\ref{PARALLTA2BUL}), only the integer values of $j$ (when ${\cal C}$ is even) and the half-integer values of $j$ (when ${\cal C}$ is odd) are selected. Thus, we must consider each case separately to find the number of normalizable zero modes.

{\bf Case I}: ${\cal C}$ is even, $C=2p\geq 0$, the gauge holonomy gives a trivial phase, and thus, one must restrict the values of $j$ to integers. In this case, the number of normalizable zero modes is
\begin{eqnarray}
 {\cal I}=\sum_{j=0,1,..}^{p-1}(2j+1)=p^2\,.
 \end{eqnarray}

{\bf Case II}: ${\cal C}$ is odd, ${\cal C}=2p+1\geq 1$, the gauge holonomy gives $(-1)$, and thus, one must restrict $j$ to half-integers. The total number of normalizable zero modes in this case is  
\begin{eqnarray}
 {\cal I}=\sum_{j=\frac{1}{2},\frac{3}{2},..}^{p-1/2}(2j+1)=p(p+1)\,.
 \end{eqnarray}

These findings agree with the results of \cite{tHooft:1988wxy} and with the Atiyah–Patodi–Singer (APS) index theorem, as discussed in detail in Appendix~\ref{The Atiyah-Patodi-Singer index theorem}.  The APS index is given by
\begin{eqnarray}
{\cal I}=\frac{1}{8\pi^2}\int_{\scriptsize\mbox{EH}}F_{(1)}\wedge F_{(1)}+\frac{1}{24\cdot 8\pi^2}\int_{\scriptsize\mbox{EH}}\mbox{tr}R\wedge R-\eta\,.
\end{eqnarray}
Unlike the Atiyah--Singer index, which applies to closed manifolds without boundary contributions, the APS index includes an additional boundary term given by the $\eta$-invariant, defined from the spectral properties of the Dirac operator on the boundary $\mathbb{RP}^3$.
When evaluating the APS index, it is crucial to account for the twist by the holonomy of the flat line bundle on $\mathbb{RP}^3$, which depends on whether ${\cal C}$ is even or odd.  
For ${\cal C}$ even, the holonomy of the line bundle is trivial, and the $\eta$-invariant takes the value $(-1/8)$. This contribution precisely cancels the bulk term from the $\hat A$-genus,  
$
\frac{1}{24 \cdot 8\pi^2}\int \mathrm{tr}\, R \wedge R \;=\; -\tfrac{1}{8},
$
where $R$ denotes the curvature $2$-form, ensuring the integrality of the index.  
For ${\cal C}$ odd, however, the flat line bundle holonomy acquires a factor of $(-1)$, shifting the $\eta$-invariant to $(+1/8)$. In this case, the bulk $\hat A$-genus and the boundary $\eta$-invariant do not cancel but instead combine to give $(-1/4)$. Remarkably, this fractional contribution is precisely compensated by the $U(1)$ topological charge,
$
\frac{1}{8\pi^2}\int F_{(1)} \wedge F_{(1)},
$
so that the full index remains an integer.

In summary, the number of the normalizable zero modes in EH space in the background of the $U(1)$ gauge field is as follows
 \begin{eqnarray}\label{counting U(1)BU}\nonumber
{\cal I}&=&p^2\,,\quad {\cal C}=2p\geq0\,,\\
{\cal I}&=&p(p+1)\,,\quad {\cal C}=2p+1\geq1\,.
\end{eqnarray}
This result will be of great importance when counting the number of normalizable zero modes in the background gauge of $\mathbb Z_N^{(1)}$ $1$-form symmetry.

%%%%%%%%%%%%%%%%%%%%%%%%%%%%%%%%%%%%%%%
\section{Background of $U(N)$ gauge bundle and fermion zero modes}
\label{The background of UN gauge bundle and}
%%%%%%%%%%%%%%%%%%%%%%%%%%%%%%%%%%%%%%%

We now turn to the case of fermions in the background field associated with the $\mathbb{Z}_N^{(1)}$ $1$-form symmetry constructed in Section~\ref{backgroundZN}.  
In general, however, fermions cannot always be consistently defined in such a background, as their behavior depends on the representation under the parent $SU(N)$ group.  

This obstruction can be understood in several ways.  
For instance, consider a fundamental fermion subject to the holonomy of the gauge field given in (\ref{gaugesun}).  
Repeating the analysis starting from (\ref{PARALLTABUL}), we find that upon traversing a contractible path near the bolt, the fermion acquires the phase
\begin{eqnarray}\label{phasePsi}\nonumber
\Psi(r \approx a,\theta,\varphi,\psi+2\pi)
&=& \exp\!\left[i\pi\, m_{(N)} \,
\mathrm{diag}_{N\times N}\!\left(1-\tfrac{1}{N},\, -\tfrac{1}{N},\, \dots,\,-\tfrac{1}{N}\right)\right] \\
&&\times \Psi(r \approx a,\theta,\varphi,\psi)\,.
\end{eqnarray}
Recalling that solutions to the Dirac equation correspond to integer or half-integer values of the total angular momentum (see the previous section), we conclude that the spinors are consistently defined if and only if the phase on the right-hand side of (\ref{phasePsi}) is equal to $\pm 1$.  One can easily check that unless this is the case, the APS index is fractional.

In Appendix \ref{Fermions in the background of PSU(2) gauge field}, we analyze the fundamental and adjoint fermion zero modes in the background of an $PSU(2)$ gauge field aligned along the $3$-direction in color space.  
Although tangential to our main work, this case is particularly instructive, as the $PSU(2)$ gauge group is of primary interest because of its simplicity, and the adjoint zero modes provide a useful setting for studying supersymmetric theories in this background.

To be able to define the fermions on EH space, in this section, we introduce a background gauge field for 
$U(N) \equiv SU(N)\times U(1)/\mathbb Z_N$ 
and analyze the fundamental fermion zero modes in this background.  
To this purpose, we switch on a field along the Cartan directions of $SU(N)$, supplemented with a $U(1)$ gauge field:
\begin{eqnarray}
A_{U(N)}\equiv A_{SU(N)}+A_{U(1)}I_{N}=\left[ m_{(N)} \bm H\cdot \bm\nu_a+\left({\cal C}+\frac{m_{(N)}}{N}\right) I_{N}\right]\frac{\sigma_z a^2}{r^2}\,.
\end{eqnarray}
By construction, this is a self-dual field. 
This particular choice of $U(1)$ background guarantees that fermions remain well-defined in the EH space.  Here, ${\cal C}$ denotes an integer $U(1)$ charge, 
$\bm H=(H_1,H_2,\dots,H_{N-1})$ are the Cartan generators, 
and $\bm\nu_a$ (with $a=1,\dots,N$) represents a weight of the fundamental representation of $SU(N)$.  
Without loss of generality, we take $a=1$, and use the identity 
$\bm \nu_a\cdot \bm\nu_b=\delta_{ab}-\tfrac{1}{N}$ 
to find that $A_{U(N)}$ reduces to a diagonal $N\times N$ matrix:
\begin{eqnarray}
A_{U(N)}= \mbox{diag}_{N\times N}\left(m_{(N)}+{\cal C},{\cal C},\dots,{\cal C}\right)\frac{\sigma_z a^2}{r^2}\,.
\end{eqnarray}
The $A_{SU(N)}$ and $U(1)$ gauge fields satisfy the following quantization conditions when their field strengths are integrated over $S^2$:
\begin{align}\nonumber
\int_{S^2} F_{(N)} &= -2\pi m_{(N)}\,\mathrm{diag}_{N\times N}\!\left(1-\tfrac{1}{N}, -\tfrac{1}{N}, \dots, -\tfrac{1}{N}\right)\,,\\
\int_{S^2} F_{(1)} &= -2\pi \left({\cal C}+\tfrac{m_{(N)}}{N}\right)\,.
\end{align}
This shows that both fields individually carry fractional fluxes.
The corresponding topological charges in the $SU(N)$ and $U(1)$ sectors are
\begin{eqnarray}
Q_{SU(N)}=\frac{m_{(N)}^2}{4}\bm\nu_a\cdot \bm \nu_a=\frac{m_{(N)}^2}{4}\left(1-\frac{1}{N}\right)\,,\quad
Q_{U(1)}=\frac{\left({\cal C}+\frac{m_{(N)}}{N}\right)^2}{4}\,.
\end{eqnarray}
In the following, we assume ${\cal C}, m_{(N)}\geq0$. 

The diagonal form of $A_{U(N)}$ makes it straightforward to construct the fundamental fermion zero modes by extending the analysis previously carried out for the $U(1)$ case (see also Appendix \ref{Fermions in the background of PSU(2) gauge field} for the treatment of fermions in the background of $PSU(2)$ bundles). The zero-mode solutions of the Dirac equation take the form
\begin{eqnarray}
\Psi=\left[\begin{array}{c}\lambda_\alpha\\0\end{array}\right]\,,\quad \quad
\lambda_{\alpha}=\left[\begin{array}{c}
\lambda_{\alpha, 1}\\
\lambda_{\alpha, 2}\\
...\\
\lambda_{\alpha,N}
\end{array}\right]\,.
\end{eqnarray}
Here, $\alpha=1,2$ denotes the spinor index (up or down), while the column vector $\lambda_\alpha$ transforms in the defining representation of $SU(N)$.  
The solutions of the Dirac equation are given by
\begin{eqnarray}\label{suNint}
\lambda_1={\cal G}_1(r)|j,m'=j,m\rangle\,, \quad 
\lambda_2={\cal G}_2(r)|j,m'=-j,m\rangle\,, 
\quad |m|\leq j\,,
\end{eqnarray}
where both ${\cal G}_1$ and ${\cal G}_2$ are $N\times 1$ column vectors.  
Substituting (\ref{suNint}) into the Dirac equation yields
\begin{eqnarray}\label{equation for lambdasuNfinal}\nonumber
\left[ \left(\frac{1}{f}\frac{\partial}{\partial r}+\frac{2f}{r}+\frac{1}{fr}\right)I_{N}
+\frac{a^2f}{r^3}\, \mbox{diag}_{N\times N}\!\left(m_{(N)}+{\cal C},{\cal C},\dots,{\cal C}\right)
+\frac{2fj}{r}I_{N}\right]{\cal G}_1&=&0\,,\\\nonumber
\left[ \left(\frac{1}{f}\frac{\partial}{\partial r}+\frac{2f}{r}+\frac{1}{fr}\right)I_{N}
-\frac{a^2f}{r^3}\, \mbox{diag}_{N\times N}\!\left(m_{(N)}+{\cal C},{\cal C},\dots,{\cal C}\right)
+\frac{2fj}{r}I_{N}\right]{\cal G}_2&=&0\,.\\
\end{eqnarray}
From the structure of these equations (in comparison with the $U(1)$ case), and assuming that both ${\cal C},m_{(N)}\geq 0$, it follows that only ${\cal G}_2$ (the spin-down component) admits normalizable zero modes. 

In order for the $SU(N)$ spinor to be globally well-defined under parallel transport along a contractible loop $\psi \to \psi+2\pi$ on $S^2$, one needs to restrict the total angular momentum $j$ of each component of the $N\times 1$ column vector $\lambda_2$ to either an integer or half-integer according to whether the entries in $\mbox{diag}_{N\times N}\!\left(m_{(N)}+{\cal C},{\cal C},\dots,{\cal C}\right)$ are even or odd.  
Thus, we have $4$ distinct cases
\begin{eqnarray}\nonumber
{\cal C}\in 2\mathbb Z,\quad m_{(N)}\in 2\mathbb Z\,,\qquad &\text{or}& \qquad
{\cal C}\in 2\mathbb Z+1,\quad m_{(N)}\in 2\mathbb Z\,,\quad \text{or} \\
{\cal C}\in 2\mathbb Z,\quad m_{(N)}\in 2\mathbb Z+1\,,\qquad &\text{or}& \qquad
{\cal C}\in 2\mathbb Z+1,\quad m_{(N)}\in 2\mathbb Z+1\,.
\end{eqnarray}
To determine the number of zero modes in each case, we simply apply the zero-mode counting result from (\ref{counting U(1)BU}) to each entry of the matrix 
$\mbox{diag}_{N\times N}\!\left(m_{(N)}+{\cal C},{\cal C},\dots,{\cal C}\right)$, 
and then sum over all contributions.

{\bf Case I}: ${\cal C}=2p_{(1)}\geq 0$, $m_{(N)}=2p_{(N)}\geq0$. 
\begin{eqnarray}
{\cal I}_\Box=\underbrace{(p_{(1)}+p_{(N)})^2}_{\mbox{1st entry}}+\underbrace{(N-1)p_{(1)}^2}_{\mbox{the rest}}=Np_{(1)}^2+p_{(N)}^2+2p_{(1)}p_{(N)}\,.
\end{eqnarray}

{\bf Case II}: ${\cal C}=2p_{(1)}+1\geq1$, $m_{(N)}=2p_{(N)}\geq 0$. 
\begin{eqnarray}
{\cal I}_\Box=\underbrace{(p_{(1)}+p_{(N)})(p_{(1)}+p_{(N)}+1)}_{\mbox{1st entry}}+\underbrace{(N-1)p_{(1)}(p_{(1)}+1)}_{\mbox{the rest}}\,.
\end{eqnarray}

{\bf Case III}: ${\cal C}=2p_{(1)}\geq0$, $m_{(N)}=2p_{(N)}+1\geq 1$. 
\begin{eqnarray}
{\cal I}_\Box=\underbrace{(p_{(1)}+p_{(N)})(p_{(1)}+p_{(N)}+1)}_{\mbox{1st entry}}+\underbrace{(N-1)p_{(1)}^2}_{\mbox{the rest}}\,.
\end{eqnarray}

{\bf Case IV}: ${\cal C}=2p_{(1)}+1\geq1$, $m_{(N)}=2p_{(N)}+1\geq 1$. 
\begin{eqnarray}
{\cal I}_\Box=\underbrace{(p_{(1)}+p_{(N)}+1)^2}_{\mbox{1st entry}}+\underbrace{(N-1)p_{(1)}(p_{(1)}+1)}_{\mbox{the rest}}\,.
\end{eqnarray}

Identical results can also be derived by applying the APS index theorem. In applying this theorem, one needs to calculate the $\eta$-invariant, which is sensitive to the holonomy of the  $U(N)$ gauge field at the boundary of EH space:
\begin{eqnarray}\label{inftyBCsun}\nonumber
\text{at}\, \partial\,(\mbox{EH}):\,
\Psi(r\rightarrow \infty, \theta, \varphi, \psi + 2\pi) 
&=&\exp\left[i\int_{\ell_\infty} A\right] \, \gamma_5 \, \Psi(r\rightarrow\infty, \theta, \varphi, \psi)\\\nonumber
&=&-\mbox{diag}_{N\times N}\!\left((-1)^{m_{(N)}+{\cal C}},(-1)^{{\cal C}},\dots,(-1)^{{\cal C}}\right)\\
&&\times  \gamma_5 \, \Psi(r\rightarrow\infty, \theta, \varphi, \psi)\,.
\end{eqnarray}
One then defines the \emph{signature} of the holonomy matrix $\mbox{Hol}=\exp\left[i\int_{\ell_\infty} A\right]$ to be the number of $(+1)$ eigenvalues 
minus the number of $(-1)$ eigenvalues, i.e.
\begin{equation}
  \mathrm{sgn}(\mathrm{Hol}) \;=\; \#(+1) - \#(-1)\,,
\end{equation}
which is gauge invariant. Then, the APS index is given by (see the derivation that leads to (\ref{indexmastermix})):
\begin{equation}\label{indexmastermixBULK}
  \mathcal{I} 
  \;=\; T_{\mathcal{R}} Q_{SU(N)}
      \;+\dim_{\mathcal{R}}Q_{U(1)}+ \frac{-\,\dim_{\mathcal{R}} + \mathrm{sgn}(\mathrm{Hol})}{8}\,.
\end{equation}
Further, using the Dynkin index and the dimension of the 
fundamental representation, $T_\Box = 1$ and $\mathrm{dim}_\Box = N$, we recover the above results of the $4$ cases.

In our treatment, we assumed that the fermion has a positive coupling to both $A_{U(1)}$ and $A_{SU(N)}$, meaning that the covariant derivative has been assumed to take the form (suppressing the coupling to gravity for notational simplicity) $D_\mu=\partial_\mu+iA_{U(1)}+iA_{SU(N)}$. In principle, we could assume different charges under $U(1)$ and $SU(N)$ fields, meaning that, now, the covariant derivatives take the form $D_\mu=\partial_\mu+iQ_{U(1)}A_{U(1)}+iQ_{SU(N)}A_{SU(N)}$. The charge $Q_{U(1)}$ can take any integer value, while $Q_{SU(N)}=\pm 1$ for the $SU(N)$ fundamental and anti-fundamental representations, respectively. This means that in this case, the normalized zero modes could have combinations of spin-up or spin-down components. However, assuming that none of the entries in the matrix $Q_{U(1)}A_{U(1)}+Q_{(SU(N))}A_{SU(N)}$ is zero, the above counting of zero modes holds\footnote{A zero entry corresponds to a free fermion component without flux, which yields no normalizable zero mode.}.

%%%%%%%%%%%%%%%%%%%%%%%%%%%%%%%%%%%%%%%%%%%%%
\section{Background of the $\mathbb Z_6^{(1)}$ $1$-form symmetry of the Standard Model: lepton- and baryon-number violation}
\label{Backgrounds for the Standard Model}
%%%%%%%%%%%%%%%%%%%%%%%%%%%%%%%%%%%%%%%%%%%%%

Now, we have all the machinery needed to apply our construction to the Standard Model  (SM) when placed in the background of EH space. 

The SM is based on the Lie algebra $\mathfrak{su}(3)\times \mathfrak{su}(2)\times \mathfrak{u}(1)$, and it exhibits two classical $0$-form symmetries $U(1)_{B-L}^{(0)}$ and $U(1)_{B+L}^{(0)}$ as well as a global $1$-form symmetry $\mathbb{Z}_6^{(1)}$. Here, $B$ and $L$ denote baryon and lepton number symmetries, respectively. We shall assume that the classical $U(1)_{B-L}^{(0)}$ and $U(1)_{B+L}^{(0)}$ symmetries are of good quality, meaning that there is no higher-dimensional operator (originating from beyond the SM physics) that breaks them explicitly. The classical $U(1)_{B+L}^{(0)}$ symmetry is broken by weak instantons to a discrete subgroup $\mathbb{Z}^{(0)}_{2n_f}$, where $n_f = 3$ is the number of SM generations.

The kinetic energy terms of the gauge field and fermion fields in the SM are\footnote{We adopt the convention that traces in the fundamental representation are normalized with the Dynkin index $T_\Box = 1$. Moreover, we include an additional factor of $1/2$ in front of the $U(1)$ kinetic term. With this choice, the standard relation between topological charges and the corresponding actions, as presented in Eq.~(\ref{SMACTIONTC}), is correctly reproduced.}
\begin{eqnarray}\label{UV KETERMS}\nonumber
S_{SM}&\supset& \frac{1}{2g_1^2}\int F_{(1)}\wedge \star F_{(1)}+ \frac{1}{g_2^2}\int\mbox{tr} F_{(2)}\wedge\star F_{(2)}+ \frac{1}{g_3^2}\int\mbox{tr} F_{(3)}\wedge\star F_{(3)}\\
&&+i\sum_{\lambda}\bar\lambda \bar \sigma^\mu D_\mu \lambda\,,
\end{eqnarray}
where we denote the gauge fields and  gauge couplings of $SU(3)$, $SU(2)$, and $U(1)$ as $A_{(3)}$,  $A_{(2)}$,  $A_{(1)}$, and  $g_3$, $g_2$, and $g_1$, respectively. The SM fermions are taken to be left-handed Weyl fermions, generally denoted by $\lambda$. The covariant derivatives (suppressing the coupling to gravity for notational simplicity) acting on the fermions $\lambda$ are given by
\begin{eqnarray}
D_\mu\lambda= (\partial_u+iQ_{(1)\lambda} A_{(1)\mu}+i Q_{(2)\lambda} A_{(2)\mu}+i Q_{(3)\lambda} A_{(3)\mu})\lambda\,,
\end{eqnarray} 
where $Q$, with appropriate subscripts, refers to the charges under $SU(3)\times SU(2)\times U(1)$.   Importantly, the $U(1)$ hypercharge coupling $g_1$ is related to the conventional hypercharge coupling $g_Y$ via \begin{eqnarray}g_1 = \frac{g_Y}{6}\,,\end{eqnarray} owing to the assumption that $U(1)$ is compact and all charges are integral. 
The charge assignments for SM matter fields are standard, except that a factor of $6$ rescales the hypercharges:
\begin{equation}
\label{charges}
\begin{array}{|c|ccc|cc|}\hline
\text{field} & SU(3) & SU(2) & U(1)  & U(1)_B & U(1)_L \\\hline
q_{L} & \square & \square & 1 & \frac{1}{3} & 0 \\
l_{L} & \mathbf{1} & \square & -3 & 0 & 1 \\
\tilde\nu_{R}& \mathbf{1} & \mathbf{1} & 0 &0&-1\\
\tilde{e}_{R} & \mathbf{1} & \mathbf{1} & 6 & 0 & -1 \\
\tilde{u}_{R} & \overline{\square} & \mathbf{1} & -4 & -\frac{1}{3} & 0 \\
\tilde{d}_{R} & \overline{\square} & \mathbf{1} & 2 & -\frac{1}{3} & 0 \\
h & \mathbf{1} & \square & 3 & 0 & 0\\\hline
\end{array}~,
\end{equation}
and $q_L$ and $l_L$ are the quark and lepton doublets
\begin{eqnarray}
q_{L}=\left(\begin{array}{c} u_{L}\\d_{L}\end{array}\right)\,, \quad l_{L}=\left(\begin{array}{c} \nu_{L}\\e_{L}\end{array}\right)\,.
\end{eqnarray}
  For reference, the last two columns display the standard baryon and lepton number assignments.   
  
It is possible to gauge the $\mathbb{Z}_6^{(1)}$ symmetry, or any of its subgroups.  
As a result, the Standard Model (SM) gauge group can be written as
\begin{equation}\label{SMGAUGEGROUP}
G^{\scriptsize\mathrm{SM}}_k \;\equiv\; 
\frac{SU(3)\times SU(2)\times U(1)}{\mathbb{Z}_k}\,, 
\qquad k = 1,2,3,6 \,,
\end{equation}
where the precise value of $k$ realized in nature remains an open question.  
In \cite{Anber:2021upc}, the gauging of  $\mathbb{Z}_6^{(1)}$ symmetry or a subgroup thereof was implemented by turning on fractional fluxes on a small four-dimensional torus $\mathbb{T}^4$ (small compared to the Higgs VEV) and summing over all such configurations.  
Here, we pursue an alternative approach: we carry out the gauging by placing the SM on EH space and turning on fractional fluxes localized at the bolt.  
This entails studying the corresponding consistency conditions and deriving explicit formulas for the fermion zero modes.

Before proceeding, let us remark that the presence of the Higgs field spoils the exact self-duality of the EH solution. This effect is particularly transparent in the absence of background gauge fields: while the EH gravitational action vanishes in that case (prior to turning on the Higgs), the inclusion of the Higgs VEV leads to a nonzero contribution to the action, dominated by the Higgs sector. This situation closely parallels the electroweak theory, where no exact Yang-Mills instanton survives once the Higgs acquires a vacuum expectation value. 
Nevertheless, in both contexts, one may consistently treat the instanton as an approximate configuration. In the electroweak setting, this is realized through the constrained (or valley) construction, which introduces a Higgs profile, renders the instanton size integral finite, preserves the fermionic zero modes underlying the ’t~Hooft operator, and reproduces the familiar $B+L$-violating processes (exponentially suppressed at zero temperature, but operative via sphalerons at high temperature); see \cite{Shifman:2012zz} for a pedagogical treatment. Whether an analogous construction can be developed for the EH geometry in the presence of the Higgs remains an open question. 
It is worth emphasizing, however, that the EH instanton contains a free parameter, $a$, constrained only to exceed the inverse Planck length\footnote{We emphasize that in this work we adopt a semi-classical treatment: all processes are studied in a fixed gravitational background, neglecting fluctuations of the spacetime geometry. Consequently, the scale parameter $a$ is treated as fixed, rather than being integrated over.}. If, in addition, one assumes $a$ to lie well below the inverse electroweak scale, then in the presence of background gauge fields, the backreaction of the Higgs VEV on the EH geometry is negligible, since the dominant contribution to the action (which comes solely from the action of the gauge fields) originates in the region near $a$.

Our starting point is to turn on a $\mathbb{Z}_6^{(1)}$ flux, which is realized by exciting fractional fluxes in the $SU(3)$, $SU(2)$, and $U(1)$ sectors.  
Following the same construction introduced in Section~\ref{The background of UN gauge bundle and}, the corresponding gauge field configurations take the form
\begin{eqnarray}\nonumber
A_{(3)}&=&m_{(3)}\,\mathrm{diag}\!\left(1-\tfrac{1}{3},-\tfrac{1}{3},-\tfrac{1}{3}\right)\frac{\sigma_z a^2}{r^2}\,, 
\qquad
A_{(2)}=m_{(2)}\,\mathrm{diag}\!\left(1-\tfrac{1}{2},-\tfrac{1}{2}\right)\frac{\sigma_z a^2}{r^2}\,,\\
A_{(1)}&=&\Bigg({\cal C}+\frac{m_{(2)}}{2}+\frac{m_{(3)}}{3}\Bigg)\frac{\sigma_z a^2}{r^2}\,.
\end{eqnarray}
Each of these background fields contributes a quantized flux through the two-sphere $S^2$ inside the EH space.:
\begin{eqnarray}\nonumber
\int_{S^2}F_{(3)}&=&-2\pi m_{(3)}\,\mathrm{diag}\!\left(\tfrac{2}{3},-\tfrac{1}{3},-\tfrac{1}{3}\right)\;\in\;2\pi\mathbb{Z}_3\,, \\ \nonumber
\int_{S^2}F_{(2)}&=&-2\pi m_{(2)}\,\mathrm{diag}\!\left(\tfrac{1}{2},-\tfrac{1}{2}\right)\;\in\;2\pi\mathbb{Z}_2\,,\\
\int_{S^2}F_{(1)}&=&-2\pi\left({\cal C}+\tfrac{3m_{(2)}+2m_{(3)}}{6}\right)\;\in\;2\pi\mathbb{Z}_6\,.
\end{eqnarray}
Finally, the corresponding topological charges associated with each sector are
\begin{eqnarray}
Q_{(3)}=\frac{m_{(3)}^2}{6}\,, 
\qquad Q_{(2)}=\frac{m_{(2)}^2}{8}\,, 
\qquad Q_{(1)}=\frac{1}{4}\left({\cal C}+\frac{3m_{(2)}+2m_{(3)}}{6}\right)^2\,.
\end{eqnarray}

The gauge field that couples to a fermion $\lambda$ via the covariant derivative takes the general form
\begin{eqnarray}\label{mostgeneralabelian}
A_\lambda=Q_{(3)\lambda}A_{(3)}\otimes I_2+Q_{(2)\lambda} I_3\otimes A_{(2)}+Q_{(1)\lambda} A_{(1)}I_3\otimes I_2\,. 
\end{eqnarray}
The explicit form of $A_\lambda$ for the SM quarks, leptons, and the Higgs is
\begin{eqnarray}\label{BGFSM}\nonumber
A_{q_L}&=&\mbox{diag}_{6\times 6}\left({\cal C}+m_{(2)}+m_{(3)},{\cal C}+m_{(2)},{\cal C}+m_{(2)}, {\cal C}+m_3, {\cal C},{\cal C}\right)\,,\\\nonumber
A_{l_L}&=&\mbox{diag}_{2\times 2}\left(-3{\cal C}-m_{(2)}-m_{(3)},-3{\cal C}-2m_{(2)}-m_{(3)}\right)\,,\\\nonumber
A_{\tilde e_R}&=&6{\cal C}+3m_{(2)}+2m_{(3)}\,,\\\nonumber
A_{\tilde u_R}&=&\mbox{diag}_{3\times 3}\left(-4{\cal C}-2m_{(2)}-2m_{(3)},-4{\cal C}-2m_{(2)}-m_{(3)},-4{\cal C}-2m_{(2)}-m_{(3)}\right)\,,\\\nonumber
A_{\tilde d_R}&=&\mbox{diag}_{3\times 3}\left(2{\cal C}+m_{(2)},2{\cal C}+m_{(2)}+m_{(3)},2{\cal C}+m_{(2)}+m_{(3)}\right)\,,\\
A_{h}&=&\mbox{diag}_{2\times 2}\left(3{\cal C}+2m_{(2)}+m_{(3)},3{\cal C}+m_{(2)}+m_{(3)} \right)\,.
\end{eqnarray}

In the next two subsections, we will first demonstrate how to introduce the background of
$\mathbb{Z}_3^{(1)} \subset \mathbb{Z}_6^{(1)}$, and then the background  of the full 
$\mathbb{Z}_6^{(1)}$ symmetry. The case of  
$\mathbb{Z}_2^{(1)} \subset \mathbb{Z}_6^{(1)}$ backgrounds follows in an analogous manner.  
We will devote considerable attention to the case of
$\mathbb{Z}_3^{(1)}$, including various sub-cases, as this will clarify many aspects 
of the procedure and highlight the essential features. Building on this example, 
the discussion of the background of the full $\mathbb{Z}_6^{(1)}$ symmetry follows the same steps. 
For brevity, we will not go through all sub-cases again, but instead present a single 
illustrative example.

%%%%%%%%%%%%%%%%%%%%%%%%%%%%%%%%%%%%
\subsection{Background of $\mathbb Z_3^{(1)}$ $1$-form symmetry}
%%%%%%%%%%%%%%%%%%%%%%%%%%%%%%%%%%%%

In this section we discuss the background of $\mathbb Z_3^{(1)}\subset\mathbb Z_6^{(1)}$ symmetry. Gauging this symmetry, i.e., summing over  $\mathbb Z_3^{(1)}$ gauge bundles,  leaves behind a global $\mathbb Z_2^{(1)}$ $1$-form symmetry. 

To restrict to backgrounds of $\mathbb Z_3^{(1)}$, it is enough to set $m_{(2)}\in 2\mathbb Z$. Yet, to further simplify the analysis, we will also set $m_{(3)}\in 2\mathbb Z$. This choice leads to two distinct possibilities:
\begin{eqnarray}\label{SMCONSI}
{\cal C}\in 2\mathbb{Z},\quad m_{(3)},m_{(2)}\in 2\mathbb{Z}\,,\qquad \text{or} \qquad
{\cal C}\in 2\mathbb{Z}+1,\quad m_{(3)},m_{(2)}\in 2\mathbb{Z}\,,
\end{eqnarray}
making the analysis less cumbersome. However, we also emphasize that more general cases are possible. 
Assuming ${\cal C}, m_{(2)}, m_{(3)} \geq 0$, one may then use (\ref{counting U(1)BU}) or the APS index to determine the number of fermion zero modes for all species. 
The restriction $m_{(2)} \in 2\mathbb{Z}$ modifies the flux quantization rules to
\begin{eqnarray}\label{importantquantizationconditions}
\int_{S^2}F_{(3)}\;\in\;2\pi\mathbb{Z}_3\,, \quad
\int_{S^2}F_{(2)}\;\in\;2\pi\mathbb{Z}\,, \quad
\int_{S^2}F_{(1)}\;\in\;2\pi\mathbb{Z}_3\,.
\end{eqnarray}
While the quantization condition on $F_{(2)}$ appears integer-valued, the associated topological charge of the weak sector turns out to be fractional.  
Writing $m_{(2)}=2p_2$ and $m_{(3)}=2p_3$, we find the topological charges
\begin{equation}
Q_{(3)}=\frac{2p_{3}^2}{3}\,, \quad
Q_{(2)}=\frac{p_2^2}{2}\,,\quad
Q_{(1)}=\frac{1}{4}\left({\cal C}+p_2+\frac{2p_3}{3}\right)^2\,.
\end{equation}

Inspecting (\ref{BGFSM}), we remark that transporting any $SU(2)$-charged field---namely $q_L$, $l_L$, and $h$---around a closed loop near the bolt results in a trivial or phase of $\pi$ (holonomy of $(-1)$), depending on whether ${\cal C}$ is even or odd. In contrast, all fields uncharged under $SU(2)$ invariably acquire a trivial phase. This distinction originates from the fact that all $SU(2)$-charged fields carry odd charges under the $U(1)$ hypercharge gauge symmetry. Therefore, in applying the APS index theorem, and whenever ${\cal C}$ is even, one must always select the boundary flat line bundle with a trivial holonomy that yields $\eta=-1/8$ in (\ref{etavalrp3}). However, when ${\cal C}$ is odd, the sign of the $\eta$-invariant will depend on the fermion species: we must take $\eta=+1/8$ in (\ref{etavalrp3}) for fields charged under $SU(2)$ (all such fields carry odd charges under $U(1)$ hypercharge gauge field) and  $\eta=-1/8$ for the rest of fields. We remind the reader that the two possible twists, $\{+1,-1\}$, on $\mathbb{RP}^3$ arise from the bulk background gauge field. When restricted to the boundary, this bulk field defines a flat line bundle whose holonomy determines the choice of twist, and hence fixes one of the two possible values of the $\eta$-invariant.

One fears that if fields that couple to each other, e.g., via mass terms, are twisted differently, the couplings may become ill-defined. However, in the SM, the fermion masses are acquired via coupling to the Higgs:
\begin{eqnarray}\label{H-FERMION COUPLING}\nonumber
{\cal L}_{HF}&=&
g_{u}^{}\epsilon_{ij}q^{a\,i}_{L\, }h^j\tilde u_{R\, a\,}+g_{d}^{}\delta_{ij}q^{a\,i}_{L\, }h^{*j}\tilde d_{R\, a\,}+g_{\nu}^{}\epsilon_{ij}L^{i}_{L\, }h^j\tilde \nu_{R\,}+g_{e}^{}\delta_{ij}L^{i}_{L\, }h^{*j}\tilde e_{R\,}+\mbox{h.c.}\,.\\
\end{eqnarray}
In this expression, we have employed the index notation to make the structure of the Higgs-fermion interactions explicit. The index $a = 1,2,3$ labels the $SU(3)$ color degrees of freedom, $i = 1,2$ runs over the components of the $SU(2)$ weak doublets. The constant $g_{u}^{}$ through $g_{e}^{}$ are the Yukawa couplings. All terms are gauge invariant under $SU(3)\times SU(2)\times U(1)$. Therefore, each term is inert to the background gauge fields (\ref{BGFSM}), and inconsistencies cannot arise.

Now, we turn to counting the fermion zero modes for the two cases. 

{\bf Case I}: We take  ${\cal C}=2p_{1}, m_{(2)}=2p_2, m_{(3)}=2p_3$. 
 The counting of the zero modes ${\cal I}_\lambda$ for the case is displayed in the following table. The counting follows from (\ref{counting U(1)BU}) by summing the contributions of the zero modes associated with each diagonal component, which couples the background to the fermions through the covariant derivatives and (\ref{BGFSM}). Equivalently, the same result can be derived from the APS index formula (\ref{indexmastermixBULK}):
\begin{equation}
\label{zerom modes SM I}
\begin{array}{|c||c|c|}\hline
\text{field} & {\cal I}_\lambda   \\\hline
q_{L} &6 p_1^2 + 6 p_1 p_2 + 3 p_2^2 + 4 p_1 p_3 + 2 p_2 p_3 + 2 p_3^2   \\
l_{L} &18 p_1^2 + 18 p_1 p_2 + 5 p_2^2 + 12 p_1 p_3 + 6 p_2 p_3 + 2 p_3^2   \\
\tilde\nu_{R} & 0\\
\tilde{e}_{R} &36 p_1^2 + 36 p_1 p_2 + 9 p_2^2 + 24 p_1 p_3 + 12 p_2 p_3 + 4 p_3^2   \\
\tilde{u}_{R} &48 p_1^2 + 48 p_1 p_2 + 12 p_2^2 + 32 p_1 p_3 + 16 p_2 p_3 + 6 p_3^2   \\
\tilde{d}_{R} &12 p_1^2 + 12 p_1 p_2 + 3 p_2^2 + 8 p_1 p_3 + 4 p_2 p_3 + 2 p_3^2  \\\hline
\end{array}~\,.
\end{equation}
An effective 't~Hooft vertex takes the form 
\begin{equation}
{\cal V} \;=\; e^{-S_{(3)}-S_{(2)}-S_{(1)}} \prod_{\lambda}\lambda^{{\cal I}_\lambda}\,,
\end{equation}
which is manifestly Lorentz invariant and gauge invariant under the full 
$SU(3)\times SU(2)\times U(1)$ symmetry group\footnote{The invariance under Lorentz and $SU(2)$ symmetries follows from the fact 
that the total number of zero modes is even. Invariance under $SU(3)$ is 
ensured by balancing the group indices carried by the left-handed and 
right-handed fermions, while invariance under the $U(1)$ hypercharge guage field can be 
verified directly by examining the transformation of ${\cal V}$ under 
the  $U(1)$ hypercharge gauge field. 
}. 

Using (\ref{UV KETERMS}), the actions $S_{(i)}$ of the self-dual gauge fields are determined from
their corresponding topological charges:
\begin{equation}\label{SMACTIONTC}
S_{(1)} \;=\; \frac{4\pi^2 |Q_{(1)}|}{g_{1}^2}\,,\quad S_{(i)} \;=\; \frac{8\pi^2 |Q_{(i)}|}{g_{i}^2}\,, 
\qquad i=2,3,
\end{equation}
while the Einstein-Hilbert action vanishes due to the self-duality of the 
EH curvature two-form. The fact that the $U(1)$ hypercharge action is half the action of the $SU(2)$ and $SU(3)$ sectors originates from the normalization of the kinetic energy terms (\ref{UV KETERMS}). 

The resulting vertex mediates baryon- and lepton-number violating processes, 
with the net violation (per family) given by\footnote{Discussions of CP violation in QED in the background of EH instantons appeared before in \cite{Arunasalam:2018eaz}; see also \cite{Deser:1980kc}.}
\begin{equation}\label{delbdellcaste2}
\Delta B \;=\; \Delta L \;=\; -18 p_1^2 - 18 p_1 p_2 - 4 p_2^2 
 - 12 p_1 p_3 - 6 p_2 p_3 - 2 p_3^2 \,.
\end{equation}

{\bf Case II}: The counting of the zero modes ${\cal I}_\lambda$ for the case ${\cal C}=2p_{1}+1, m_{(2)}=2p_2, m_{(3)}=2p_3$ is displayed in the following table. 
\begin{equation}
\label{zerom modes SM II}
\begin{array}{|c||c|c|}\hline
\text{field} & {\cal I}_\lambda   \\\hline
q_{L} &6 p_1 + 6 p_1^2 + 3 p_2 + 6 p_1 p_2 + 3 p_2^2 + 2 p_3 + 4 p_1 p_3 + 2 p_2 p_3 + 
 2 p_3^2 \\
l_{L} &4 + 18 p_1 + 18 p_1^2 + 9 p_2 + 18 p_1 p_2 + 5 p_2^2 + 6 p_3 + 12 p_1 p_3 + 
 6 p_2 p_3 + 2 p_3^2   \\
\tilde\nu_{R} & 0\\
\tilde{e}_{R} & 9 + 36 p_1 + 36 p_1^2 + 18 p_2 + 36 p_1 p_2 + 9 p_2^2 + 12 p_3 + 24 p_1 p_3 + 
 12 p_2 p_3 + 4 p_3^2  \\
\tilde{u}_{R} & 12 + 48 p_1 + 48 p_1^2 + 24 p_2 + 48 p_1 p_2 + 12 p_2^2 + 16 p_3 + 
 32 p_1 p_3 + 16 p_2 p_3 + 6 p_3^2  \\
\tilde{d}_{R} & 3 + 12 p_1 + 12 p_1^2 + 6 p_2 + 12 p_1 p_2 + 3 p_2^2 + 4 p_3 + 8 p_1 p_3 + 
 4 p_2 p_3 + 2 p_3^2  \\\hline
\end{array}~\,.
\end{equation}
Again, a Lorentz and gauge-invariant effective 't Hooft vertex mediates baryon-number and lepton-number-violating processes:
\begin{eqnarray}\label{delbdelboddc}\nonumber
\Delta B=\Delta L=-5 - 18 p_1 - 18 p_1^2 - 9 p_2 - 18 p_1 p_2 - 4 p_2^2 - 6 p_3 - 12 p_1 p_3 - 
 6 p_2 p_3 - 2 p_3^2\,.\\
\end{eqnarray}

The fact that $\Delta B=\Delta L$ in both cases, even and odd ${\cal C}$, is attributed to the fact that the EH spaces do not support fermion zero modes as we switch off the background fields. This means that $U(1)_{B-L}$ symmetry is an exact symmetry of the SM in the EH space, even in the presence of $1$-form fluxes.

 We use the amplitude of the ’t~Hooft vertex  
$
|{\cal V}|\sim e^{-S_{(1)} - S_{(2)} - S_{(3)}}
$
to estimate the strength of baryon- and lepton-number violation in the background of an EH instanton. We assume that the instanton scale $a$ is much smaller than the electroweak symmetry-breaking scale. The dominant contributions to the actions $S_{(i)}$ arise from the region near the bolt; consequently, the parameter $a$ effectively acts as an infrared cutoff for the running of the gauge couplings.  

To evaluate the couplings $g_{i}(a)$ at the scale $a$, we evolve them from the weak scale, taken to be the $Z$-boson mass $M_Z$, up to $a$. The running is governed by  
\begin{eqnarray}
\nonumber
\frac{8\pi^2}{g_{i}^2(a)} &=& \frac{8\pi^2}{g_{i}^2(M_Z)} + b_i \log\!\left(\frac{1}{aM_Z}\right),
\qquad i = 1,2,3 \\[6pt]
b_1 &=& -(80 n_f + 6 n_H), \qquad 
b_2 = \frac{22}{3} - \frac{4}{3}n_f - \frac{n_H}{6}, \qquad 
b_3 = 11 - \frac{4}{3}n_f,
\label{running of couplings}
\end{eqnarray}  
for $n_f$ fermion families and $n_H$ Higgs doublets.  
The coefficients $b_2$ and $b_3$ coincide with the standard results (see, e.g., \cite{Martin:1997ns}), while the $U(1)$ coupling $g_1$ and its $\beta$-function differ from the conventional hypercharge normalization through the relation  
$
g_1 = \frac{g_Y}{6}
$.
For numerical input, we use  
$
\frac{g_3^2(M_Z)}{4\pi} \simeq 0.118, 
\frac{g_{\text{EM}}^2(M_Z)}{4\pi} \simeq \frac{1}{128},
$
together with  
$
g_Y(M_Z) = \frac{g_{\text{EM}}(M_Z)}{\cos\theta_W(M_Z)}, 
\qquad 
g_2(M_Z) = \frac{g_{\text{EM}}(M_Z)}{\sin\theta_W(M_Z)}, 
\qquad 
\sin^2\theta_W(M_Z) \simeq 0.23
$ , 
where $\theta_W(M_Z)$ denotes the Weinberg angle at the $Z$-mass scale.  

Taking $n_f = 3$ and $n_H = 1$, and setting ${\cal C} = m_{(2)} =0, m_{(3)} = 2$ (as an example) with $a^{-1}\sim 10^{16}$ GeV,  Eq.~(\ref{delbdellcaste2})  gives  
$\Delta_B = \Delta_L = -2,
$
which corresponds to a ’t~Hooft vertex with amplitude  
$
|{\cal V}| \sim 10^{-550},
$
an extraordinarily small value. By varying $m_{(2)}$, $m_{(3)}$, and ${\cal C}$, the amplitude can vary slightly, but it remains extremely suppressed. This strong suppression arises from the smallness of the $U(1)$ hypercharge gauge coupling constant.  

%%%%%%%%%%%%%%%%%%%%%%%%%%%%%%%%%%%%
\subsection{Background of the full $\mathbb Z_6^{(1)}$ $1$-form symmetry}
%%%%%%%%%%%%%%%%%%%%%%%%%%%%%%%%%%%%

Analyzing the general case of the full $\mathbb Z_6^{(1)}$ $1$-form symmetry encompasses many cases. Instead, we shall consider a single example by using the minimal fluxes\footnote{The fluxes with ${m_{(2)}}=m_{(3)}=-{\cal C}=1$ do not lead to any normalizable zero modes.}, setting ${m_{(2)}}=m_{(3)}=1,{\cal C}=0$. The zero modes in this case are
\begin{equation}
\label{zerom modes SM II full gauge horizontal}
\begin{array}{|c||c|c|c|c|c|c|}\hline
\text{field} & q_{L} & l_{L} & \tilde\nu_{R} & \tilde{e}_{R} & \tilde{u}_{R} & \tilde{d}_{R} \\\hline
{\cal I}_\lambda & 1 & 3 & 0 & 6 & 8 & 2 \\\hline
\end{array}\,,
\end{equation}
and the 't Hooft vertex is
\begin{eqnarray}
{\cal V}=e^{-S_{(3)}-S_{(2)}-S_{(1)}} q_Ll_L^{3}\tilde e_R^{6}\tilde u_R^{8}\tilde d_R^{2}\,,
\end{eqnarray}
which is invariant under $SU(3)\times SU(2)\times U(1)$ and mediates baryon- and lepton-number violations $\Delta B=\Delta L=-3$. Taking $a^{-1}\sim 10^{16}$ GeV, we find that the amplitude of the vertex is extremely small $\sim 10^{-5268}$.

%%%%%%%%%%%%%%%%%%%%%%%%%%%%%%
\section{EH spaces as saddles in the path integral: gauging the SM $1$-form symmetry}
\label{EH spaces as saddles in the path integral}
%%%%%%%%%%%%%%%%%%%%%%%%%%%%%%

Although Eguchi–Hanson instantons are among the most thoroughly explored spaces in mathematical physics, their precise role in the semiclassical description of quantum gravity remains elusive—largely because no clear Lorentzian interpretation arises from a straightforward Wick rotation. Yet,
't Hooft provided a physical interpretation of these instantons \cite{tHooft:1988wxy}, and building on this perspective, 
we will argue how their inclusion in the path integral effectively gauges the SM $1$-form symmetry. Our approach will be based on the semi-classical description of gravity \cite{Hawking:1978ghb}, i.e., we are working in the limit $G_N\rightarrow 0$. In this limit, we are dealing with a local quantum field theory in the background of a classical curved spacetime, and we do not account for the spacetime fluctuations; see \cite{Held:2025mai} for a recent review of this approach and discussion of a few salient points.

Let us work in the coordinates $(x,y,z,t)$ introduced in (\ref{hoopfxyzt}). As we discussed in Section \ref{Eguchi-Hanson instantonsbulk} (see also Appendix \ref{The Eguchi-Hanson geometry}),  restricting the range of $\psi$ to $0\leq \psi\leq 2\pi$ implies that we are imposing the anti-podal identification $(x,y,z,t)\sim-(x,y,z,t)$. Thus,  the space surrounding the EH instanton is really a half-space.  Under this identification the background $1$-form gauge potential flips its sign \begin{equation}\label{IdentA}A^{(B)}_\mu(x,y,z,t)=-A^{(B)}_\mu(-x,-y,-z,-t)\,,\end{equation} as can be easily seen from expression (\ref{form of ABUL}) written in the $(x,y,z,t)$ coordinates; see Eq. (\ref{form of A}). Notice that we used the superscript $(B)$ to distinguish the background from general field configurations we discuss later. Also,  to avoid clutter, here we use $A$ to represent all the abelian fields: the $U(1)$ hypercharge gauge field as well as abelian components along the Cartan generators of $SU(3)\times SU(2)$ as in (\ref{mostgeneralabelian}).   The field strength of the given background, on the other hand, is invariant under this identification: \begin{eqnarray}\label{IdentF}F^{(B)}_{\mu\nu}(x,y,z,t)=F^{(B)}_{\mu\nu}(-x,-y,-z,-t)\,,\end{eqnarray} see, e.g., Eq. (\ref{selfdual F}).

Next, let us recall how we think about the path integrals in field theory: they give the transition amplitude from some initial state, at some initial time, which we may take to be $t=t_0$, to a final state at a later time, and it is up to us to specify the states we are interested in; these are boundary conditions on the path integral. The states in Hilbert space live on hyper-surfaces corresponding to constant-time slices. Let us take the final state (the out state) to be the vacuum $|\Omega\rangle$, while the initial state (the state that corresponds to the EH instanton) will be prepared to respect the identification $(x,y,z,t)\sim-(x,y,z,t)$, which sets at $t=t_0$ the antipodal constraint 
\begin{equation}\label{initialA}
A_\mu(x,y,z,t_0)=-A_\mu(-x,-y,-z,t_0)\,.
\end{equation}
$A_\mu$ are the fluctuations around the EH background $A_{\mu}^{(B)}$, and both respect the antipodal constraint\footnote{\label{aboutparity}In the EH geometry, the anti-podal identification is 
$
(x,y,z,t) \;\sim\; (-x,-y,-z,-t)
$. In our path integral treatment, however, it is sufficient to impose a constraint on the initial data at some Euclidean time slice $t=t_0$. Specifically, for the 
gauge potential, we require
$
A_\mu(x,y,z,t_0) \;=\; - A_\mu(-x,-y,-z,t_0)
$, which enforces the correct parity under spatial inversion. The Euclidean time evolution then propagates this condition consistently into the bulk.}. 
We shall assume $t_0>a$, since the region $x^2+y^2+z^2+t^2<a^2$ is excluded\footnote{This, however, does not imply that the spacetime ends abruptly at the bolt; EH space is smooth and geodesically complete, and elliptic self-adjoint operators in this space, e.g., the Laplacian, the Dirac operator, etc., furnish a complete set of eigenstates without imposing further boundary conditions at ``the would be end of space".}. Condition (\ref{initialA})  reflects the fact that the gauge bundle on EH has a nontrivial $\mathbb Z_2$ transition function (of geometric origin) across the antipodal map, which can be made explicit in the chosen gauge. This constraint is not arbitrary; it is a physical requirement, up to a gauge transformation, if we want a path integral over $A_\mu$ to describe the transition amplitude in the background of the EH instanton\footnote{To add to Footnote~\ref{aboutparity}, one could in principle choose the 
initial slice at $t_0=0$, in which case the antipodal identification would 
imply
$
A_\mu(x,y,z,0) \;=\; -\,A_\mu(-x,-y,-z,0)
$. However, the state defined at $t_0=0$, namely $|\mathrm{EH}\rangle_0$, is not 
well defined because the geometry requires $x^2+y^2+z^2+t^2 \geq a^2$ 
(see~\cite{tHooft:1988wxy} for further discussion). 
The resolution is to evolve this formal state forward in Euclidean time 
using the Hamiltonian $H$, leading to the well-defined state
$
|\mathrm{EH}\rangle_{t_0} \;=\; e^{-t_0 H}\,|\mathrm{EH}\rangle_0
$
.}.

We divide the space into two regions. The first region is covered by the coordinates $\bm x\equiv (x,y,z)$, let us call it region $R$, and the second is the diagonally opposite region, covered by the coordinates $-\bm x\equiv (-x,-y,-z)$, let us call it region $L$. If we prepare the initial field configuration at $t=t_0$ to respect the identification (\ref{initialA}), then field values in the region covered by $\bm{x}$ and those in the opposite region $-\bm{x}$ are correlated accordingly. The resulting state on the full Hilbert space is globally pure but not separable; it is not of the form $|R\rangle\otimes|L\rangle$, but instead an entangled state of the form\footnote{\label{discretized}Expression (\ref{EHMAINSTATE}) is a condensed notation for the entangled state in the continuum. It may help to write the corresponding discretized state.  On a lattice with $N$ sites in region $L$ and $N$ sites in region $R$ (in
each of the spatial $3$ directions $\mu=1,2,3$), this state can be written as  
 \begin{equation}
|\mathrm{EH}\rangle_{t_0}
\;\propto\;
\prod_{i=1}^{N}\prod_{\mu=1}^{3}
\left( \frac{1}{N_\theta}\sum_{n_{\mu,i}=0}^{N_\theta-1}
   \bigl|U_{\mu,i}^{(n_{\mu,i})}\bigr\rangle_R 
   \otimes 
   \bigl|U_{\mu,i}^{(N_\theta-n_{\mu,i})}\bigr\rangle_L
\right)\,.
\end{equation}
For each site $i$ and direction $\mu$, the discretized link variable is defined by  
\begin{equation}
   U_{\mu,i}^{(n_{\mu,i})} 
   \;=\; \exp\!\left( i\,\theta_{\mu,i}^{(n_{\mu,i})} \right),
   \qquad 
   \theta_{\mu,i}^{(n_{\mu,i})} \;=\; \frac{2\pi}{N_\theta}\, n_{\mu,i},
\end{equation}
with 
\begin{equation}
   n_{\mu,i} \in \{0,1,\ldots,N_\theta-1\}.
\end{equation}
The left link variable is related to the right one by inversion, 
$(U_{\mu,i}^{(n_{\mu,i})})^{-1}=
U_{\mu,i}^{(N_\theta-n_{\mu,i})}$, reflecting the continuum condition 
$A_\mu(-x)=-A_\mu(x)$. In the continuum limit $N_\theta \to \infty$, this reduces to the original integral expression (\ref{EHMAINSTATE}). However, notice that we have not yet imposed any gauge conditions on the state in (\ref{EHMAINSTATE}); more on that at the end of this section.
}
\begin{equation}\label{EHMAINSTATE}
|\mathrm{EH}\rangle_{t_0}={\cal N}\; \prod_{\bm x, \mu=1,..,4}\int  dA_\mu \;\; |A_\mu(\bm{x})\rangle_R \otimes |A_\mu(-\bm{x})\rangle_L\,,
\end{equation}
where ${\cal N}$ is a normalization constant, and the antipodal constraint (\ref{initialA}) is imposed. The ket $|A_\mu(\bm{x})\rangle$ stands for an eigenvector of the field operator $\hat A(\bm x)$ such that $\hat A(\bm x)|A_\mu(\bm{x})\rangle=A(\bm x)|A_\mu(\bm{x})\rangle$, for both the $R$ and $L$ regions.    Then, the conditional path integral in the background of the EH instanton is given by (ignoring the fermions for now)
\begin{eqnarray}\label{mainEHPI}
\langle \Omega|\mathrm{EH}\rangle_{t_0}^{m_{(3)}, m_{(2)},{\cal C}}\sim\int [\mathcal D  A_\mu]\Bigg|_{\substack{A_\mu(\bm x, t_0)=A_\mu(\bm x)\\A_\mu(-\bm x, t_0)=-A_\mu(\bm x)} }\;e^{-S}\,.
\end{eqnarray}
The superscripts $m_{(3)}, m_{(2)},{\cal C}$ indicate that this is the transition amplitude in the presence of the background fields of the $\mathbb Z_6^{(1)}$ $1$-form symmetry. 
Here, $S$ stands for the sum of the actions of the various self-dual background gauge fields $A^{(B)}$ as well as the quantum fluctuations. The path integral projects the vacuum wave-functional onto the subspace of configurations satisfying the antipodal constraint. The tilde means that the equality holds up to a normalization constant, which we shall not concern ourselves with in what follows. It is enough to point out that the transition amplitude $|\langle \Omega|\mathrm{EH}\rangle_{t_0}|\sim e^{-S_{B}}=e^{-S_{(3)}-S_{(2)}-S_{(1)}}$, assuming the absence of fermions. It is also important to point out that an actual calculation of the path integral (\ref{mainEHPI}) is a cumbersome task, involving gauge fixing and integrating over the bosonic moduli space of the gauge field.

To ensure that the path integral respects  locality (cluster decomposition), one must also sum over all integer topological sectors\footnote{This assertion stems from our experience with Yang-Mills theory, see, e.g.,\cite{Weinberg:1996kr,Donnelly:2013tia,Tanizaki:2019rbk,Anber:2025vjo}.}. For the $U(1)$ hypercharge gauge field, this entails a sum over ${\cal C}\in \mathbb{Z}$ (as well as sums over the integer topological charges of the $SU(3)\times SU(2)$ sectors, which we do not discuss here). 
Thus, in a fixed $m_{(3)}, m_{(2)}$ background, summing over ${\cal C}$, and gauge-fixing via the Fadeev–Popov procedure, the physical amplitude takes the form
\begin{equation}\label{semifinalampl}
\langle \Omega|\mathrm{EH}\rangle_{t_0}^{m_{(3)}, m_{(2)}} 
\;\sim\; e^{-S_{(3)}(m_{(3)})-S_{(2)}(m_{(2)})} \sum_{{\cal C}\in \mathbb{Z}} e^{-S_{(1)}({\cal C})} 
\int [\mathcal D A_\mu] \Bigg|_{\substack{A_\mu(\bm x, t_0)=A_\mu(\bm x)\\A_\mu(-\bm x, t_0)=-A_\mu(\bm x)}} 
e^{-S_{\scriptsize\rm eff}}\,,
\end{equation}
where $S_{\scriptsize\rm eff}$ is the quadratic effective Euclidean action for the fluctuations around a given flux sector $m_{(3)}, m_{(2)},{\cal C}$. 
This amplitude represents the transition from an initial state prepared as in \eqref{EHMAINSTATE} to the vacuum in a fixed background flux of the $\mathbb{Z}_6^{(1)}$ 1-form symmetry labeled by $(m_{(2)},m_{(3)})$. Allowing a sum over all $(m_{(2)},m_{(3)})$, i.e., summing over all fractional topological charges so that:
\begin{eqnarray}\label{finalampl}
\langle\Omega|\mathrm{EH}\rangle_{t_0}=\sum_{m_{(2)},m_{(3)}}\langle \Omega|\mathrm{EH}\rangle_{t_0}^{m_{(3)}, m_{(2)}} 
\end{eqnarray}
 effectively gauges the $\mathbb{Z}_6^{(1)}$ 1-form symmetry. Restricting the sum to $m_{(3)}$ or  $m_{(2)}$ leads to the gauging of $\mathbb Z_3^{(1)}$ and $\mathbb Z_{2}^{(1)}$ $1$-form symmetries, respectively. This concludes our quest.

Turning on the SM fermions forces the amplitude 
$
  \langle \Omega \,|\, \mathrm{EH}\rangle^{\,m_{(3)},\, m_{(2)},\, \mathcal{C}}_{t_0}
$
to vanish in any background flux sector $(m_{(3)}, m_{(2)}, \mathcal{C})$ that supports fermion zero modes. 
Throughout, we assume that 
$
  a \langle h \rangle \ll 1,
$
so that the EH scale $a$ is much smaller than the inverse Higgs VEV. In this regime, the Higgs remains in its vacuum, and the fermions effectively stay massless.  
Nevertheless, once fermion sources are inserted, the amplitude becomes nonzero:
$
   \mathcal{A} \;\sim\; e^{-S_{(3)} - S_{(2)} - S_{(1)}} \,,
$
which is precisely the amplitude associated with the  't~Hooft vertex responsible for baryon- and lepton-number violation discussed in the previous section.

Let us emphasize an important point in the above setup. 
In writing (\ref{semifinalampl}, \ref{finalampl}), we assumed a fixed EH background and included a sum over all $\mathbb{Z}_6^{(1)}$ flux sectors supported by the bolt. 
It is useful to contrast this with the case of $U(1)$ gauge theory: there, summing over all integer flux sectors ${\cal C}\in \mathbb{Z}$ is \emph{mandatory}, since otherwise the path integral would violate cluster decomposition. 
By comparison, summing over distinct $(m_{(3)},m_{(2)})$ is a matter of physical choice: performing the sum corresponds to gauging the discrete $\mathbb{Z}_6^{(1)}$ one-form symmetry. 
If one does \emph{not} perform this sum, the theory is defined in a fixed background of the $\mathbb{Z}_6^{(1)}$ symmetry; if one \emph{does} perform the sum, the symmetry is dynamically gauged. 
Thus, the correct procedure for computing physical observables (e.g., correlation functions or condensates) is as follows: first, compute the path integral in a given topological sector, then
\begin{itemize}
    \item always sum over ${\cal C} \in \mathbb{Z}$ to enforce  cluster decomposition in the semi-classical sense; the latter ensures that widely separated operators approximately factorize at large distances. We emphasize that in full quantum gravity, cluster decomposition may not be strictly valid due to effects such as topology change or wormholes, but it provides a guiding principle in the semi-classical regime;
    \item optionally sum over $(m_{(3)},m_{(2)})$ depending on whether one wishes to gauge the $\mathbb{Z}_6^{(1)}$ symmetry. Summing over all $\mathbb{Z}_6^{(1)}$ flux sectors ensures the absence of the 1-form global symmetry in the semi-classical picture. We stress that this is a semi-classical realization showing \emph{how} the symmetry is effectively absent, without claiming a rigorous proof that gravity must force us to sum over these sectors.
\end{itemize}

One may also envision that if spacetime fluctuations can nucleate EH/anti-EH saddles with suppressed density, a dilute-gas picture analogous to Yang–Mills instanton gases could arise. Establishing this requires (i) a computation of the one-loop determinant around EH and proper integration over the moduli space of the fluctuating spacetime and (ii) control over gravitational backreaction and interactions between instantons. A systematic treatment of this instanton-gas scenario lies beyond the scope of the present work.

Before closing this section, let us acknowledge two important points that we have not addressed in detail. 
First, we did not discuss how gauge transformations act on states in the Hilbert space. 
Second, we have not touched upon the role of Gauss’ law and its proper implementation. 
Both of these aspects involve subtle issues. 
In Appendix~\ref{Phase space path integral}, we provide a preliminary exploration of the second point within the phase-space path integral formalism. 
However, it should be emphasized that these questions remain open and merit further, more systematic investigation in the future.

%%%%%%%%%%%%%%%%%%%%%%%%
\section{Outlook and future directions}
\label{Outlook and future directions}
%%%%%%%%%%%%%%%%%%%%%%%%

In this work, we have examined the fate of the SM $\mathbb{Z}_6^{(1)}$ $1$-form symmetry in the presence of gravitational instantons, focusing on EH geometries. These spaces provide a natural semiclassical setting in which nontrivial flux sectors can be consistently supported, thanks to the existence of a unique normalizable self-dual $2$-form that allows gauge fields to localize at the bolt without backreacting on the geometry. We demonstrated how turning on such fluxes induces fractional topological charges and modifies the boundary conditions of fermions, leading to the appearance of localized zero modes. 

Embedding the SM gauge group into this background and summing over the fractional fluxes in $SU(3)\times SU(2)\times U(1)$ sectors has the effect of gauging the $\mathbb{Z}_6^{(1)}$ symmetry of the SM. This mechanism provides a dynamical setup by which one avoids exact global higher-form symmetries in the semi-classical limit of quantum gravity. Moreover, we argued that the resulting backgrounds mediate baryon- and lepton-number violating processes, albeit with amplitudes exponentially suppressed by the square inverse of the hypercharge gauge coupling.

Altogether, our analysis points toward novel nonperturbative phenomena in the SM when embedded in curved backgrounds. A few comments, including lessons learned and proposals for future directions, are in order.

\begin{enumerate}

\item  Gauging the Standard Model $1$-form symmetry was also carried out  on $\mathbb{T}^4$ in \cite{Anber:2021upc}. Yet, there are important differences between the two setups. On $\mathbb{T}^4$, self-duality of the gauge fields imposes nontrivial constraints on the ratio of the torus periods. When all fluxes are embedded along the Cartan generators, the resulting field strengths are constant throughout the $\mathbb{T}^4$ volume. This stands in sharp contrast to the EH case, where the self-dual gauge fields are localized on the bolt. Consequently, EH instantons provide more physically (and phenomenologically) relevant semiclassical backgrounds for supporting fractional fluxes. Nevertheless, in both cases the presence of these fluxes induces baryon- and lepton-number violating processes, with amplitudes suppressed in a parametrically similar way. From a practical standpoint, quantum field theory on $\mathbb{T}^4$ is technically simpler to analyze than on EH spaces, and thus toroidal backgrounds can still yield valuable insights into nonperturbative Standard Model dynamics, even though they are less realistic than the gravitational instanton configurations considered here.

 \item  It was recently emphasized in \cite{Choi:2023pdp,Reece:2023iqn} that gauging the $\mathbb{Z}_6^{(1)}$ symmetry imposes striking quantization conditions on the Chern--Simons couplings of the QCD axion to photons and gluons. The arguments in those works were purely kinematic, relying on the transformation of the SM action under large gauge transformations in the background of fractional topological charges. In this work, we uncover a dynamical origin of these constraints: EH instantons with $\mathbb{Z}_6$ fluxes provide explicit semiclassical realizations that enforce some quantization conditions, though the exact quantization conditions in the EH space should be worked out, an exercise left for the future. Thus, what previously appeared as a formal requirement of consistency now emerges as a genuine nonperturbative mechanism rooted in gravitational dynamics.

\item Our analysis of baryon- and lepton-number violation in the presence of $1$-form fluxes revealed effects that are extremely suppressed and, in all likelihood, phenomenologically irrelevant. Nevertheless, EH instantons may still have important phenomenological implications. In particular, their nontrivial geometry permits the existence of an electromagnetic $\theta_{\mathrm{em}}$ angle that cannot be rotated away. This raises the prospect of $\cal{CP}$-violating effects associated with $\theta_{\mathrm{em}} \neq 0$, which merit further investigation. In connection to the discussion above, it would also be valuable to study the role of $\theta_{\mathrm{em}}$ in the Standard Model under toroidal compactifications, in analogy with the recent analysis of such effects in mass-deformed super Yang-Mills theory presented in \cite{Anber:2025vjo}.

\item Our study suggests several open directions for further investigation of EH instantons and their role in semi-classical gravity, a subject that remains only partially understood. In particular, developing a clearer picture of the entangled state (\ref{EHMAINSTATE}) would be highly valuable. This includes a more careful treatment of gauge fixing and Gauss' law, which may in turn illuminate the precise meaning of the entanglement between the left and right regions of space. It would also be interesting to compute the entanglement entropy of such a state (see, e.g., \cite{Casini:2013rba,Donnelly:2016auv}), which could shed new light on the physical interpretation of EH geometries and their potential relevance in quantum gravity.

\item Since EH spaces are the simplest examples of asymptotically locally Euclidean (ALE) gravitational instantons, it is natural to ask how our analysis extends to more general ALE spaces. By Kronheimer’s classification \cite{Kronheimer:1989zs,Kronheimer:1989pu}, ALE spaces are in one-to-one correspondence with finite subgroups $\Gamma \subset SU(2)$, leading to a rich family of hyper-K\"ahler manifolds with nontrivial topology. In particular, while EH corresponds to the $\Gamma=\mathbb{Z}_2$ case, higher-rank discrete subgroups yield more intricate spaces with multiple independent $2$-cycles and a lattice of harmonic $2$-forms. This suggests the existence of a larger variety of flux sectors, fractional charges, and entanglement structures than those uncovered in the EH case.  

A systematic study of $1$-form fluxes, fermion zero modes, and path integrals in these more general ALE backgrounds may reveal qualitatively new mechanisms. Moreover, such an extension would place the EH analysis in a broader mathematical and physical framework, clarifying to what extent the phenomena we observed are special to the $\mathbb{Z}_2$ case or are generic features of gravitational instantons. Ultimately, exploring these directions may deepen our understanding of the interplay between topology, entanglement, and nonperturbative dynamics in the Standard Model coupled to gravity.

\item A notable bonus of our study is the demonstration that fermions can be consistently placed on EH spaces supporting fractional instantons. This opens a powerful new avenue: the framework developed here can be used to uncover novel anomalies and probe the dynamics of strongly coupled theories.

\end{enumerate}

{\bf {\flushleft{Acknowledgments:}}} We thank M. Dierigl for discussions that inspired our interest in the topic of EH instantons. We also thank A. Braun, F. Gagliano, I. Garci\'a-Etxebarria,  S. Lee, E. Poppitz, and O. Sakhelashvili for many interesting discussions, and E. Poppitz for comments on the manuscript.   We acknowledge CERN for the warm hospitality extended to us during the initial stage of this work.  We are supported by STFC through the grant ST/X000591/1.

\appendix

%%%%%%%%%%%%%%%%%%%%%%%%%%%%%%
\section{The Eguchi-Hanson geometry}
\label{The Eguchi-Hanson geometry}
%%%%%%%%%%%%%%%%%%%%%%%%%%%%%

%%%%%%%%%%%%%%%%%%%%%%
\subsection{Mathematical preliminaries}
%%%%%%%%%%%%%%%%%%%%%%%

In this section, we review the mathematical tools needed to understand the EH geometries. See \cite{Eguchi:1980jx} for a review. 

We introduce the vielbeins $e^a = e^a_{\mu} \, dx^{\mu}$, which define a local orthonormal frame at each point in spacetime. The curved spacetime metric $ g_{\mu\nu}$ is related to the vielbeins via
\begin{equation}
g_{\mu\nu} = \eta_{ab} \, e^a_{\mu} \, e^b_{\nu},
\end{equation}
where $\eta_{ab}$ is the Euclidean metric in the local inertial frame. Latin indices $a, b = 0,1,2,3$ refer to components in the locally flat (tangent) spacetime, while Greek indices $\mu, \nu = 0,1,2,3$ refer to components in the curved spacetime manifold.
We use the flat spacetime metric $\eta_{ab}$ to raise and lower Latin indices, and the curved spacetime metric $g_{\mu\nu}$ for Greek indices. 
The inverse vielbeins $E_a^{\mu}$ are defined such that they satisfy the relation
\begin{equation}
E_a^{\mu} \, e^b_{\mu} = \delta_a^b.
\end{equation}

We also introduce the spin connection $1$-forms $\omega^a_{\;\; b} \equiv \omega^a_{\;\; b\mu} \, dx^{\mu}$, and we restrict our analysis to torsion-free geometries. In this case, the torsion-free condition is expressed as
\begin{equation}\label{tortionfree}
de^a + \omega^a_{\;\; b} \wedge e^b = 0.
\end{equation}
Notice that $\omega^a_{\;\; b}=-\omega^b_{\;\; a}$. 
The corresponding Riemann curvature $2$-form is given by
\begin{equation}
R^a_{\;\; b} = d\omega^a_{\;\; b} + \omega^a_{\;\; c} \wedge \omega^c_{\;\; b},
\end{equation}
where $R^a_{\;\; b}=\frac{1}{2}R^a_{\;\; b\mu\nu} dx^\mu\wedge dx^\nu=\frac{1}{2}R^a_{\;\;bcd} e^c\wedge e^d$. We also define the dual $2$-form
\begin{eqnarray}
\tilde R^a_{\;\; b}\equiv \frac{1}{2}\epsilon_{abcd}R^c_{\;\; d}\,.
\end{eqnarray}
We shall adopt the convention (this defines our convention for the frame orientation) \begin{equation}\epsilon_{0123}=1\end{equation} with cyclic permutations. 
The curvature $2$-form is said to be (anti)-self-dual if it satisfies the relation
\begin{eqnarray}
\tilde R^a_{\;\; b}=\pm  R^a_{\;\; b}\,,
\end{eqnarray}
where the positive sign stands for self-dual and the negative sign is for anti-self-dual $2$-forms. It can be easily shown that (anti)-self-dual $2$-form curvatures automatically satisfy the vacuum Einstein's equations:
\begin{eqnarray}
R^{ac}_{\;\;bc}=0\,.
\end{eqnarray} 
Likewise, one says that the connection $1$-form is (anti)self-dual if it satisfies the relation
\begin{eqnarray}
\tilde\omega^a_{\;\; b}\equiv  \frac{1}{2}\epsilon_{abcd}\omega ^c_{\;\; d}=\pm \omega^a_{\;\; b}\,.
\end{eqnarray}
It can be shown that (anti)self-dual connection $1$-forms imply that the curvature $2$-forms are (anti)self-dual.

%%%%%%%%%%%%%%%%%%%%%%%%%%
\subsection{The Eguchi-Hanson instanton}
%%%%%%%%%%%%%%%%%%%%%%%%%%%

Next, we consider the following general metric ansatz introduced by Hawking and Gibbons \cite{Gibbons:1978tef}:
\begin{equation}\label{GHmetric}
ds^2 = V^{-1}(d\tau + \bm{w} \cdot d\bm{x})^2 + V\, d\bm{x} \cdot d\bm{x}\,,
\end{equation}
where $\bm{x} = (x_1, x_2, x_3)$ are Cartesian coordinates, and $\tau$ may be interpreted as a Euclidean time coordinate. The functions $V(\bm{x})$ and $\bm{w}(\bm{x}) = (w_1(\bm{x}), w_2(\bm{x}), w_3(\bm{x}))$ can be viewed, respectively, as graviscalar potential and gravivector potential sourced by a distribution of dyons carrying the same electric and magnetic charges $(e,m)=(1,\pm1)$. Specifically, we take the scalar potential to be
\begin{equation}
V = \sum_{i=1}^M \frac{1}{|\bm{x} - \bm{x}_i|}\,,
\end{equation}
where $\bm{x}_i$ denote the positions of the dyons. We shall return to the discussion of the vector potential $\bm{w}$ shortly, after reviewing some geometric preliminaries.

The vielbeins corresponding to the metric \eqref{GHmetric} are given by
\begin{equation}\label{xijkveil}
e^0 = V^{-1/2}(d\tau + \bm{w} \cdot d\bm{x})\,, \qquad e^i = V^{1/2} dx^i\,, \quad i = 1,2,3\,,
\end{equation}
and the non-vanishing components of the spin connection are
\begin{align}
\omega^0_{\;\;j} &= \frac{V^{-3/2}}{2} \left( \partial_j V\, e^0 - (\partial_j w_i - \partial_i w_j)\, e^i \right)\,, \nonumber \\
\omega^j_{\;\;k} &= \frac{V^{-3/2}}{2} \left( (\partial_j w_k - \partial_k w_j)\, e^0 + \partial_j V\, e^k - \partial_k V\, e^j \right)\,,
\end{align}
which can be derived directly from the torsion-free condition:
\begin{equation} \label{tortionfree}
de^a + \omega^a_{\;\;b} \wedge e^b = 0\,.
\end{equation}

Furthermore, if the scalar and vector potentials satisfy the relation
\begin{equation}\label{antisdr}
\partial_i V = \pm \epsilon_{ijk} \partial_j w_k\,, \qquad \text{or equivalently} \qquad \bm{\nabla} V = \pm \bm{\nabla} \times \bm{w}\,,
\end{equation}
with $\epsilon_{123} = 1$, etc., then the connection $1$-form is (anti-)self-dual; that is, it satisfies identities such as $\omega_{01} = \pm \omega_{23}$, and so on. 

Now, we comment on the construction of the vector potential $\bm w$. Recall that the $1$-form magnetic potential of a unit monopole located at the origin of the coordinates $\bm x$ is given by 
\begin{eqnarray}
w=\bm w\cdot d\bm x= w_1dx^1+w_2dx^2+w_3 dx^3=\mp (1-\cos\theta) d\psi\,,
\end{eqnarray}
where $\theta$ is the angle between a general position vector $\vec x$ and the $x_3$ axis, while $\psi$ is the azimuthal angle. By means of a gauge transformation $w\rightarrow w\pm d\psi$, we can bring $w$ to the form
\begin{eqnarray}
w=\pm\cos\theta d\psi\,.
\end{eqnarray}
Next, consider this unit monopole displaced along the $x_3$-axis at $x_3=z_0$. Then, using the coordinate parametrization $x_1=r\sin\theta\cos\psi$, $x_2=r\sin\theta\sin\psi$, and $x_3=z_0+r\cos\theta$, the potential takes the form
\begin{eqnarray}
w=\pm\frac{x_3-z_0}{\sqrt{x_1^2+x_2^2+(x_3-z_0)^2}}d\left[ \tan^{-1}\left(\frac{x_2}{x_1}\right)\right]\,,
\end{eqnarray}
or in terms of the components:
\begin{eqnarray}\label{magnetic potential}\nonumber
w_1=\mp \frac{x_3-z_0}{\sqrt{x_1^2+x_2^2+(x_3-z_0)^2}}\frac{x_2}{x_1^2+x_2^2}\,,\quad w_2=\pm \frac{x_3-z_0}{\sqrt{x_1^2+x_2^2+(x_3-z_0)^2}}\frac{x_1}{x_1^2+x_2^2}\,.\\
\end{eqnarray}
It can be easily checked that $\bm w$ and $V$ satisfy the (anti)-self-dual relation (\ref{antisdr}). In the gauge used in (\ref{magnetic potential}), we find that there is a Dirac string emanating from the monopole at both the north and south poles, $x_3>z_0$ and $x_3<z_0$, as evident from the singularity at $x_1=x_2=0$.

We can also repeat the analysis for $M$ distinct dyons by applying the principle of superposition, owing to the linearity of the (anti)-self-dual relation (\ref{antisdr}). In the following, however, we shall focus on the case of two dyons, taking $M=2$. We place the two dyons, of the exact same electric and magnetic charges, along the $x_3$ axis at $\pm z_0$. Thus, the scalar and vector potentials take the form
\begin{eqnarray}
V=\frac{1}{R_+}+\frac{1}{R_-}\,,\quad w=\pm\left(\frac{Z_+}{R_+}+\frac{Z_-}{R_-}\right)d\left[ \tan^{-1}\left(\frac{x_2}{x_1}\right)\right]\,,
\end{eqnarray}
where
\begin{eqnarray}
R_{\pm}=\sqrt{x_1^2+x_2^2+(x_3\pm z_0)^2}\,, \quad Z_\pm=x_3\pm z_0\,.
\end{eqnarray}
Notice that with two dyons, Dirac strings are emanating from the two dyons at $x_3<z_0$ and $x_3>z_0$, while the strings in the intermediate region $|x_3|<z_0$ cancel out. 

Next, we use the elliptic coordinates, as suggested by Prasad \cite{Prasad:1979kg}, to put the metric (\ref{GHmetric}) in the standard EH form. Take
\begin{eqnarray}\nonumber
x_1&=&z_0\sinh\alpha\sin\theta\cos\psi\,,\\\nonumber
x_2&=&z_0\sinh\alpha\sin\theta\sin\psi\,,\\\nonumber
x_3&=&z_0\cosh\alpha\cos\theta\,,\\
\tau&=&2\varphi\,,
\end{eqnarray}
with coordinate ranges 
\begin{eqnarray}
0\leq\alpha<\infty\,,\quad 0\leq \theta \leq \pi\,,\quad 0\leq \psi \leq 2\pi\,,\quad 0\leq \varphi \leq 2\pi \,.
\end{eqnarray}
The rationale behind the choice of this range of coordinates will be discussed momentarily. 
 We may also introduce the coordinate $r$ defined via
\begin{eqnarray}
a^2\cosh\alpha=r^2\,, \quad a^2\equiv 8z_0\,, \quad r^2\geq a^2\,.
\end{eqnarray}
Surfaces of constant $\alpha$ are ellipsoids with foci at $\left(x_3,\sqrt{x_1^2+x_2^2}\right)=(z_0,0)$ and $(-z_0,0)$, while surfaces of constant $\theta$ are hyperboloids. 
In the new coordinates, the dyons are located at $(r,\theta,\psi)=(a,0,0)$ and $(a,\pi,0)$, while the Dirac strings are located at $\theta=0$ and $\theta=\pi$ for $r>a$. 
With such a change of coordinates, we find
\begin{eqnarray}\label{trans1}
V=\frac{16 r^2}{r^4-a^4\cos^2\theta}\,, \quad w=\bm w\cdot d\bm x=\pm\frac{(r^4-a^4)\cos\theta}{r^4-a^4\cos^2\theta}d\psi\,,
\end{eqnarray}
and
\begin{eqnarray}\label{trans2}
d\bm x\cdot d\bm x=\frac{1}{64}(r^4-a^4\cos^2\theta)\left(d\theta^2+\frac{4r^2dr^2}{r^4-a^4}\right)+\frac{\sin^2\theta}{64}(r^4-a^4)d\psi^2\,.
\end{eqnarray}
Substituting (\ref{trans1}, \ref{trans2}) into (\ref{GHmetric}), we readily obtain the EH instanton in its original form \cite{Eguchi:1978xp,Eguchi:1978gw}
\begin{eqnarray}\label{EHORigin}\nonumber
ds^2_{EH}=\frac{r^2}{4}\left(1-\frac{a^4}{r^4}\right)\left(\pm d\psi+\cos\theta d\varphi\right)^2+\left(1-\left(\frac{a}{r}\right)^4\right)^{-1} dr^2+\frac{r^2}{4}\left(d^2\theta+\sin^2\theta d^2\varphi\right)\,.\\
\end{eqnarray}

Let us now study the metric (\ref{EHORigin}) in various limits. To understand the behavior near $r = a$, we use the change of variables
\begin{eqnarray}
u^2=r^2\left(1-\left(\frac{a}{r}\right)^4\right)\,,
\end{eqnarray}
and the metric becomes
\begin{eqnarray}\label{canEHAPP}
ds^2_{EH}=\frac{du^2}{\left(1+\frac{a^4}{r^4}\right)^2}+\frac{u^2}{4}\left(\pm d\psi+\cos\theta d\varphi\right)^2+\frac{r^2}{4}\left(d^2\theta+\sin^2\theta d^2\varphi\right)\,.
\end{eqnarray}
In the limit $r \rightarrow a$, we have $u \cong 0$, so the metric reduces to
\begin{eqnarray}\label{EHRA}
\mbox{Lim}_{r\rightarrow a}\,ds^2_{EH}\cong \frac{du^2}{4}+\frac{u^2}{4}\left(\pm d\psi+\cos\theta d\varphi\right)^2+\frac{a^2}{4}\left(d^2\theta+\sin^2\theta d^2\varphi\right)\,,
\end{eqnarray}
showing that the metric is regular at $r=a$.
 For fixed $\theta$ and $\varphi$, we obtain
the metric $\frac{1}{4}(du^2+u^2d^2\psi)$. This is simply flat $\mathbb{R}^2$, provided that $\Delta\psi = 2\pi$, so we avoid conical singularities. The $\mathbb R^2$ metric shrinks to a point at $r=a$ or $u=0$, and thus the $4$-D metric describes a  $S^2$ at $r=a$.
We therefore conclude that, near $r = a$, the Eguchi-Hanson space has the topology of $S^2 \times \mathbb{R}^2$. 

The two-sphere $S^2$ at $r=a$ is a non-contractible $2$-cycle in the manifold 
and is often referred to as the \emph{bolt} of the EH space. The 
radial $\mathbb{R}^2$ factor describes the smooth transverse directions into the 
bulk.

To analyze the behavior of the EH space in the limit $r \rightarrow \infty$, we rewrite equation (\ref{EHORigin}) in terms of the left-invariant one-forms $\sigma_x, \sigma_y, \sigma_z$:
\begin{eqnarray}\label{EHsigma}
ds_{EH}^2 = f^2(r)\, dr^2 + r^2 \left( \sigma_x^2 + \sigma_y^2 + f^{-2}(r)\, \sigma_z^2 \right),
\end{eqnarray}
where \begin{equation}f^{-2}(r) = 1 - \left( \frac{a}{r} \right)^4\,,\end{equation} and the $1$-forms $\sigma_x$, $\sigma_y$, and $\sigma_z$ are defined as (from here on, we chose the upper sign in (\ref{EHORigin}))
\begin{eqnarray}\label{sigmasinewh} \nonumber
\sigma_x &=& \frac{1}{2} \left( \sin\psi\, d\theta - \sin\theta \cos\psi\, d\varphi \right)\,, \\\nonumber
\sigma_y &=& \frac{1}{2} \left( -\cos\psi\, d\theta - \sin\theta \sin\psi\, d\varphi \right)\,, \\
\sigma_z &=& \frac{1}{2} \left(  d\psi + \cos\theta\, d\varphi \right)\,,
\end{eqnarray}
with vielbeins given by
\begin{eqnarray}\label{velbspher}
e^a=(e^0,e^1,e^2,e^3)=\left(f(r)dr,r\sigma_x,r\sigma_y,rf^{-1}(r)\sigma_z\right)\,,
\end{eqnarray}
keeping in mind that these are not the direct coordinate transformations of the ones given in (\ref{xijkveil}). The $1$-forms satisfy the relations
\begin{eqnarray}\nonumber
d\sigma_x&=&2\sigma_y\wedge \sigma_z\,,\quad d\sigma_y=2\sigma_z\wedge \sigma_x\,,\quad d\sigma_z=2\sigma_x\wedge \sigma_y\,,\\
\sigma_x\wedge\sigma_x&=&\sigma_y\wedge\sigma_y=\sigma_z\wedge\sigma_z=0\,.
\end{eqnarray}

For comparison, flat Euclidean space is described by the metric
\begin{eqnarray}
ds^2_{\mathbb{R}^4} = dr^2 + r^2 \left( \sigma_x^2 + \sigma_y^2 + \sigma_z^2 \right),
\end{eqnarray}
with coordinate ranges given by
\begin{eqnarray}
0 \leq r < \infty,\quad 0 \leq \theta \leq \pi,\quad 0 \leq \varphi \leq 2\pi,\quad 0 \leq \psi \leq 4\pi\,.
\end{eqnarray}
In particular, at large $r$, flat $\mathbb{R}^4$ approaches the 3-sphere $S^3$, parametrized by the Euler angles $(\theta, \varphi, \psi)$ over the specified domain.
Comparing this to the EH space as $r \rightarrow \infty$, we observe that the angular part of the metric approaches that of a  3-sphere. However, in the EH geometry, the coordinate $\psi$ has half the range it does in flat space, namely $\Delta\psi = 2\pi$ instead of $4\pi$.
This implies that the asymptotic boundary of the EH space is not $S^3$ but rather the quotient $S^3/\mathbb Z^2=\mathbb {RP}^3$. We note that the volume $3$-form on $\mathbb {RP}^3$ is given by
\begin{eqnarray}\label{volume3form}
{\cal V}_3=\sigma_x\wedge \sigma_y\wedge \sigma_z=-\frac{ 1}{8} \sin\theta d\theta\wedge d\varphi\wedge d\psi\,.
\end{eqnarray}

To reiterate the understanding of the identifications of the opposite points on $S^3$, we also note that one can use the coordinates $(x,y,z,t)$ to cover the flat spacetime:
\begin{eqnarray}
x+iy=r\cos\left(\frac{\theta}{2}\right)e^{\frac{i}{2}(\psi+\varphi)}\,, \quad z+it=r\sin\left(\frac{\theta}{2}\right)e^{\frac{i}{2}(\psi-\varphi)}\,,
\end{eqnarray}
such that \begin{equation}\label{definerxyz}r^2=x^2+y^2+z^2+t^2\,,\quad-\infty<x,y,z,t<\infty\,,\end{equation} and  \begin{equation}ds^2_{\mathbb R^4}=dx^2+dy^2+dz^2+dt^2\,.\end{equation} 
Under $\psi\rightarrow\psi+2\pi$ we have $(x,y,z,t)=-(x,y,z,t)$, and thus, restricting the range of $\psi$ to $0\leq \psi\leq 2\pi$ implies that we are identifying $(x,y,z,t)$ with $-(x,y,z,t)$. We may also use the coordinates $(x,y,z,t)$ to write the EH metric in the form (\ref{EHsigma}), with $\sigma_{x},\sigma_{y},\sigma_z$ in (\ref{sigmasinewh}) are given by
\begin{eqnarray}\label{1-invarintformsxyzt}\nonumber
\sigma_x&=&\frac{1}{r^2}\left(xdt-tdx+ydz-zdy\right)\,,\\\nonumber
\sigma_y&=&\frac{1}{r^2}\left(ydt-tdy+zdx-xdz\right)\,,\\
\sigma_z&=&\frac{1}{r^2}\left(zdt-tdz+xdy-ydx\right)\,.
\end{eqnarray}
and $r^2$ is given by (\ref{definerxyz}), keeping in mind that we must make the identification $(x,y,z,t)=-(x,y,z,t)$ and exclude the region with $r^2<a^2$.

To summarize, the constant hypersurfaces of the metric (\ref{EHsigma}) have the topology of $\mathbb {RP}^3$, and at the boundary $r\rightarrow\infty$ the metric becomes the canonical metric on $\mathbb{ RP}^3$. However, as we approach $r=a$, the manifold shrinks to $S^2$.

Finally, we may use equation (\ref{sigmasinewh}) to compute the spin connection $1$-forms by imposing the torsion-free condition. This yields
\begin{eqnarray}\label{omegaEH}
\omega_{10} = \omega_{23} = \frac{\sigma_x}{f}\,, \quad \omega_{20} = \omega_{31} = \frac{\sigma_y}{f}\,, \quad \omega_{30} = \omega_{12} = (2 - f^{-2}) \sigma_z\,.
\end{eqnarray}
From these, we compute the corresponding curvature $2$-forms:
\begin{eqnarray} \label{CT2EH} \nonumber
R_{10} &=& R_{23} = -\frac{2a^4}{r^6}(e^1 \wedge e^0 + e^2 \wedge e^3)\,, \\\nonumber
R_{20} &=& R_{31} = -\frac{2a^4}{r^6}(e^2 \wedge e^0 + e^3 \wedge e^1)\,, \\
R_{30} &=& R_{12} = \frac{4a^4}{r^6}(e^3 \wedge e^0 + e^1 \wedge e^2)\,.
\end{eqnarray}
Remarkably, in the standard EH coordinates given by the metric (\ref{EHsigma}), we find that the spin connection $1$-forms and curvature $2$-forms satisfy the anti-self-duality condition, for example $\tilde{R}_{12} = -R_{12}$, and similarly for the others.
The Pontryagin number is given by
\begin{eqnarray}
p = -\frac{1}{8\pi^2} \int_{\scriptsize\mbox{EH}} \text{tr}\left[ R \wedge R \right] = \frac{1}{8\pi^2} \int_{\scriptsize\mbox{EH}}  R_{ab} \wedge R_{ab} \,.
\end{eqnarray}
Substituting the expressions from (\ref{CT2EH}), we obtain
\begin{eqnarray}
p = -\frac{4a^8(16 + 32)}{8\pi^2} \int \frac{ e^0\wedge e^1 \wedge e^2 \wedge e^3 }{r^{12}} = 3\,,
\end{eqnarray}
and we use a convention where the wedge product  $e^0\wedge e^1 \wedge e^2 \wedge e^3$ is positive since for the orientation frame we chose $\epsilon_{0123}=1$. With the convention that $e^0\wedge e^1 \wedge e^2 \wedge e^3>0$, the positive volume form in the $(r,\theta,\varphi,\psi)$ coordinates is $-dr\wedge d\theta \wedge d\varphi\wedge \psi$; thus, the coordinate orientation we choose is opposite to the frame orientation. This choice makes it easier to apply the index theorem without having to carry an overall negative sign.

%%%%%%%%%%%%%%%%%%%%%%%%
\subsection{Turning on a $U(1)$ gauge field}
\label{Turning on a U1 gauge field}
%%%%%%%%%%%%%%%%%%%%%%%%

We may also introduce a $U(1)$ gauge field on the Eguchi-Hanson (EH) background without inducing any backreaction on the geometry. To achieve this, we consider a gauge field of the form
\begin{eqnarray}\label{form of A}
A_{U(1)} = {\cal C} a^2 \frac{\sigma_z}{r^2} = {\cal C} a^2 \frac{z\,dt - t\,dz + x\,dy - y\,dx}{(x^2 + y^2 + z^2 + t^2)^2}\,,
\end{eqnarray}
where ${\cal C}$ is a constant parameter. The corresponding field strength is anti-self-dual and is given by
\begin{eqnarray}\label{form of A2}
F_{(1)} = \frac{2{\cal C} a^2}{r^4} (e^3 \wedge e^0 + e^1 \wedge e^2)\,.
\end{eqnarray}
Using the $(x,y,z,t)$ coordinates, one can easily show that (we use $\epsilon_{xyzt}=1$)
\begin{equation}\label{selfdual F}
-\tilde F_{tx}=F_{tx}=F_{yz}=\frac{4{\cal C} a^2 (t y + x z)}{(t^2 + x^2 + y^2 + z^2)^3}\,,
\end{equation}
 etc., meaning that  $F_{(1)}$ is anti-iself-dual. Thus,  the associated energy-momentum tensor vanishes identically, and therefore the gauge field does not backreact on the geometry. Other self-dual $U(1)$ gauge field ansatzes are possible, though they are singular. 

The $U(1)$ flux through the two-sphere $S^2$ at $r = a$ can be computed by integrating $F$ over  $S^2$:
\begin{eqnarray}
\int_{S^2} F_{(1)} = \frac{2{\cal C}}{a^2} \int_{S^2} e^1 \wedge e^2 = -2\pi {\cal C}\,.
\end{eqnarray}
The topological charge carried by the $U(1)$ field is given by
\begin{eqnarray}
Q_T &=& \frac{1}{8\pi^2} \int_{\scriptsize\mbox{EH}} F_{(1)} \wedge F_{(1)} = -\frac{8{\cal C}^2 a^4}{8\pi^2} \int \frac{1}{r^8} e^0\wedge  e^1 \wedge e^2 \wedge e^3
= \frac{{\cal C}^2}{4}\,.
\end{eqnarray}
%

%%%%%%%%%%%%%%%%%%%%%%%%%%%
\section{Fermions in the EH background: fermion zero modes}
\label{Fermions in the EH background fermion zero modes}
%%%%%%%%%%%%%%%%%%%%%%%%%%%%%

We now turn to the problem of coupling fermions to the Eguchi-Hanson (EH) background. Specifically, we aim to solve for the fermionic zero modes in this geometry by analyzing the Dirac equation,
\begin{eqnarray}
\slashed D\Psi = 0\,, \quad \slashed D \equiv \gamma^a E_a^\mu \left( \partial_\mu + \frac{1}{2}\Omega_\mu + i A_\mu \right)\,,
\end{eqnarray}
where the spin connection term is defined by\footnote{The covariant derivative can be envisaged from the transformation of $\Psi$ under the Lorentz group: $\Psi\rightarrow e^{-\frac{1}{8}\omega_{ab}[\gamma^a,\gamma^b]}\Psi$.}
\begin{eqnarray}
\Omega_\mu = \omega_{ab\,\mu} \, \Sigma^{ab}\,, \quad \Sigma^{ab} = \frac{1}{4}[\gamma^a, \gamma^b]\,,
\end{eqnarray}
and the Dirac matrices $\gamma^a$ satisfy the Clifford algebra with $\gamma^a\gamma^b+\gamma^b\gamma^a=2\delta^{ab}$.

The inverse vielbeins $E^a_\mu$ are obtained via the relation $E_a^{\mu} \, e^b_{\mu} = \delta_a^b$:
\begin{eqnarray} \nonumber
E_0^\mu &=& \left( \frac{1}{f}, 0, 0, 0 \right)\,, \quad
E_1^\mu = \left( 0,  \frac{2\sin\psi}{r}, -\frac{2\csc\theta \cos\psi}{r},  \frac{2\cot\theta \cos\psi}{r} \right)\,, \\ \nonumber
E_2^\mu &=& \left( 0, -\frac{2\cos\psi}{r}, - \frac{2\csc\theta \sin\psi}{r}, \frac{2\cot\theta \sin\psi}{r} \right)\,, \quad
E_3^\mu = \left( 0, 0, 0,  \frac{2f}{r} \right)\,.
\end{eqnarray}

Substituting into the Dirac operator, we obtain
\begin{eqnarray} \nonumber
\slashed D &=& \gamma^0 \left( \frac{1}{f} \right) D_r + \left[ \gamma^1 \left(  \frac{2\sin\psi}{r} \right) + \gamma^2 \left( -\frac{2\cos\psi}{r} \right) \right] D_\theta \\ \nonumber
&& + \left[ \gamma^1 \left( -\frac{2\csc\theta \cos\psi}{r} \right) + \gamma^2 \left( - \frac{2\csc\theta \sin\psi}{r} \right) \right] D_\varphi \\ 
&& + \left[ \gamma^1 \left(  \frac{2\cot\theta \cos\psi}{r} \right) + \gamma^2 \left( \frac{2\cot\theta \sin\psi}{r} \right) + \gamma^3 \left(  \frac{2f}{r} \right) \right] D_\psi\,,
\end{eqnarray}
where the covariant derivatives are given by
\begin{eqnarray} \nonumber
D_r &=& \frac{\partial}{\partial r}\,,\quad D_\psi = \frac{\partial}{\partial \psi} + \frac{1}{2}(\Sigma^{30} + \Sigma^{12})(2 - f^{-2}) + i \frac{{\cal C} a^2}{2r^2}\,. \\ \nonumber
D_\theta &=& \frac{\partial}{\partial \theta} +  \frac{  (\Sigma^{10} + \Sigma^{23}) \sin\psi - (\Sigma^{20} + \Sigma^{31}) \cos\psi }{2f}\,, \\ \nonumber
D_\varphi &=& \frac{\partial}{\partial \varphi} - (\Sigma^{10} + \Sigma^{23}) \frac{\sin\theta \cos\psi}{2f} -  (\Sigma^{20} + \Sigma^{31}) \frac{\sin\theta \sin\psi}{2f} \\ \nonumber
&& + \frac{1}{2}(\Sigma^{30} + \Sigma^{12})(2 - f^{-2}) \cos\theta + i \frac{{\cal C} a^2}{2r^2} \cos\theta\,, 
\end{eqnarray}

To proceed further, we recall that the Dirac matrices $\gamma^a$ are given in terms of the Pauli matrices as
\begin{eqnarray}
\gamma^a = \left[ \begin{array}{cc}
0 & \hat\sigma^a \\
\bar{\hat\sigma}^a & 0
\end{array} \right]\,,
\end{eqnarray}
where\footnote{The hats are used to distinguish between the Pauli matrices and the left-invariant $1$-forms defined in (\ref{sigmasinewh}).} $\hat\sigma^a = (\hat\sigma^0, \hat\sigma^1, \hat\sigma^2, \hat\sigma^3) = (1, i\vec{\hat\sigma})$ and $\bar{\hat\sigma}^a = (1, -i\vec{\hat\sigma})$.
 Then\footnote{In particular, we note that because of the self-dual nature of EH space, i.e., $\omega_{10}=\omega_{23}$, etc., the spin connection one form part of the covariant derivative contributes terms of the form $(\Sigma^{10}+\Sigma^{23})\omega_{10}$ etc., where
\begin{eqnarray}\label{Sigmama}
\Sigma^{10}+\Sigma^{23}&=&i\left[\begin{array}{cc}\hat\sigma_x&0\\0&0\end{array}\right]\,,\quad \Sigma^{20}+\Sigma^{31}=i\left[\begin{array}{cc}\hat\sigma_y&0\\0&0\end{array}\right]\,,\quad
\Sigma^{30}+\Sigma^{12}=i\left[\begin{array}{cc}\hat\sigma_z&0\\0&0\end{array}\right]\,.
\end{eqnarray}
The fact that these matrices are asymmetric means that left-handed and right-handed fermions couple differently to gravity, thanks to the self-dual nature of EH space.
 }

\begin{eqnarray}
\Sigma^{ab}=\left[\begin{array}{cc}\frac{1}{4}\left(\hat\sigma^a\bar{\hat\sigma}^b-\hat\sigma^b\bar{\hat\sigma}^a\right)&0\\0&\frac{1}{4}\left(\bar{\hat\sigma}^a\sigma^b-\bar{\hat \sigma}^b\hat\sigma^a\right)\end{array}\right]\,,
\end{eqnarray}
Further, let us define the angular momentum operators $\hat {\cal J}_x$, $\hat{\cal J}_y$, and $\hat{\cal J}_z$ as
\begin{eqnarray}\nonumber
\hat{\cal J}_x&=&i\left(\sin\psi\frac{\partial}{\partial\theta}-\cos\psi\csc\theta\frac{\partial}{\partial\varphi}+\cot\theta\cos\psi\frac{\partial}{\partial\psi}\right)\,,\\\nonumber
\hat{\cal J}_y&=&i\left(-\cos\psi\frac{\partial}{\partial\theta}-\csc\theta\sin\psi\frac{\partial}{\partial\varphi}+\cot\theta\sin\psi\frac{\partial}{\partial\psi}\right)\,,\\
\hat{\cal J}_z&=&- i\frac{\partial}{\partial\psi}\,,
\end{eqnarray} 
which satisfy the commutation relation $\left[\hat{\cal J}_x,\hat{\cal J}_y\right]=i\hat{\cal J}_z$, etc. Using $\hat{\cal J}_x,\hat{\cal J}_y,\hat{\cal J}_z$, and after some algebra, we can write the full Dirac equation in the form
\begin{eqnarray}
\slashed D\Psi=\left[\begin{array}{cc}0&{\cal  A} \\ {\cal  B} &0 \end{array}\right]\Psi=0\,,
\end{eqnarray}
where
\begin{eqnarray}\label{ANABDBDI}\nonumber
{\cal  A}&\equiv& \frac{1}{f}\frac{\partial}{\partial r}I_2-\frac{a^2{\cal C} f}{r^3}\hat\sigma_z+\frac{2}{r}\left(\hat {\cal J}_x\hat\sigma_x+\hat {\cal J}_y\hat\sigma_y-f\hat {\cal J}_z\hat\sigma_z\right)\,,\\
{\cal B}&\equiv& \left(\frac{1}{f}\frac{\partial}{\partial r}+\frac{2f}{r}+\frac{1}{fr}\right)I_2+\frac{a^2{\cal C} f}{r^3}\hat\sigma_z-\frac{2}{r}\left(\hat {\cal J}_x\hat\sigma_x+\hat {\cal J}_y\hat\sigma_y-f\hat {\cal J}_z\hat\sigma_z\right)\,. 
\end{eqnarray}
Writing $\Psi=\left[\begin{array}{c}\lambda_\alpha\\\bar\chi^{\dot\alpha}\end{array}\right]$\,, we obtain the two equations
\begin{eqnarray}\nonumber
\left[ \frac{1}{f}\frac{\partial}{\partial r}I_2-\frac{a^2{\cal C} f}{r^3}\hat\sigma_z+\frac{2}{r}\left(\hat {\cal J}_x\hat\sigma_x+\hat {\cal J}_y\hat\sigma_y-f\hat {\cal J}_z\hat\sigma_z\right)\right]\bar\chi=0\,,\\
\left[ \left(\frac{1}{f}\frac{\partial}{\partial r}+\frac{2f}{r}+\frac{1}{fr}\right)I_2+\frac{a^2{\cal C} f}{r^3}\hat\sigma_z-\frac{2}{r}\left(\hat {\cal J}_x\hat\sigma_x+\hat {\cal J}_y\hat\sigma_y-f\hat {\cal J}_z\hat\sigma_z\right)\right]\lambda=0\,.
\end{eqnarray}

We start by solving for $\lambda$. After simple manipulations, we find
\begin{eqnarray}\label{equation for lambda}\nonumber
\left[ \frac{1}{f}\frac{\partial}{\partial r}+\frac{2f}{r}+\frac{1}{fr}+\frac{a^2{\cal C}f}{r^3}+\frac{2f}{r}\hat{\cal J}_z\right]\lambda_1-\frac{2}{r}\hat{\cal J}_-\lambda_2&=&0\,,\\
-\frac{2}{r}\hat{\cal J}_+\lambda_1+\left[ \frac{1}{f}\frac{\partial}{\partial r}+\frac{2f}{r}+\frac{1}{fr}-\frac{a^2{\cal C}f}{r^3}-\frac{2f}{r}\hat{\cal J}_z\right]\lambda_2&=&0\,,
\end{eqnarray}
and $\hat{\cal J}_{\pm}=\hat{\cal J}_x\pm i \hat{\cal J}_y$.
The solutions to these equations can be decomposed as radial eigenstates times angular eigenstates on $S^3$:
\begin{eqnarray}\label{formalsol}
\lambda_1=g_1(r)|j_1,m_1',m_1\rangle\,,\quad \lambda_2=g_2(r)|j_2,m_2',m_2\rangle\,,
\end{eqnarray}
where $|j,m',m\rangle$ is the Wigner  D-matrix \cite{Sakurai:2011zz}:
\begin{eqnarray}\nonumber
|j,m',m\rangle\equiv D_{m',m}^j(\theta,\varphi,\psi)\equiv d^j_{m',m}(\theta)e^{im\varphi}e^{im'\psi}\,,\quad j=0,\frac{1}{2}, 1,...\,, \quad  |m|, |m'|\leq j\,,\\
\end{eqnarray}
and $d^j_{m',m}(\theta)$ is the Wigner (small) d-matrix:
\begin{eqnarray}\nonumber
d^j_{m',m}(\theta)&=&\sum_{k}(-1)^{k-m+m'}\frac{\sqrt{(j+m)!(j-m)!(j+m')!(j-m')!}}{(j+m-k)!k!(j-k-m')!(k-m+m')!}\\
&&\times\left(\cos\frac{\theta}{2}\right)^{2j-2k+m-m'}\left(\sin\frac{\theta}{2}\right)^{2k-m+m'}\,,
\end{eqnarray}
and the sum over $k$ includes all integers whenever none of the arguments of the factorials in the denominator is negative. These functions satisfy the following algebraic relations:
\begin{eqnarray}\label{updown}\nonumber
\hat{\cal J}_z |j,m',m\rangle=m' |j,m',m\rangle\,,\quad \hat{\cal J}_\pm |j,m',m\rangle=\sqrt{j(j+1)-m'(m'\pm 1)} |j,m'\pm1,m\rangle\,.\\
\end{eqnarray}
The states $\{|j,m',m\rangle\}$ form a complete set of orthonormal bases and satisfy the orthogonality condition
\begin{eqnarray}
\int_0^{2\pi}d\psi \int_0^{\pi}d\theta \sin\theta\int_0^{2\pi} d\varphi\langle j_1,m_1',m_1|j_2,m_2',m_2\rangle=\frac{8\pi^2}{2j_1+1}\delta_{j_1j_2}\delta_{m_1'm_2'}\delta_{m_1m_2}\,.
\end{eqnarray}

Substituting (\ref{formalsol}) into (\ref{equation for lambda}) and using (\ref{updown}), we find
\begin{eqnarray}\label{equation for lambdaalmost}\nonumber
&&\left[ \frac{1}{f}\frac{\partial}{\partial r}+\frac{2f}{r}+\frac{1}{fr}+\frac{a^2{\cal C}f}{r^3}+\frac{2fm_1'}{r}\right]g_1(r)|j_1,m_1',m_1\rangle\\
&&\quad\quad\quad-\frac{2}{r}\sqrt{j_2(j_2+1)-m_2'(m_2'- 1)}g_2(r)|j_2,m_2'-1,m_2\rangle=0\,,
\end{eqnarray}
and
\begin{eqnarray}\label{equation for lambdaalmost2}\nonumber
&&-\frac{2}{r}\sqrt{j_1(j_1+1)-m_1'(m_1'+ 1)}g_1(r)|j_1,m_1'+1,m_1\rangle\\
&&\quad\quad\quad+\left[ \frac{1}{f}\frac{\partial}{\partial r}+\frac{2f}{r}+\frac{1}{fr}-\frac{a^2{\cal C}f}{r^3}-\frac{2fm_2'}{r}\right]g_2(r)|j_2,m_2',m_2\rangle=0\,.
\end{eqnarray}
Inspection of the structure of (\ref{equation for lambdaalmost}) reveals that there are two possibilities in which the equation can be satisfied. First, we can assume that the states $|j_1,m_1',m_1\rangle$ and $|j_2,m_2'-1,m_2\rangle$ are identical. In this case, we must set $j_1=j_2=j$ and $m_1'=m_2'-1$. However, we know that the minimum value of $m_2'$ is $-j$, and thus, we find that there will be a state with $m_1'=-j-1<-j$, which is a contradiction. So, this cannot happen. The other option is that the states $|j_1,m_1',m_1\rangle$ and $|j_2,m_2'-1,m_2\rangle$ are distinct. Since they are orthogonal, each term in the l.h.s. of  (\ref{equation for lambdaalmost}) must vanish identically. This forces us to choose the state with $m_2'=-j_2$. Repeating the same reasoning to Eq. (\ref{equation for lambdaalmost2}), we find that the only possible solutions are given by
\begin{eqnarray}\nonumber
\lambda_1=g_1(r)|j_1,m'_1=j_1,m_1\rangle\,, \quad \lambda_2=g_2(r)|j_2,m'_2=-j_2,m_2\rangle\,,\quad |m_1|\leq j_1\,,|m_2|\leq j_2\,.\\
\end{eqnarray}
Using $\hat {\cal J}_-|j,m'=-j,m\rangle=\hat {\cal J}_+|j,m'=j,m\rangle=0$, Eq. (\ref{equation for lambda}) simplifies to (we do not distinguish between $j_1$ and $j_2$, referring to both total angular momentum by $j$)
\begin{eqnarray}\nonumber
\left[ \frac{1}{f}\frac{\partial}{\partial r}+\frac{2f}{r}+\frac{1}{fr}+\frac{a^2{\cal C}f}{r^3}+\frac{2f j}{r}\right]g_1&=&0\,,\\
\left[ \frac{1}{f}\frac{\partial}{\partial r}+\frac{2f}{r}+\frac{1}{fr}-\frac{a^2{\cal C}f}{r^3}+\frac{2f j}{r}\right]g_2&=&0\,.
\end{eqnarray}

The normalizability condition is
\begin{eqnarray}\label{normalizofpsi}
\int |\Psi|^2 r^3 dr\wedge \sigma_x\wedge \sigma_y\wedge \sigma_z<\infty\,.
\end{eqnarray}
For ${\cal C} > 0$, only $g_2$ yields normalizable solutions on the Eguchi--Hanson space, while for ${\cal C} < 0$, only $g_1$ is normalizable. When ${\cal C}=0$, it is easy to check that none of the equations yield normalizable zero modes.  In the following, and without loss of generality, we perform our analysis assuming ${\cal C} > 0$. In this case, the normalizable solutions of $g_2$ exist for ${\cal C}>2j$. They are given by
\begin{eqnarray}
g_2(r)=\frac{1}{r}(r^2-a^2)^{\frac{{\cal C}-2j-2}{4}}(r^2+a^2)^{-\frac{{\cal C}+2j+2}{4}}\,, \quad C>2j\,,\quad j\geq 0\,.
\end{eqnarray}
Near $r=a$ we have $g_2^2\sim (r-a)^{\frac{{\cal C}-2j-2}{2}}$, and thus, the integral (\ref{normalizofpsi}) converges at $r=a$ provided that ${\cal C}>2j$. On the other hand, as $r\rightarrow \infty$, we have $r^3g_2^2\sim r^{-4j-3}$, and the integral (\ref{normalizofpsi})  is convergent at $r\rightarrow \infty$ for $j\geq0$.

We must distinguish between even and odd values of ${\cal C}$. When ${\cal C}$ is even, ${\cal C} = 2,4,6,\ldots$, the following values of $j$ yield normalizable zero modes:
\begin{equation}
j = 0, \frac{1}{2}, 1, \ldots, \frac{{\cal C} - 1}{2}.
\end{equation}
When ${\cal C}$ is odd, ${\cal C} = 1,3,5,\ldots$, the values of $j$ that yield normalizable zero modes are
\begin{equation}
j = 0, \frac{1}{2}, 1, \ldots, \frac{{\cal C} - 1}{2}.
\end{equation}
In both cases, even or odd ${\cal C} > 0$, the values of $m$ run between $-j, -j+1, \ldots, j-1, j$. Thus, one would conclude, in both cases, that the total number of the zero modes is ${\cal C}({\cal C}+1)/2$, a result that references \cite{Franchetti:2017ftp,Franchetti:2023wcw,Catinari:2024zon} arrived at. This, however, is not correct.

The idea is that not all these normalizable zero modes are physical in the sense that some of the spinor zero modes must be eliminated as their parallel transport does not yield a single-valued element in the lift of $\mathrm{SO}(4)$ to $\mathrm{Spin}(4)$. Let us elaborate on this statement in more detail. The original range of the angle $\psi$ over $S^3$, e.g., in flat $4$-D space, is $0 \leq \psi \leq 4\pi$. In the Eguchi--Hanson space, however, one identifies
\begin{equation}
(x,y,z,t) \sim (-x,-y,-z,-t),
\end{equation}
and thus, the range of $\psi$ is reduced to the interval $0 \leq \psi \leq 2\pi$. A spin structure in this space does exist provided that $\Psi(x)$ is continuous along a path from $(x,y,z,t) $ to $(-x,-y,-z,-t)$ and this path defines an element of the covering group of $\mathrm{SO}(4)$, i.e. $\mathrm{Spin}(4)$, unambiguously.
Recalling that the identification of sections of the spinor bundle at $(x,y,z,t)$ with those at $(-x,-y,-z,-t)$ amounts to identifying the spinors at $\psi$ with those at $\psi+2\pi$, one must ensure that the spinors are single-valued under this identification.

To gauge whether the spinors at $\psi$ and $\psi+2\pi$ are identical, one must first parallel transport a fermion located at $\psi$ to $\psi+2\pi$ on a contractible path. The fermion acquires $U(1)$ as well as gravitational holonomies upon  parallel transporting along a contractible path $\ell$:
\begin{align}\label{PARALLTA}
\Psi(r,\theta,\varphi, \psi+2\pi) &= \exp\left[i\int_{\ell}A\right] \exp\left[-\frac{1}{2}\int_{\ell}\omega_{ab}\Sigma^{ab}\right] \Psi(r,\theta,\varphi, \psi)\,.
\end{align}
 We take the path $\ell_{r \approx a}$ that connects two points $(r \approx a,\theta=0,\varphi=0, \psi)$ and $(r \approx a,\theta=0,\varphi=0, \psi+2\pi)$ near the bolt $\mathbb {S} ^2$, which is contractible.
Using (\ref{omegaEH}, \ref{Sigmama}, \ref{form of A2}), and $f^{-1}(r \approx a)=0$, yields
\begin{eqnarray}\nonumber
\exp\left[-\frac{1}{2}\int_{\ell_{r \approx a}}\omega_{ab}\Sigma^{ab}\right]&=&\exp\left[-i\int_{\ell_{r \approx a}}\left[\begin{array}{cc}\frac{\sigma_x\hat\sigma_x}{f}+\frac{\sigma_y\hat\sigma_y}{f}+(2-f^{-2})\sigma_z\hat\sigma_z&0\\0&0\end{array}\right]\right]=I_4\,, \\
 \exp\left[i\int_{\ell_{r\approx a}}A\right] &=& e^{i\pi{\cal C}}\,.
\end{eqnarray}
Notice that the gravitational holonomy is a trivial phase upon parallel transporting the spinor along $\ell_{\approx a}$. Thus, under gravity and gauge fields, parallel transporting the spinor along the path $\ell_{\approx a}$ gives 
\begin{align}\label{PARALLTA2}
\Psi(r \approx a,\theta,\varphi, \psi+2\pi) &= e^{i\pi{\cal C}} \Psi(r \approx a,\theta,\varphi, \psi)\,.
\end{align}
On the other hand, from the solution of the zero modes, we can directly calculate $\Psi$ at $\psi+2\pi$:
\begin{align}\label{aretheysame}\nonumber
\Psi(r,\theta,\varphi,\psi) &= \left[\begin{array}{c}
0\\
g_2(r)e^{-ij\psi}e^{im\varphi}d_{m',m}^j(\theta)\\
0\\
0
\end{array}\right] \longrightarrow\\
\Psi(r,\theta,\varphi,\psi+2\pi)&= e^{-i2\pi j}\left[\begin{array}{c}
0\\
g_2(r)e^{-ij\psi}e^{im\varphi}d_{m',m}^j(\theta)\\
0\\
0
\end{array}\right]\,.
\end{align}
The fermion zero modes are well-defined on EH space, i.e., they are unambiguously defined on a contractible path, if and only if the r.h.s. of (\ref{PARALLTA2}) and (\ref{aretheysame}) are identical. 

We further study the behavior of the fermions as $r\rightarrow \infty$, where EH space approaches its boundary $\mathbb {RP}^3$. In particular, we need to find the correct boundary conditions at infinity consistent with (\ref{PARALLTA2}). To this end,  let $\ell_\infty$ be a path that takes us from $\psi$ to $\psi+2\pi$ at a fixed $ r\rightarrow \infty$ and fixed $\theta,\varphi$. Under parallel transport along $\ell_\infty$, and using $f^{-1}(r\rightarrow \infty)\rightarrow 1$, we find 
\begin{eqnarray}\label{holatinfty}\nonumber
&&\Psi(r\gg a, \theta,\varphi, \psi+2\pi)= \underbrace{\exp\left[i\int_{\ell_\infty}A\right]}_{=(-1)^{{\cal C}}} \exp\left[-\frac{1}{2}\int_{\ell_\infty}\omega_{ab}\Sigma^{ab}\right] \Psi(r\gg a,\theta,\varphi, \psi)\\\nonumber
&&=(-1)^{{\cal C}}\exp\left[-i\int_{\ell_{\infty}}\left[\begin{array}{cc}\frac{\sigma_x\hat\sigma_x}{f}+\frac{\sigma_y\hat\sigma_y}{f}+(2-f^{-2})\sigma_z\hat\sigma_z&0\\0&0\end{array}\right]\right] \Psi(r\gg a,\theta,\varphi, \psi)\\\nonumber
&&=(-1)^{{\cal C}}\exp\left[\left[\begin{array}{cc}-i\pi\hat\sigma_z&0\\0&0_2\end{array}\right]\right] \Psi(r\gg a,\theta,\varphi, \psi)=-(-1)^{{\cal C}}\gamma_5  \Psi(r\gg a,\theta,\varphi, \psi)\,,\\
\end{eqnarray}
where 
\begin{eqnarray}
\gamma^5\equiv\gamma^0\gamma^1\gamma^2\gamma^3=\left[\begin{array}{cc}I_2&0\\0&-I_2\end{array}\right]\,.
\end{eqnarray}
The computation that gives $\exp\left[i\int_{\ell_\infty}A\right]=(-1)^{{\cal C}}$ is discussed in the bulk of the paper.

The following remark is in order. 
The relations \eqref{PARALLTA2} and \eqref{holatinfty} have different logical status.
First, \eqref{PARALLTA2} is imposed on a \emph{contractible} loop near the bolt. In this region, the gravitational holonomy is trivial, so the only non–trivial phase is the abelian gauge holonomy, which is a scalar. Matching this with the explicit $\psi$--dependence of the modes, $\Psi \sim e^{-ij\psi}$, fixes the allowed values of the total angular momentum $j$.
By contrast, \eqref{holatinfty} describes the spin$\otimes$gauge holonomy on the \emph{non–contractible} $\mathbb{Z}_2$ cycle of $\mathbb{RP}^3$ at infinity. The operator $(-1)^{C+1}\gamma_5$ acts on the spinor indices (and on the $U(1)$ charge), and should not be identified with the scalar phase $e^{-i2\pi j}$ coming from the Wigner $D$–functions. In particular, there is no further quantization condition of the form
$
  e^{-i2\pi j} = (-1)^{C+1}
$
at $r\to\infty$; the only restriction on $j$ comes from \eqref{PARALLTA2}. Equation \eqref{holatinfty} instead provides the input for the $\eta$–invariant and the APS index theorem, encoding the non–trivial spin structure and flat boundary line bundle.

In conclusion, we use the boundary conditions near the bolt (\ref{PARALLTA2}) to determine the number of globally well-defined zero modes. We distinguish two cases.

{\bf Case I}: ${\cal C}$ is even, $C=2p\geq 0$, the gauge holonomy near the bolt gives a trivial phase, and thus, one must restrict the values of $j$ to integers. In this case, the number of normalizable zero modes is
\begin{eqnarray}
 {\cal I}_{U(1)}=\sum_{j=0,1,..}^{p-1}(2j+1)=p^2\,.
 \end{eqnarray}

{\bf Case II}: ${\cal C}$ is odd, ${\cal C}=2p+1\geq 1$, the gauge holonomy near the bolt gives an $e^{i\pi}$ phase, and thus, one must restrict $j$ to half-integers. The total number of normalizable zero modes in this case is  
\begin{eqnarray}
 {\cal I}_{U(1)}=\sum_{j=\frac{1}{2},\frac{3}{2},..}^{p-1/2}(2j+1)=p(p+1)\,.
 \end{eqnarray}
 
These findings match the results in \cite{tHooft:1988wxy} as well as the Atiyah-Patodi-Singer index theorem.

Before concluding this section, let us briefly comment on $\bar\chi$. 
Repeating the same analysis, one finds that these fermions admit no 
normalizable zero modes. The difference compared to $\lambda$ arises from 
their distinct gravitational couplings to the EH geometry. In particular, 
$\chi$ is acted upon by the operator 
$\tfrac{1}{f}\tfrac{\partial}{\partial r}$ within ${\cal A}$ in 
(\ref{ANABDBDI}), whereas $\lambda$ is acted upon by 
$\tfrac{1}{f}\tfrac{\partial}{\partial r} + \tfrac{2f}{r} + \tfrac{1}{fr}$ 
within ${\cal B}$ in (\ref{ANABDBDI}). This asymmetry is natural, since the 
EH space is self-dual: the coupling of fermions to gravity through the spin 
connection $\Sigma^{ab}\omega_{ab}$ singles out one chirality. If instead we 
considered the EH space with the opposite orientation (self-dual rather 
than anti-self-dual), the situation would be reversed and $\bar\chi$ would possess 
normalizable zero modes.

%%%%%%%%%%%%%%%%%%%%%%%%%%%%%%%
\section{The Atiyah-Patodi-Singer index theorem}
\label{The Atiyah-Patodi-Singer index theorem}
%%%%%%%%%%%%%%%%%%%%%%%%%%%%%%%

To verify the counting of the zero modes, we apply the Atiyah-Patodi-Singer (APS) index theorem \cite{Atiyah:1975jf,Atiyah:1976qjr,Atiyah:1976jg,Eguchi:1980jx}, which gives the number of chiral-zero modes in the background of a twisted spin complex (twisted by a gauge bundle $V$) with boundaries\footnote{We note that the sign in front of the term $\frac{T_{\cal R}}{8\pi^2}\int_{\mathbb M}\mbox{tr}F\wedge F$ in \cite{Eguchi:1980jx} is opposite to our sign since they use anti-hermitian gauge fields.}:
\begin{eqnarray}\label{APSINDEXTHEO}\nonumber
{\cal I}=\frac{T_{\cal R}}{8\pi^2}\int_{\mathbb M}\mbox{tr}F\wedge F+\mbox{dim}_{ {\cal R}}\left[\frac{1}{24\cdot 8\pi^2}\int_{\mathbb M}\mbox{tr}R\wedge R-\int_{\partial\mathbb M}\mbox {tr}\, \Theta\wedge R\right]-\eta_V(\partial\mathbb M)\,.\\
\end{eqnarray}
Here, $\Theta$ is the second fundamental form computed at the boundary of $\mathbb M$ and $\eta$ is the $\eta$-invariant defined as
\begin{eqnarray}\label{etainvdef}
\eta_V(\partial\mathbb M)=\frac{1}{2}\mbox{lim}_{T\rightarrow 0}\sum_{\lambda_i\neq 0}\mbox{sign}(\lambda_i)|\lambda_i|^{-T}\,.
\end{eqnarray}
The subscript $V$ means that in general, the $\eta$-invariant depends on the vector bundle $V$, and
the sum is over all the non-zero eigenvalues of the Dirac operator in the presence of the gauge-bundle twist computed at the boundary of $\mathbb M$.   The term $\int_{\partial\mathbb M}\mbox{tr}\, \Theta\wedge R$ is the Chern-Simons boundary term, which vanishes for the EH space \cite{Eguchi:1978gw}. Finally, $T_{\cal R}$ and $\mbox{dim}_{ {\cal R}}$ are the Dynkin index and the dimension of the group representation ${\cal R}$ of the gauge bundle $V$, respectively. 

Although the EH space is noncompact, its asymptotic region has the
structure of a cylindrical end $\mathbb{R}_+ \times \mathbb{RP}^3$. To apply the
APS index theorem in this ALE background, we introduce a large radial cutoff
$r = R$ (a regulator) and restrict to the compact manifold $\mathbb{M}_R$ with boundary
$\partial \mathbb{M}_R \simeq \mathbb{RP}^3$ at $r=R$. On $\partial \mathbb{M}_R$, we keep fixed
the induced metric and the pullback of the background gauge connection,
including the $\mathbb{Z}_2$ holonomy information, and we choose
the compatible (twisted) spin structure. The APS boundary condition, which is a non-local condition, restricts
the boundary value of the spinor to lie in the subspace spanned by eigenmodes
of the three-dimensional Dirac operator on $\mathbb{RP}^3$ (with flat
bundle) whose eigenvalues have a prescribed sign, and ensures that the Dirac operator is self-adjoint at the boundary. 
 The index computed from
(\ref{APSINDEXTHEO}) on $\mathbb{M}_R$ is independent of $R$ (since it is a topological invariant independent of the regulator), so we can take the limit
$R \to \infty$ and regard the result as the index of the Dirac operator on the
full EH space with an asymptotically cylindrical end. Notice that the use of the APS non-local boundary conditions is totally acceptable when it comes to counting the zero modes, as it only provides a convenient mathematical device to guarantee that the Dirac operator is self-adjoint.
 For a physicist's friendly review of this procedure on a general manifold with boundaries, we refer the reader to \cite{Ninomiya:1984ge}.

The contribution from the $U(1)$ and gravitational backgrounds to the APS index were already computed above:
\begin{eqnarray}
\frac{1}{8\pi^2}\int_{\scriptsize\mbox{EH}}F_{(1)}\wedge F_{(1)}=\frac{{\cal C}^2}{4}\,, \quad \frac{1}{24\cdot 8\pi^2}\int_{\scriptsize\mbox{EH}}\mbox{tr}R\wedge R=-\frac{1}{8}\,.
\end{eqnarray}
 The $\eta$-invariant gives a non-trivial contribution at the boundary $\mathbb {RP}^3$ of EH space. Its value depends on whether the $U(1)$ charge ${\cal C}$ is even or odd (recall the discussion near Eq.  (\ref{inftyBC}) in Section \ref{Spinors in EH space in the presence of abelian fields} about the behavior of fermions at $\mathbb {RP}^3$) :
\begin{eqnarray}\label{etavalrp3}\nonumber
\eta(\mathbb{RP}^3)&=&-\frac{1}{8}\,, \quad {\cal C}\in 2\mathbb Z\,,\\
\eta(\mathbb{RP}^3)&=&+\frac{1}{8}\,, \quad {\cal C}\in 2\mathbb Z+1\,.
\end{eqnarray}
The two different values of $\eta$ are attributed to the fact that there are two choices of twisting by a flat line bundle $\{+1,-1\}$, which are selected by whether ${\cal C}$ is even or odd, as we explain in the next section. 
Using this information, we readily find
\begin{eqnarray}\label{counting U(1)}\nonumber
{\cal I}&=&p^2\,,\quad {\cal C}=2p\geq0\,,\\
{\cal I}&=&p(p+1)\,,\quad {\cal C}=2p+1\geq1\,,
\end{eqnarray}
matching the number of zero modes obtained by directly constructing them.

When applying the APS index to a non-abelian bundle, one must pay special attention to the holonomy on $\mathbb{RP}^3$. 
Since the holonomy is defined modulo $2$, the holonomy of a well-defined general gauge bundle 
(such that fermions can be globally defined) takes the form, e.g., 
\begin{equation}
  \mathrm{Hol}(A) \;\equiv\; \exp\!\left(i \int_{\ell_\infty} A\right) 
  \;=\; \mathrm{diag}(+1, -1, -1, \ldots)\,.
\end{equation}
We define the \emph{signature} of the holonomy matrix to be the number of $(+1)$ eigenvalues 
minus the number of $(-1)$ eigenvalues, i.e.
\begin{equation}
  \mathrm{sgn}(\mathrm{Hol}) \;=\; \#(+1) - \#(-1)\,.
\end{equation}
Since gauge transformations act on the holonomy by conjugation, $\mbox{Hol}\rightarrow U\;\mbox{Hol}\;U^{-1}$, the multiset of eigenvalues
of the holonomy matrix is preserved. In particular, if the holonomy has eigenvalues 
$\{ \lambda_1, \dots, \lambda_n \}$, these are unchanged (up to reordering) under gauge transformations. Thus,  $ \mathrm{sgn}(\mathrm{Hol}) $ is gauge invariant. 
With this definition, the APS index takes the form
\begin{equation}\label{indexmaster}
  \mathcal{I} 
  \;=\; \frac{T_{\mathcal{R}}}{8\pi^2} \int_{\scriptsize\mbox{EH}} \mathrm{tr}\, F \wedge F 
      \;+\; \frac{-\,\dim_{\mathcal{R}} + \mathrm{sgn}(\mathrm{Hol})}{8}\,.
\end{equation}

In the presence of both a nonabelian $F$ and an abelian $f$ gauge fields, the APS index is easily generalized to
\begin{equation}\label{indexmastermix}
  \mathcal{I} 
  \;=\; \frac{T_{\mathcal{R}}}{8\pi^2} \int_{\scriptsize\mbox{EH}} \mathrm{tr}\, F \wedge F 
      \;+\frac{\dim_{\mathcal{R}}}{8\pi^2}\int_{\scriptsize\mbox{EH}}f\wedge f+ \frac{-\,\dim_{\mathcal{R}} + \mathrm{sgn}(\mathrm{Hol})}{8}\,.
\end{equation}

%%%%%%%%%%%%%%%%%%%%%%%%%%%%%%%%%%%%
\subsection{An explicit calculation of the $\eta$-invariant on $L^2(n)$}
\label{An explicit calculation of eta on L2}
%%%%%%%%%%%%%%%%%%%%%%%%%%%%%%%%%%%

In this part, we briefly sketch the calculations of the $\eta$-invariant on the Lens space $L^2(n)\equiv S^3/\mathbb Z_n$, where $\mathbb {RP}^3\equiv S^3/\mathbb Z_2$ is a special case.  We closely follow the steps in \cite{Pope:1981jx} that calculated the invariant on $S^3$. We mod out by $\mathbb Z_n$ at the end. One important technical point is that the action of $\mathbb Z_n$ on $S^3$ must be fixed-point free, otherwise the resulting quotient space becomes singular at the fixed points. Therefore, a care must be exercised when choosing the coordinates that cover $S^3$. It turns out that one can avoid this technical difficulty by working with Hopf coordinates, where $S^3 \cong S^2\times S^1$. The metric on $S^3$ reads
\begin{eqnarray}
ds^2=\sigma_1^2+\sigma_2^2+\sigma_3^2\,,
\end{eqnarray}
where $\sigma_{1,2,3}$ are the left-invariant $1$-form on $S^3$ defined in (\ref{sigmasinewh}).
The eigenvalues and eigenstates of the Dirac operator in the background of this metric were explicitly constructed in  \cite{Pope:1981jx}. The positive and negative eigenvalues and their degeneracies are given by: 
\begin{eqnarray}
\nonumber
\lambda_+&=&\frac{3}{2}+m\,,\quad d=(m+1)(m+2)\,,\\
\lambda_-&=&-\frac{1}{2}-m\,,\quad d=m(m+1)\,,
\label{eigenvalues on s3}
\end{eqnarray}
for $m=0,1,...$.
 Before modding out by $\mathbb Z_n$, we use the definition (\ref{etainvdef}) to calculate the $\eta$-invariant. We directly sum over the signs of the eigenvalues (\ref{eigenvalues on s3}), introduce a regulator $T>0$ to render the sum finite and well defined, and take the limit $T\rightarrow 0$ at the end of calculations (the regulator we use here is more friendly than the one used in the original definition in  (\ref{etainvdef})):
\begin{eqnarray}
\nonumber
&&\eta(i\slashed D_{S^3})=\\\nonumber
&&\mbox{lim}_{T\rightarrow 0}\left\{\frac{1}{2}\sum_{m=0}^{\infty}(m+1)(m+2)e^{-T(m+3/2)}-\frac{1}{2}\sum_{m=0}^{\infty}m(m+1)e^{-T(m+1/2)}\right\}=0\,,\\
\label{eta sum}
\end{eqnarray}
which is the expected result on a symmetric  $S^3$. 

Next, we compute $\eta$ on the Lens space $L^2(n)$ by modding out $S^3$ by the discrete group $\mathbb Z_n$. In terms of the coordinates on $S^3$, this means that
Euler angle $\psi$ is the identified modulo $4\pi/n$. To this end,  we introduce the $n$-th root of unity $\omega\equiv e^{i\frac{2\pi}{n}}$ and define the projection operator:
\begin{eqnarray}
P_{m,s}=\frac{1}{n}\sum_{j=1}^n \omega^{j(m+s+1)}=\left\{\begin{array}{cc} 1\,,&\quad m+s+1=0 \,\mbox{mod}\, n\\ 0\,,&\quad \mbox{otherwise}
\end{array}\right.\,.
\end{eqnarray}
When applied to each term in the sum (\ref{eta sum}), this operator retains the eigenvalues that satisfy the condition $m+s+1=np$ for some integer $p$. Setting $s=0$,  we find that only the eigenvalues that satisfy $m+1=np$ contribute to the $\eta$-invariant; these eigenvalues correspond to the eigenstates that respect the identification  $\psi\sim\psi+\frac{4\pi}{n}$ under $\mathbb Z_n$. By definition, we declare that fermions whose eigenvalues obey the relation $m+1=np$ have charge $s=0$ under $\mathbb Z_n$. Higher values of $s$ correspond to fermions that have higher charges under $\mathbb Z_n$. Notice also the equivalence of charges $s\sim s+n$. Finally, by inserting $P_{m,s}$ into the sum  (\ref{eta sum}) we obtain
\begin{eqnarray}
\nonumber
&&\eta(i\slashed D_{\mathbb L^2(n)})=\\
\nonumber
&&\mbox{lim}_{T\rightarrow 0}\left\{\frac{1}{2}\sum_{m=0}^{\infty}(m+1)(m+2)P_{m,s}e^{-T(m+3/2)}-\frac{1}{2}\sum_{m=0}^{\infty}m(m+1)P_{m,s}e^{-T(m+1/2)}\right\}\\
&&\quad\quad=\frac{1}{n}\sum_{j=1}^{n-1}\omega^{js}\frac{\omega^j}{(\omega^j-1)^2}\,.
\label{eta sum on Lens 2}
\end{eqnarray}

In the special case of $\mathbb{RP}^3$, we have $\omega=-1$, and there are two possible values of $s=0,1$. The value $s=0$ corresponds to a trivial twist when ${\cal C}\in 2\mathbb Z$ in (\ref{inftyBC}). This gives  $\eta(\mathbb{RP}^3)=-1/8$. The second value $s=1$ corresponds to a twist by $(-1)$ when  ${\cal C}\in 2\mathbb Z+1$ in (\ref{inftyBC}). This gives  $\eta(\mathbb{RP}^3)=1/8$.

%%%%%%%%%%%%%%%%%%%%%%%%%%%%%%%%%%%%%%%%
\section{Fermions in the background of $PSU(2)$ gauge field}
\label{Fermions in the background of PSU(2) gauge field}
%%%%%%%%%%%%%%%%%%%%%%%%%%%%%%%%%%%%%%%%

Here, we study the fundamental- and adjoint-fermion zero modes in the background of the $PSU(2)\equiv SU(2)/\mathbb Z_2$ gauge field. This topic, albeit being tangential to the main theme of this paper, provides valuable lessons, especially since the background is simple enough. Moreover, the construction of the adjoint zero modes can be of interest in studying supersymmetric gauge theories.

We start by turning on a background along the Cartan generator of $SU(2)$:
\begin{eqnarray}
A_{SU(2)}=m_{(2)}H\nu  \frac{\sigma_z a^2}{r^2}=m_{(2)}\left[\begin{array}{cc}\frac{1}{2}&0\\0&-\frac{1}{2}\end{array}\right]\frac{\sigma_z a^2}{r^2}\,,
\end{eqnarray}
where $H$ is the Cartan generator of $SU(2)$, $\nu$ is the weight of the fundamental representation, $m_{(2)}$ is an integer, and the subscript reminds us that we are dealing with the $SU(2)$ gauge group. The field strength of this background gauge field (the background $2$-form field) is
\begin{eqnarray}
F_{(2)} = \frac{2m_{(2)}a^2}{r^4}\left[\begin{array}{cc}\frac{1}{2}&0\\0&-\frac{1}{2}\end{array}\right] (e^3 \wedge e^0 + e^1 \wedge e^2)\,,
\end{eqnarray}
with flux given by
\begin{eqnarray}
\int_{S^2} F_{(2)}=-2\pi m_{(2)} \left[\begin{array}{cc}\frac{1}{2}&0\\0&-\frac{1}{2}\end{array}\right]\in 2\pi\mathbb Z_2\,,
\end{eqnarray}
showing that this is the flux of a $\mathbb Z_2^{(1)}$ $1$-form gauge field that pierces through the bolt of EH space. The topological charge of this background gauge field is
\begin{eqnarray}
Q_{(2)}=\frac{1}{8\pi^2}\int_{\scriptsize\mbox{EH}} \mbox{tr}F_{(2)}\wedge F_{(2)}=\frac{m_{(2)}^2}{8}\,.
\end{eqnarray}

\subsection{Fundamental fermions}
We start with the fundamental fermions. Without loss of generality, we take $m_{(2)}\in\mathbb Z^+$. Repeating the same analysis we carried out in the $U(1)$ case, we find that the Dirac equation is:
\begin{eqnarray}\nonumber
&&\left[ \frac{1}{f}\frac{\partial}{\partial r}I_2\otimes I_{SU(2)}-\frac{a^2 m_{(2)} f}{r^3}\hat\sigma_z\otimes \left[\begin{array}{cc}\frac{1}{2}&0\\0&-\frac{1}{2}\end{array}\right]+\frac{2}{r}\left(\hat {\cal J}_x\hat\sigma_x+\hat {\cal J}_y\hat\sigma_y-f\hat {\cal J}_z\hat\sigma_z\right)\otimes I_{SU(2)}\right]\bar\chi=0\,,\\\nonumber
&&\left[ \left(\frac{1}{f}\frac{\partial}{\partial r}+\frac{2f}{r}+\frac{1}{fr}\right)I_2\otimes I_{SU(2)}+\frac{a^2m_{(2)} f}{r^3}\hat\sigma_z\otimes \left[\begin{array}{cc}\frac{1}{2}&0\\0&-\frac{1}{2}\end{array}\right]\right.\\
&&\left.-\frac{2}{r}\left(\hat {\cal J}_x\hat\sigma_x+\hat {\cal J}_y\hat\sigma_y-f\hat {\cal J}_z\hat\sigma_z\right)\otimes I_{SU(2)}\right]\lambda=0\,.
\end{eqnarray} 
$I_{SU(2)}$ is a unit $2\times 2$ matrix, and the subscript is used to distinguish it from the unit matrix in spin space. 
Again, it is easy to show that $\chi$ does not have normalizable zero modes, while the equation of $\lambda$ gives 
\begin{eqnarray}\label{equation for lambdasu2}\nonumber
\left[ \left(\frac{1}{f}\frac{\partial}{\partial r}+\frac{2f}{r}+\frac{1}{fr}\right)I_{SU(2)}+\frac{a^2m_{(2)}f}{r^3}\left[\begin{array}{cc}\frac{1}{2}&0\\0&-\frac{1}{2}\end{array}\right]+\frac{2f}{r}\hat{\cal J}_zI_{SU(2)}\right]\lambda_1-\frac{2}{r}\hat{\cal J}_-I_{SU(2)}\lambda_2&=&0\,,\\\nonumber
-\frac{2}{r}\hat{\cal J}_+I_{SU(2)}\lambda_1+\left[ \left(\frac{1}{f}\frac{\partial}{\partial r}+\frac{2f}{r}+\frac{1}{fr}\right)I_{SU(2)}-\frac{a^2m_{(2)}f}{r^3}\left[\begin{array}{cc}\frac{1}{2}&0\\0&-\frac{1}{2}\end{array}\right]-\frac{2f}{r}\hat{\cal J}_zI_{SU(2)}\right]\lambda_2&=&0\,,\\
\end{eqnarray}
where both $\lambda_1$ and $\lambda_2$ must be understood as a fundamental fermion in $SU(2)$, i.e., a $2\times 1$ column vector:
\begin{eqnarray}
\lambda_{\alpha}=\left[\begin{array}{c}\lambda_{\alpha,1}\\\lambda_{\alpha, 2}\end{array}\right]\,, 
\end{eqnarray}
and $\alpha=1,2$ are the up or down spinor indices. The solution of (\ref{equation for lambdasu2}) is given by setting
\begin{eqnarray}\label{su2intm}
\lambda_1={\cal G}_1(r)|j,m'=j,m\rangle\,, \quad \lambda_2={\cal G}_2(r)|j,m'=-j,m\rangle\,,\quad |m|\leq j\,,
\end{eqnarray}
where both ${\cal G}_1$ and ${\cal G}_2$ are $2\times 1$ column vectors.  Substituting (\ref{su2intm}) into (\ref{equation for lambdasu2}), we obtain
\begin{eqnarray}\label{equation for lambdasu2final}\nonumber
\left[ \left(\frac{1}{f}\frac{\partial}{\partial r}+\frac{2f}{r}+\frac{1}{fr}\right)I_{2}+\frac{a^2m_{(2)}f}{r^3}\left[\begin{array}{cc}\frac{1}{2}&0\\0&-\frac{1}{2}\end{array}\right]+\frac{2fj}{r}I_{2}\right]{\cal G}_1&=&0\,,\\
\left[ \left(\frac{1}{f}\frac{\partial}{\partial r}+\frac{2f}{r}+\frac{1}{fr}\right)I_{2}+\frac{a^2m_{(2)}f}{r^3}\left[\begin{array}{cc}-\frac{1}{2}&0\\0&\frac{1}{2}\end{array}\right]+\frac{2fj}{r}I_{2}\right]{\cal G}_2&=&0\,.
\end{eqnarray}
For $m_{(2)} > 0$, normalizable zero modes arise in the lower color-space component of $\lambda_{1}$ and in the upper color-space component of $\lambda_{2}$. For a given $m_{(2)}>0$, there exists a finite set of normalizable states with total angular momentum $j\geq 0$ and magnetic angular momentum $m$ satisfying
\begin{eqnarray}
m_{(2)}>4j\,, \quad |m|\leq j\,,\quad\mbox{for}\, j=0,\frac{1}{2},1,...
\end{eqnarray}
However, as in the $U(1)$ case, not all normalizable modes possess a spin structure upon parallel transport along closed paths. Under parallel transport from the point $(r\approx a,\theta=0,\varphi=0,\psi)$ to $(r\approx a,\theta=0,\varphi=0,\psi+2\pi)$, the Dirac spinor acquires a $SU(2)$ holonomy:
\begin{eqnarray}\label{su2holo}\nonumber
\Psi(r,\theta,\varphi,\psi+2\pi)=\exp\left[i \int_{\ell}A_{SU(2)}\right]\Psi(r,\theta,\varphi,\psi)=\exp\left[\begin{array}{cc}\frac{i\pi m_{(2)}}{2}&0\\0&-\frac{i\pi m_{2}}{2}\end{array}\right]\Psi(r,\theta,\varphi,\psi)\,.\\
\end{eqnarray}
On the other hand, assuming that a zero mode of $\Psi$ exists, then one can use the solution to directly compute
\begin{eqnarray}
\Psi(r,\theta,\varphi,\psi+2\pi)=e^{-i 2\pi j}\Psi(r,\theta,\varphi,\psi)\,,
\end{eqnarray}
where $j$ are integers or half-integers.
A consistent zero mode can be defined provided that the phase on the r.h.s. of (\ref{su2holo}) matches the phase $e^{-i 2\pi j}$.
From this, we immediately observe that:

{\bf Case I}: when $m_{(2)}$ is odd, $m_{(2)}=2p+1$, $p\geq 0$ (this includes the minimal value $m_{(2)}=1$), the holonomy in (\ref{su2holo}) yields a $\pm i$ and there exists no angular momentum $j$ that matches the holonomy. We conclude that there exists no globally well-defined fermion zero-mode in the background of a $\mathbb Z_2^{(1)}$ center flux. The APS index yields a fraction. 

When $m_{(2)}$ is even, then the fermions are well-defined in the background of the gauge field. However, we must distinguish between two cases.

{\bf Case II}: $m_{(2)}= 4p$, $p\geq 0$. Here, the gauge holonomy yields a trivial phase, and $j$ must be an integer. Setting $m_{(2)}/2={\cal C}=2p$, as we did in the $U(1)$ case, we immediately conclude that there are $p^2$ zero modes in the lower color-space component of $\lambda_{1}$ and $p^2$ zero modes in the upper color-space component of $\lambda_{2}$, giving a total of
 \begin{equation}
 {\cal I}_\Box=2p^2
 \end{equation}
 zero modes. In applying the APS index (\ref{indexmaster}), we point out that the signature of the holonomy $\mbox{sgn(Hol)}=2$. Using $T_\Box=1$, $\mbox{dim}_\Box=2$, the APS index then gives the same result. 
 
  {\bf Case III}:  $m_{(2)}=4p+2$, $p\geq 0$. Here, the gauge holonomy yields an $e^{i\pi}$ phase, which forces the angular momentum $j$ to be a half-integer. Recalling the $U(1)$ case, setting $m_{(2)}/2={\cal C}=2p+1$, we find that there are $p^2+p$ zero modes in the lower color-space component of $\lambda_{1}$ and $p^2+p$ zero modes in the upper color-space component of $\lambda_{2}$. In total, we have
   \begin{equation}
 {\cal I}_\Box=2p(p+1)
 \end{equation}
  zero modes. In applying the APS index (\ref{indexmaster}), we point out that the signature of the holonomy $\mbox{sgn(Hol)}=-2$. The APS index then gives the same result.

\subsection{Adjoint fermions}
Next, we turn to the adjoint fermions. The Dirac equation we need to solve is
\begin{eqnarray}
 \gamma^a E_a^\mu \left( \partial_\mu + \frac{1}{2}\Omega_\mu + i [A_{SU(2)\mu}, \quad ]\right)\Psi=0\,.
\end{eqnarray}
To this end, we expand $\Psi$ in the Cartan-Weyl basis:
\begin{eqnarray}
\Psi=\Psi_{H}H+\Psi_+E_++\Psi_-E_-\,,
\end{eqnarray}
from which we find
\begin{eqnarray}
 [A_{SU(2)\mu},\Psi]=m_{(2)}\Psi_+E_+-m_{(2)}\Psi_-E_-\,.
\end{eqnarray}
Then, as before, we find that the right-hand component of $\Psi$ does not yield any normalizable zero modes, while the left-handed component gives the system of equations
\begin{eqnarray}\label{equation for lambdaadj1}\nonumber
\left[ \frac{1}{f}\frac{\partial}{\partial r}+\frac{2f}{r}+\frac{1}{fr}+\frac{a^2m_{(2)}f}{r^3}+\frac{2f}{r}\hat{\cal J}_z\right]\lambda_{1+}-\frac{2}{r}\hat{\cal J}_-\lambda_{2+}&=&0\,,\\
-\frac{2}{r}\hat{\cal J}_+\lambda_{1+}+\left[ \frac{1}{f}\frac{\partial}{\partial r}+\frac{2f}{r}+\frac{1}{fr}-\frac{a^2m_{(2)}f}{r^3}-\frac{2f}{r}\hat{\cal J}_z\right]\lambda_{2+}&=&0\,,
\end{eqnarray}
\begin{eqnarray}\label{equation for lambdaadj2}\nonumber
\left[ \frac{1}{f}\frac{\partial}{\partial r}+\frac{2f}{r}+\frac{1}{fr}-\frac{a^2m_{(2)}f}{r^3}+\frac{2f}{r}\hat{\cal J}_z\right]\lambda_{1-}-\frac{2}{r}\hat{\cal J}_-\lambda_{2-}&=&0\,,\\
-\frac{2}{r}\hat{\cal J}_+\lambda_{1-}+\left[ \frac{1}{f}\frac{\partial}{\partial r}+\frac{2f}{r}+\frac{1}{fr}+\frac{a^2m_{(2)}f}{r^3}-\frac{2f}{r}\hat{\cal J}_z\right]\lambda_{2-}&=&0\,,
\end{eqnarray}
and 
\begin{eqnarray}\label{equation for lambdaadjH}\nonumber
\left[ \frac{1}{f}\frac{\partial}{\partial r}+\frac{2f}{r}+\frac{1}{fr}+\frac{2f}{r}\hat{\cal J}_z\right]\lambda_{1H}-\frac{2}{r}\hat{\cal J}_-\lambda_{2H}&=&0\,,\\
-\frac{2}{r}\hat{\cal J}_+\lambda_{1H}+\left[ \frac{1}{f}\frac{\partial}{\partial r}+\frac{2f}{r}+\frac{1}{fr}-\frac{2f}{r}\hat{\cal J}_z\right]\lambda_{2H}&=&0\,,
\end{eqnarray}
Since the Cartan components $\lambda_{1,2H}$ do not couple to the background gauge field, they cannot yield any zero modes. On the other hand, the solutions of both $\lambda_{1,2\pm}$ take the form
\begin{eqnarray}\label{su2int}
\lambda_{1\pm}=g_{1\pm}(r)|j,m'=j,m\rangle\,, \quad \lambda_{2\pm}=g_{2\pm}(r)|j,m'=-j,m\rangle\,,\quad |m|\leq j\,,
\end{eqnarray}
which simplifies (\ref{equation for lambdaadj1}, \ref{equation for lambdaadj1}) to
\begin{eqnarray}\nonumber
\left[ \frac{1}{f}\frac{\partial}{\partial r}+\frac{2f}{r}+\frac{1}{fr}\pm\frac{a^2 m_{(2)}f}{r^3}+\frac{2f j}{r}\right]g_{1\pm}&=&0\,,\\
\left[ \frac{1}{f}\frac{\partial}{\partial r}+\frac{2f}{r}+\frac{1}{fr}\mp\frac{a^2 m_{(2)}f}{r^3}+\frac{2f j}{r}\right]g_{2\pm}&=&0\,.
\end{eqnarray}
From the above treatment, it is clear that the structure of these equations reveals the existence of normalizable zero modes for $\lambda_{1-}$ and $\lambda_{2+}$. For a given $m_{(2)}$, the normalizable states are designated by $j$ and $m$ such that
\begin{equation}
j = 0, \frac{1}{2}, 1, \ldots, \frac{m_{(2)} - 1}{2}\,, \quad |m|\leq j.
\end{equation}
To see whether such zero modes are consistent with a spin structure, we need to study the parallel transport of the adjoint fermion between two points on the bolt at angles $\psi$ and $\psi+\pi$, which is given by the gauge transformation
\begin{eqnarray}
\Psi(r,\theta,\varphi,\psi+2\pi)=U^\dagger \Psi(r,\theta,\varphi,\psi) U\,,
\end{eqnarray}
where $U=\exp\left[i\int_{\ell} A_{SU(2)}\right]$, and $\ell$ is the path connecting two points on the bolt at $\psi$ and $\psi+2\pi$. A simple calculation shows that $\Psi$ acquires a phase $e^{i \pi m_{(2)}}$ upon parallel transporting, meaning that for $m_{(2)}$ even, we need to restrict $j$ to integer values, while for odd $m_{(2)}$, we need to restrict $j$ to half integers.  Comparing with the analysis of the $U(1)$ field, we find that the two cases are

{\bf Case I}: When $m_{(2)} = 2p+1$ is odd, by analogy with the $U(1)$ case, we obtain
\begin{eqnarray}
{\cal I}_{\text{adj}} = 2p(p+1)\,.
\end{eqnarray}
In applying the APS index (\ref{indexmaster}), we point out that the signature of the holonomy $\mbox{sgn(Hol)}=-1$. Using $T_{\text{adj}}=4$, $\mbox{dim}_{\text{adj}}=3$, the APS index then gives the same result. 

{\bf Case II}: $m_{(2)}=2p$ is even:
\begin{eqnarray}
{\cal I}_{\scriptsize\mbox{adj}}=2p^2\,,
\end{eqnarray}
also consistent with the calculations of the APS index.
In particular, for $m_{(2)}=2$, a background gauge field carrying a topological charge $Q_{(2)}=\frac{1}{2}$, there are $2$ adjoint zero modes.

%%%%%%%%%%%%%%%%%%%%%%%%%%
\section{Phase-space path integral}
\label{Phase space path integral}
%%%%%%%%%%%%%%%%%%%%%%%%%%

Even though the path integral \eqref{mainEHPI} captures the essential physics, it requires further justification and clarification. It is well known that the temporal component of the gauge field, $A_t$, is not dynamical; rather, it enters the action as a Lagrange multiplier. To enforce this point,  we perform the path integral in phase space coordinates $A_i, \Pi_i$, the gauge field and its conjugate momentum. To this end, one foliates space into hypersurfaces $\Sigma$ using the Arnowitt--Deser--Misner  (ADM) formalism \cite{Arnowitt:1962hi}. Essentially, applying this formalism, one uses the Hamiltonian prescription in a  Lorentzian signature space. Yet, EH spaces have no well-defined continuation into Minkowski space. One then might want to adapt a Euclidean version of the ADM formalism. Yet, doing so, one faces a problem with the signs of the quadratic terms $(\nabla \times\bm A)^2$ and $\Pi_i^2$. 

We were not able to find a totally satisfactory fix to this problem. However, one way out is to use the standard Lorentzian signature, pretending that we are not performing the calculations specifically in the background of a self-dual solution. Only after performing the path integral over $\Pi_i$, do we analytically continue to the Euclidean space.

The ADM metric foliation is given by \cite{Arnowitt:1962hi}
\begin{equation}
ds^2 = -N^2 dt^2 + h_{ij}\big(dx^i + N^i dt\big)\big(dx^j + N^j dt\big)\,,
\end{equation}
where $N(\bm x,t)$ and $N^i(\bm x,t)$ are the lapse and shift functions (with $i,j,k\in\{x,y,z\}$) and $h_{ij}(\bm x,t)$ is the induced metric on the hypersurface $\Sigma$. The path integral \eqref{mainEHPI}, expressed in terms of the spatial gauge field $A_i$, its conjugate momentum $\Pi^i$, and the non-dynamical field $A_t$, then takes the form\footnote{For a standard formulation of Maxwell theory in curved space using the ADM formalism, see, e.g.,~\cite{Jha:2022svf}.} 
\begin{equation}\label{mainEHPIHam}
\begin{aligned}
\langle \Omega|\mathrm{EH}\rangle_{t_0}^{m_{(3)}, m_{(2)},{\cal C}} \;\sim\; 
&\int [\mathcal D A_i]\,[\mathcal D \Pi^i]\,[\mathcal D A_t]\;
\Bigg|_{\substack{
A_i(\bm x)|_{\Sigma_{t_0}}=A_i(\bm x) \\
A_i(-\bm x)|_{\Sigma_{t_0}}=-A_i(\bm x)
}}
\\
&\times 
\exp\!\left[i
\int_{t_0}^\infty dt \int_\Sigma d^3x 
\;\big(\Pi^i \partial_t A_i - \mathcal H(A_i,\Pi^i,A_t)\big)
\right],
\end{aligned}
\end{equation}
where $\mathcal H$ denotes the Hamiltonian density,
\begin{equation}\label{hameh}
\begin{aligned}
\mathcal H \;&=\; 
N\left(
\frac{1}{2\sqrt{h}}\,\Pi^i\Pi_i 
+ \frac{1}{4}\sqrt{h}\,F_{ij}F^{ij}
\right)
+ N^i \Pi^j F_{ij} 
+ \Pi^i \nabla_i A_t, 
\\[6pt]
\Pi^i \;&=\; 
-\frac{1}{N}\,\sqrt{h}\,
\Big(F_{kt} - N^j F_{jk}\Big)\,h^{ki},
\end{aligned}
\end{equation}
$\Sigma_{t_0}$ denotes the initial surface at $t=t_0$, and $\nabla_i$ is the covariant derivative on $\Sigma$. 
For notational simplicity, in the action we use $(A_i,\Pi^i,A_t)$ to denote the sum of the background fields and fluctuations, whereas in the path integral measure they should be understood (at this level) as the fluctuations around the chosen background. We are also setting the gauge field coupling constants $g_1=g_2=g_3=1$, as we are not trying to perform detailed calculations.

From \eqref{hameh}, it is clear that $A_t$ carries no conjugate momentum and therefore acts as a Lagrange multiplier. 
Integrating the last term in $\mathcal H$ by parts and discarding the boundary contribution\footnote{All boundary topological data are in fact encoded in the holonomy $e^{i\int_{\ell_\infty}A}$ on $\mathbb{RP}^3$, as discussed extensively in the bulk of the paper.}, 
the path integral over $A_t$ produces a functional Dirac delta,
\begin{equation}
\delta\!\left(\nabla_i \Pi^i\right),
\end{equation}
which enforces Gauss’ law, $\nabla_i \Pi^i=0$, as a constraint within the path integral.

Far from the bolt, the Eguchi–Hanson space asymptotes to flat space (modulo a $\mathbb{Z}_2$ identification), so we can set
$
N \simeq 1, \quad N^i \simeq 0, \quad h_{ij} \simeq \delta_{ij},
$
and hence $\Pi^i \simeq F^{it}$ corresponds to the electric field. 

At this stage, the path integral over $\Pi^i$ can be performed directly since the action is quadratic in $\Pi^i$\footnote{We, however, note that $\Pi^i$ contains both background and fluctuation contributions, and only the fluctuation part is integrated in the Gaussian approximation.}. This is particularly simple in the asymptotic flat region, $N\simeq1$, $N^i\simeq0$, $h_{ij}\simeq \delta_{ij}$; see, e.g., \cite{Weinberg:1995mt}. Now, one continues from Minkowski to the Euclidean space, performing the path integral over $A_\mu$ in the Euclidean sense.  We note that the electric field satisfies 
$
F_{it}(\bm x, t) = F_{it}(-\bm x, t),
$
and therefore is continuous across the surfaces $x_i=0$. There is no induced charge or current from the antipodal identification, as expected.

%\bigskip

  \bibliography{RefEHB.bib}
  
  \bibliographystyle{JHEP}
  \end{document}